\documentclass[]{aa}
\usepackage[arrow=cyan]{callouts}
\usepackage{amsmath}
\usepackage{txfonts} 
\usepackage{graphicx}
\usepackage{color}
\usepackage{bm}
\usepackage{natbib}
\usepackage{url}
\usepackage{tabularx}
\usepackage{hhline}
\usepackage{arydshln}
\usepackage{lscape}
\usepackage{longtable}
\usepackage{array}
\usepackage{rotating}
\usepackage{afterpage}
\usepackage{lipsum}
\usepackage{hyperref}

\newcommand{\msun}{$M_\odot$}

\newcommand{\mic}{$\mu$m}

\newcolumntype{R}[1]{>{\raggedleft\arraybackslash }b{#1}}
\newcolumntype{L}[1]{>{\raggedright\arraybackslash }b{#1}}
\newcolumntype{C}[1]{>{\centering\arraybackslash }b{#1}}

\newlength{\pointwidth}
\begin{document}

\title{Observational correlation between magnetic field, angular momentum, and fragmentation in the envelopes of Class 0 protostars?}

\author{Maud Galametz 
               \inst{1},
              Ana\"elle Maury
               \inst{1,2},
              Josep M. Girart 
               \inst{3,4},
              Ramprasad Rao 
               \inst{2},
              Qizhou Zhang
               \inst{2},\\
              Mathilde Gaudel
               \inst{5},
              Valeska Valdivia
               \inst{1}, 
                Patrick Hennebelle
                \inst{1},
                Victoria Cabedo-Soto
                \inst{1},
                Eric Keto
                \inst{2},
                Shih-Ping Lai
                \inst{6}
}

\institute{
Astrophysics department, CEA/DRF/IRFU/DAp, Universit\'{e} Paris Saclay, UMR AIM, F-91191 Gif-sur-Yvette, France, \\
\email{maud.galametz@cea.fr} 
\and
Center for Astrophysics $\vert$ Harvard \& Smithsonian, 60 Garden street, Cambridge, MA 02138, USA 
\and
Institut de Ci\`encies de l'Espai (ICE, CSIC), Can Magrans, S/N, E-08193 Cerdanyola del Vall\`es, Catalonia, Spain
\and
Institut d'Estudis Espacials de de Catalunya (IEEC), E-08034 Barcelona, Catalonia, Spain
\and
LERMA, Observatoire de Paris, 61 av. de l'Observatoire F-75014 Paris, France
\and 
Institute of Astronomy and Department of Physics, National Tsing Hua University, Hsinchu 30013, Taiwan
}
    
\abstract
{}
{Our main goal in this analysis is to assess the potential role of magnetic fields in regulating the envelope rotation, 
the formation of disks, and the fragmentation of Class 0 protostars in multiple systems.}
{We used the Submillimeter Array to carry out observations of the dust polarized emission at 0.87 mm in the envelopes of a large sample of 
20 Class 0 protostars. We estimated the mean magnetic field orientation over the central 1000 au envelope scales to characterize the 
orientation of the main component of the organized magnetic field at the envelope scales in these embedded protostars. This direction was 
compared to that of the protostellar outflow in order to study the relation of their misalignment and the kinematics of the circumstellar gas. 
The latter is traced through the velocity gradient that is observed in the molecular line emission (mainly N$_2$H$^+$) of the gas at intermediate envelope scales.}
{We discover that the misalignment of the magnetic field orientation is strongly related to the outflow and the amount of 
angular momentum observed at similar scales in the protostellar envelope. This reveals a potential link between the kinetic and the 
magnetic energy at envelope scales. The relation could be driven by favored B-misalignments in more dynamical envelopes or by 
a dependence of the envelope dynamics on the initial large-scale B configuration. By comparing the trend with the presence of 
fragmentation, we observe that single sources are mostly associated with conditions of low angular momentum in the inner 
envelope and good alignment of the magnetic field with protostellar outflows at intermediate scales. Our results suggest that 
the properties of the magnetic field in protostellar envelopes are tightly related to the rotating infalling gas that is directly 
involved in the formation of stars and disks: we find that it may not only affect the fragmentation of 
protostellar cores into multiple stellar systems, but also sets the conditions that establish the pristine properties of planet-forming disks.}
{}


\authorrunning{M. Galametz et al}
\titlerunning{Polarized-dust emission in solar-type Class 0 protostars}
\maketitle

%%%%%%%%%%%%%%%%%%%%%%%%%%%%%%%%%%%
%       Introduction                            %
%%%%%%%%%%%%%%%%%%%%%%%%%%%%%%%%%%%

\section{Introduction} 

A majority of stars in our Galaxy are found in multiple stellar systems, and a significant fraction of solar-type stars will host planetary systems \citep{Duchene2013,Danley2019}. 
Most of the final stellar mass is collected during a short but vigorous accretion phase. 
During this so-called protostellar phase, the star forms at the center of an infalling rotating core, concomitantly with a surrounding disk of gas in circular orbits around the star:
while the star will inherit the majority of the accreted mass, most of the angular momentum contained in the protostellar envelope is expected to be expelled or stored in the 
protostellar disk. This evolution will eventually lead to protoplanetary systems \citep{Zhao2020}. 
Class 0 objects are the youngest accreting protostars and are surrounded by a dense envelope that is accreted onto the central protostellar embryo during a short (t $<$ 5$\times$10$^4$ yr) 
accretion phase \citep{Andre2000,Evans2009}. Characterizing the dynamics of the gas and the physical processes of these youngest protostars is crucial for understanding the 
efficiency of the star formation process, the global properties of stars in our Galaxy, and the setting of the conditions that allow disks and planets to form around them. 

Magnetic fields (hereafter B) are ubiquitous in the Universe \citep{Vallee2004} and have been observed to permeate the interstellar material deep down into star-forming cores and 
protostellar environments \citep{Girart2006, Hull2019}. From a theoretical point of view, the presence of B in star-forming cores has been shown to significantly alter the dynamics of 
the gas participating in the building of stars during the accretion phase, and it affects the resulting properties of these stars and associated circumstellar disks \citep{Terebey1984,Wurster2018,HennebelleInutsuka2019}. 
The mechanism for evacuating angular momentum from the infalling gas through magnetic torques applied by Alfv\'{e}n waves is called magnetic braking. 
The initial global collapse, driven by gravity, drags the field lines, which leads to an hourglass morphology of the field lines and amplifies the magnetic intensity. 
The magnetic braking seems to be particularly enhanced by the pinching of field lines, which 
lengthens the magnetic lever arms and efficiently transports the angular momentum from the inner envelope toward the outer one \citep{Galli2006}. 
As less angular momentum is transported toward the forming star, only small protoplanetary disks form while the star grows \citep{Allen2003,HennebelleFromang2008,Masson2016,Hirano2019}. 

The importance of the misalignment between magnetic field and rotation axis has been stressed by several authors
\citep{CiardiHennebelle2010,Joos2012,GrayMcKee2018}. They found that in ideal magnetohydrodynamic (MHD) calculations, only small disks form or even no disk forms in the aligned 
configurations when the field is strong enough; and it is comparatively far easier to form a disk in the misaligned case. 
Numerical simulations taking into account non-ideal MHD effects (such as ambipolar diffusion or Hall effect) were 
able to overcome the magnetic braking catastrophe, leading to the formation of disks similar to those observed \citep{Hennebelle2016,Zhao2018,Wurster2019}. 
Studies by \citet{Hennebelle2020} or \citet{Wurster2020} appear to also predict that the 
misalignment of the magnetic field with the envelope rotation axis directly affects the protostellar disk formation, for instance, leading to the formation of 
larger planet-forming disks in the misaligned cases that were investigated compared to the smaller disks that are observed in aligned cases. This is particularly clear when the field 
intensity is such that the mass-to-flux ratio is on the order of 10. Another key prediction of magnetized models is that a strong, organized B partly alters the 
ability of the envelope to fragment. This suggests that B is one of the regulating agents that drive the birth of the multiple stellar systems we commonly 
observe in the Galaxy \citep{HennebelleTeyssier2008}.

%%%%%%% Sample %%%%%%%%%%%%%%%%%%%%%%%%%%%%%%%%%%%%%%%%%%%%%%%%
\begin{table*}
\centering
\caption{Characteristics of the full sample.}
\begin{tabular}{ccccccc}
\hline
\hline
\vspace{-5pt}
&\\
Name    & $\alpha$ (J2000) & $\delta$ (J2000) & Cloud & Distance & M$_{env}$ & References\\%& Observation \\
                &&&          & (pc) && \\%& Date\\
\hline
\vspace{-5pt}
&\\
Per B1-bS$^a$                   & 03:33:21.35   & +31:07:26.4 & Perseus / Barnard & 230 & 3.0 & [2, 12] \\ %& 20181127 \\
Per B1-c$^a$                    & 03:33:17.88   & +31:09:32.0 & Perseus / Barnard & 230 & 2.1 & [2, 11]\\ %& 20181127, 20181206\\
B335                                    & 19:37:00.90   & +07:34:09.6 & isolated        & 100 & 1.3$^b$ & [3, 15]\\
BHR7-MMS$^a$            & 08:14:23.33   & -34:31:03.7 & Gum / Vela              & 400 & 1.0 & [5, 13]\\ %& 201811(27,28,29), 20181205\\
CB230                                   & 21:17:40.00   & +68:17:32.0 & Cepheus         & 352 & 3.4 & [4, 14]\\
HH25-MMS$^a$            & 05:46:07.40   & -00:13:43.4 & Orion / L1630   & 400 & 0.5 & [6, 16]\\ %& 20181207\\
HH211-mm$^a$            & 03:43:56.52   & +32:00:52.8 & Perseus / IC348   & 320 & 1.5 & [1, 16]\\ %& 201811(28-29), 20181205\\
HH212$^*$                       & 05:43:51.40   & -01:02:53.0 & Orion / L1630         & 400 & 0.2 & [6, 17]\\ %& 20181129, 201812(05,06)\\
HH797                                   & 03:43:57.10   & +32:03:05.6 & Perseus / IC348   & 320 & 1.1 & [1, 19]\\
IRAS03282                       & 03:31:20.40   & +30:45:24.7 & Perseus         & 293 & 2.2 & [1, 18]\\
IRAS16293-A                     & 16:32:22.9    & -24:28:36.0 & Ophiuchus       & 150 & 2.3 & [8, 16] \\
L1157                                   & 20:39:06.3    & +68:02:15.8 & Cepheus         & 352 & 3.0 & [4, 10]\\
L1448C                                  & 03:25:38.9 & +30:44:05.4 & Perseus            & 293 & 2.0 & [1, 10]\\
L1448N-B                        & 03:25:36.3 & +30:45:14.9   & Perseus          & 293 & 4.8 & [1, 10]\\
L1448-2A                                & 03:25:22.4    & +30:45:13.0 & Perseus         & 293 & 1.9 & [1, 10]\\
L483-mm$^a$                     & 18:17:29.94   & -04:39:39.3 & Serpens Cirrus  & 250 & 1.8 & [7, 18]\\ %& 201904(15-16)\\
NGC~1333 IRAS4A                 & 03:29:10.5    & +31:13:31.0 & Perseus         & 293 & 12.3 & [1, 10]\\
NGC~1333 IRAS4B         & 03:29:12.0    & +31:13:08.0 & Perseus         & 293 & 4.7 & [1, 10]\\
Serpens South MM18$^a$ & 18:30:04.12 & -02:03:02.55 & Serpens South & 350 & 5  & [9, 10]\\ %& 201904(15-16)\\
SVS13-B                                 & 03:29:03.1    & +31:15:52:0 & Perseus         & 293 & 2.8 & [1, 10]\\
\vspace{-5pt}
&\\
\hline
\end{tabular}
\begin{list}{}{}
\vspace{5pt}
\item[$^a$] Sources whose SMA polarization observations are described in this paper. 
The polarization results for the remaining sources are described in \citet{Galametz2018}, \citet{Girart2014}, and \citet{Rao2009}.
\item[$^b$] The total globule mass is probably a factor of 3-5 higher \citep{Stutz2008}.
\vspace{5pt}
\item[References for the distances and M$_{env}$ -]
[1] \citet{Ortiz-Leon2018},
[2] \citet{Cernis2003},
[3] \citet{OlofssonOlofsson2009},
[4] \citet{Zucker2019},
[5] \citet{Woermann2001},
[6] \citet{Anthony-Twarog1982}
[7] \citet{Herczeg2019}
[8] \citet{Ortiz-Leon2018_2},
[9] Ongoing work re-analyzing the Gaia data toward Serpens South suggests a high extinction layer up to distances of 350 pc (Palmeirim, Andr\'{e} et al. in prep). We use this reevaluated distance.
[10] \citet{Maret2020},
[11] \citet{Matthews2006},
[12] \citet{Andersen2019},
[13] \citet{Tobin2018},
[14] \citet{Massi2008},
[15] \citet{Launhardt2013},
[16] \citet{Andre2000},
[17] \citet{Wiseman2001},
[18] \citet{Tobin2011},
[19] \citet{Sadavoy2014}.
\end{list}
\label{Sources}
\end{table*}
%%%%%%% Sample %%%%%%%%%%%%%%%%%%%%%%%%%%%%%%%%%%%%%%%

The effect of the various characteristics of the B~field (orientation with the collapse direction and strength), however, is poorly quantified observationally speaking. 
Only a few studies have attempted to test the predicted relation that links the B-field orientation in protostellar cores to the magnitude of the angular momentum of the gas responsible for 
disk properties and the formation of multiple stellar systems \citep[see, e.g., the works of][]{Chapman2013,Hull2013,Zhang2014}. 
Because it is difficult to trace B in these small embedded astrophysical structures, it has so far been difficult to reach the statistical significance that 
would allow us to draw firm conclusions about the role of magnetic braking in the formation of stars and disks \citep{Yen2015,Maury2018}. 
In order to statistically investigate the B-field orientation, we carried out a SubMillimeter Array (SMA) survey of 20 low-mass Class 0 protostars, using 0.87 mm polarized dust emission. Because asymmetric dust particles of the interstellar medium align themselves with their minor axis parallel to the B-field lines \citep{Andersson2015}, the 
observed polarized angle provides us with a robust proxy of the direction perpendicular to the magnetic field orientation.  
Class 0 objects were ideally suited for this analysis because most of the mass that collapses onto the central embryo still resides in the envelope, allowing us to trace the 
B orientation at envelope (1000-2000 au) scales. 
\citet{Galametz2018} presented results for a first subsample of 12 sources, focusing on the properties of polarization fractions and general alignment 
between B and the outflow at envelope scales. We reported the detection of linearly polarized dust emission in all the objects of the sample. By comparing the 
B orientation with that of the outflow axis, which is commonly used as a proxy for the rotational axes of these systems, we noted that 
at the scales traced in our analysis, the B-field lines were preferentially misaligned in sources for which large equatorial 
velocity gradients were reported in the literature. A potential link between the envelope dynamics and the B orientation 
might be an additional signature that B has a strong effect on the collapse and fragmentation, as has also been suggested by the 
analysis of massive cores by \citet{Zhang2014}. 

%%%%%%% Details on the Observations %%%%%%%%%%%%%%%%%%%%%%%%%%%%%%%%%%%
\begin{table*}
\centering
\caption{Details of the observations $^{a}$}
\begin{tabular}{ccccc}
\hline
\hline
\vspace{-5pt}
&\\
Date   & Gain           & Polarization & Antenna & Flux cal.             \\
       & calibrator     & calibrator   & used    & scaling factor $^{b}$         \\
\vspace{-5pt}
&\\
\hline
\vspace{-5pt}
&\\
Nov 27 2018     & 1925+211, 0336+323, 3C 84, 0747-331   & 3C 454.3, 3C 84 
&       1, 2, 3, 4, 5, 6, 7, 8   & 1.7 \\ %& Neptune 
Nov 28 2018     & 0336+323, 3C 84, 0747-331 & 3C 454.3, 3C 84 
& 2, 3, 4, 5, 6, 8  & 1.5 \\ %& Neptune 
Nov 29 2018     & 0336+323, 0747-331, 0607-085, 0532+075 & 3C 454.3 3C 84 
& 3, 4, 5, 6, 7, 8  & 1.4 \\ %& Neptune 
Dec 05 2018     & 0336+323, 0607-085, 0532+075, 0747-331 & 3C 84, 3C 279, 3C 454.3 
& 1, 3, 4, 5, 6, 8  & 1.2  \\ %& Neptune 
Dec 06 2018     & 0336+323, 0607-085, 0532+075 & 3C 454.3, 3C 279 
& 1, 3, 4, 5, 6, 8  &  1.3 \\ %& Neptune 
Dec 07 2018     & 0532+075, 0607-085 & 3C 279 
& 1, 2, 3, 4, 5, 6, 8  & 1.4 \\  %-
Apr 15 2019     & 1733-130, 1751+096, 1924-292, mwc349a & 3C 279, 3C 273 
&       1, 2, 3, 5, 7, 8  & 0.9 \\ %& Callisto 
Apr 16 2019     & 1733-130, 1751+096 & 3C 279, 3C 273 
& 1, 2, 3, 5, 7, 8 & 0.9  \\ %& Callisto 
\vspace{-5pt}
&\\
\hline
\end{tabular}
\begin{list}{}{}
\item[$^a$] Details of the observations for the first half of the sample are presented in \citet{Galametz2018}.
\item[$^b$] Scaling factors derived by comparing the quasars observed with the SMA with their fluxes at similar dates in the ALMA calibration source catalog.
\end{list}
\label{Observations}
\end{table*}
%%%%%%% Observations %%%%%%%%%%%%%%%%%%%%%%%%%%%%%%%%%%%%%%%%%%%%%%%%%%

%%%%%%% Characteristics of the final SMA maps %%%%%%%%%%%%%%%%%%%%%%%%
\begin{table*}
\centering
\caption{Characteristics of the SMA maps $^{a}$}
\begin{tabular}{ccccccccccc}
\hline
\hline
\vspace{-5pt}
&\\
Name            &       \multicolumn{2}{c}{Synthesized beam}    & \multicolumn{3}{c}{rms}  && \multicolumn{3}{c}{in the 0.87mm reconstructed map} \\

\cline{4-6} 
\cline{8-10} 
\vspace{-5pt}
&\\
&&& I $^{b}$    &       Q       & U             && Peak intensity & Peak P$_i$ & p$_{frac}$$^c$  \\      
&&& \multicolumn{3}{c}{(mJy/beam)} && (Jy/beam) & (mJy/beam) & \% \\
\vspace{-5pt}
&\\
\hline
\vspace{-5pt}
&\\
Per B1-bS & 2\farcs1$\times$1\farcs2 & \hspace{-10pt}(-54$\degr$)   &   
4.0 & 1.2 & 1.3 &&
0.46 & 5.1 & 5.3\\
Per B1c & 1\farcs8$\times$1\farcs3 & \hspace{-10pt}(-58$\degr$)   &     
4.0     & 2.3 & 2.4 &&
0.44 & 8.2 & 7.5        \\
BHR7-MMS& 2\farcs7$\times$1\farcs3 & \hspace{-10pt}(-29$\degr$)   &     
2.5 & 1.1       & 1.1 &&
0.53 & 4.5 & 0.8        \\ 
HH25-MMS& 1\farcs7$\times$1\farcs5 & \hspace{-10pt}(-67$\degr$)   &     
3.4 & 1.9       & 1.9 &&
0.25 &6.2 & -   \\
HH211-mm& 1\farcs5$\times$1\farcs4 & \hspace{-10pt}(71$\degr$)   &      
1.2 & 0.7       & 0.7    &&
0.17 &2.3 & 3.3 \\ 
HH212& 1\farcs$7\times$1\farcs3 & \hspace{-10pt}(-81$\degr$)   &        
0.6 & 0.2       & 0.2 &&
0.19 & 2.1 & 3.0        \\
L483-mm& 1\farcs$9\times1$\farcs5 & \hspace{-10pt}(30$\degr$)   &       
1.6 & 1.3       & 1.2 &&
0.10 & 2.7 & 13.6       \\ 
Serpens SMM18& 1\farcs$9\times$1\farcs5         & \hspace{-10pt}(34$\degr$)   &  
11.3    & 1.9   & 2.0 &&
0.86 &19.4 & 3.8        \\
\vspace{-5pt}
&&&&\\
\hline
\end{tabular}
\begin{list}{}{}
\item[$^a$] Details of the SMA maps for the first half of the sample are presented in \citet{Galametz2018}.
\item[$^b$] after self-calibration.
\item[$^c$] Mean polarization fraction defined as the unweighted ratio between the mean polarization over total flux.
\end{list}
\label{SMAmapsrms}
\vspace{-20pt}
\end{table*}
%%%%%% Characteristics of the final SMA maps  %%%%%%%%%%%%

We complement the \citet{Galametz2018} observations with eight additional Class 0 envelopes observed with the SMA at 0.87mm
at comparable scales. The full SMA B measurements are combined with gas kinematics information obtained 
homogeneously from N$_2$H$^+$ observations of velocity gradients in the envelopes, either from the Continuum And 
Lines in Young ProtoStellar Objects survey \citep[CALYPSO;][seven sources]{Maury2019,Gaudel2020} or from observations published 
in the literature that we reprocessed when required (12 sources, see Table \ref{AnglesVelGrad}).
Our goal is to observationally test the theoretical predictions of the conditions required for magnetic braking affecting 
the collapse and assess the potential role of B in regulating the matter infall and envelope rotation, the formation 
of disks, and the fragmentation into multiple systems.

%%%%%%%%%%%%%%%%%%%%%%%%%%%%%%%%%%%
%       Observations                            %
%%%%%%%%%%%%%%%%%%%%%%%%%%%%%%%%%%%

\section{Observations}

\subsection{Sample description}

\citet{Galametz2018} presented 12 low-mass Class 0 protostars observed in polarization at 345 GHz with the SMA. 
 We complement this first subsample with 8 additional low-mass protostars. The 20 sources cover a wide 
 range of protostellar properties: isolated, binary, triple, or quadruple systems form in cores whose masses 
 range from 0.2 to 12\msun. Details of the full sample are provided in Table~\ref{Sources}.

\subsection{SMA dust polarization observations}

The observations (taken in both compact and sub-compact configurations), data reduction and polarization 
maps of the first 12 low-mass Class 0 protostars observed are presented in \citet{Galametz2018}. 
The observations of the 8 additional low-mass protostars were obtained with the SMA settled in compact configuration in 
the 345 GHz band (Project 2018B-S015, PI: A. Maury).  
The antennas used for each observation date are listed in Table~\ref{Observations}. 
The polarimeter on the SMA makes use of a quarter-wave plate (QWP)
in order to convert the linear polarization into circular polarization. The antennas are 
switched between polarizations (QWP are rotated at various angles) in a coordinated 
temporal sequence to sample the various combinations of circular polarizations on each baseline.  
A variety of observational modes (single- and dual-receiver polarization modes) were used for the 
observations of \citet{Galametz2018}. The dual-receiver full polarization mode, fully commissioned, was then 
the only mode we used to observe the additional 8 targets.
The new observations were also taken using the SMA Wideband Astronomical ROACH2 Machine (SWARM)
rather than the Application-Specific Integrated Circuit (ASIC) correlator: the added bandwidth has helped increase the SMA sensitivity.
A detailed description of the SMA polarimeter system is provided by \citet{Marrone2006} 
and \citet{Marrone2008}. Frequent observations of various calibrators were taken between
the target observations to ensure the future gain and polarization calibration. Flux
calibrators (Callisto and Neptune) were also observed, but were not used when we performed 
the flux calibration of the observations (see \S~\ref{DataReduc}).

\subsection{Data reduction, self-calibration, and flux calibration}
\label{DataReduc}

We performed the data reduction on the raw visibilities using the IDL-based software MIR (for Millimeter Interferometer Reduction). The calibration includes an initial flagging of high 
system temperatures $T_\mathrm{sys}$ and other incorrect visibilities, a bandpass 
calibration, a correction of the cross-receiver delays, 
and a gain and flux calibration. The various calibrators observed for each of 
these steps and the list of antennas used for the observations 
are summarized in Table~\ref{Observations}. Data were then exported to \texttt{MIRIAD} 
\citep{Sault1995}\footnote{https://www.cfa.harvard.edu/sma/miriad/}
for additional processing (i.e., additional flagging) and in particular to 
perform the instrumental polarization calibration.
Quasars were observed to calculate the leakage terms. The continuum data of the targets were used to perform an iterative
self-calibration of the Stokes I visibilities. The process was repeated with deeper cleans and
shorter intervals until it converged (no rms improvement).
We finally used the Atacama Large Millimeter/submillimeter Array (ALMA) calibration source catalog\footnote{https://almascience.eso.org/sc/} 
to gather the fluxes of quasars we also observed with the SMA (3C 84, 3C 454.3, 3C 273, and 3C 279) 
at similar dates as those of our observations. These fluxes were compared to the SMA amplitudes
in order to derive the multiplying factors that were to be applied to the target visibility amplitude in order to
flux-calibrate the dataset. These scaling factors are reported for each date in Table~\ref{Observations}. 
The visibilities covered by the observations range from 15 to 85 k$\lambda$.

\subsection{Deriving the continuum and polarization maps}

The Stokes parameters are defined as
\begin{equation}
   \overrightarrow{S} = \begin{bmatrix}
                         I \\
                        Q \\
                        U \\
                        V \\
         \end{bmatrix}
,\end{equation}

\noindent with Q and U the linear polarization and V the circular polarization.
We used a robust weighting of 0.5 to transform the visibility data into a dirty map
(using the \texttt{MIRIAD} {\it invert} task). We produced cleaned images of the 
various Stokes parameters (using the \texttt{MIRIAD} {\it clean} task).
Finally, maps of the polarized intensity (debiased) $\mathrm{P_i}$, polarization fraction $\mathrm{p_{frac}}$ , and
polarization angle $\mathrm{P.A.}$ were produced (using the \texttt{MIRIAD} {\it impol} task) as follows:

%%%%%%%%%%%% SMA maps with B overlaid %%%%%%%%%%%%%%
\begin{figure*}
\begin{tabular}{p{4.2cm}p{4.2cm}p{4.2cm}p{4.2cm}}
\hspace{-0.5cm}\begin{annotate}{\includegraphics[height=4.8cm, angle=-90]{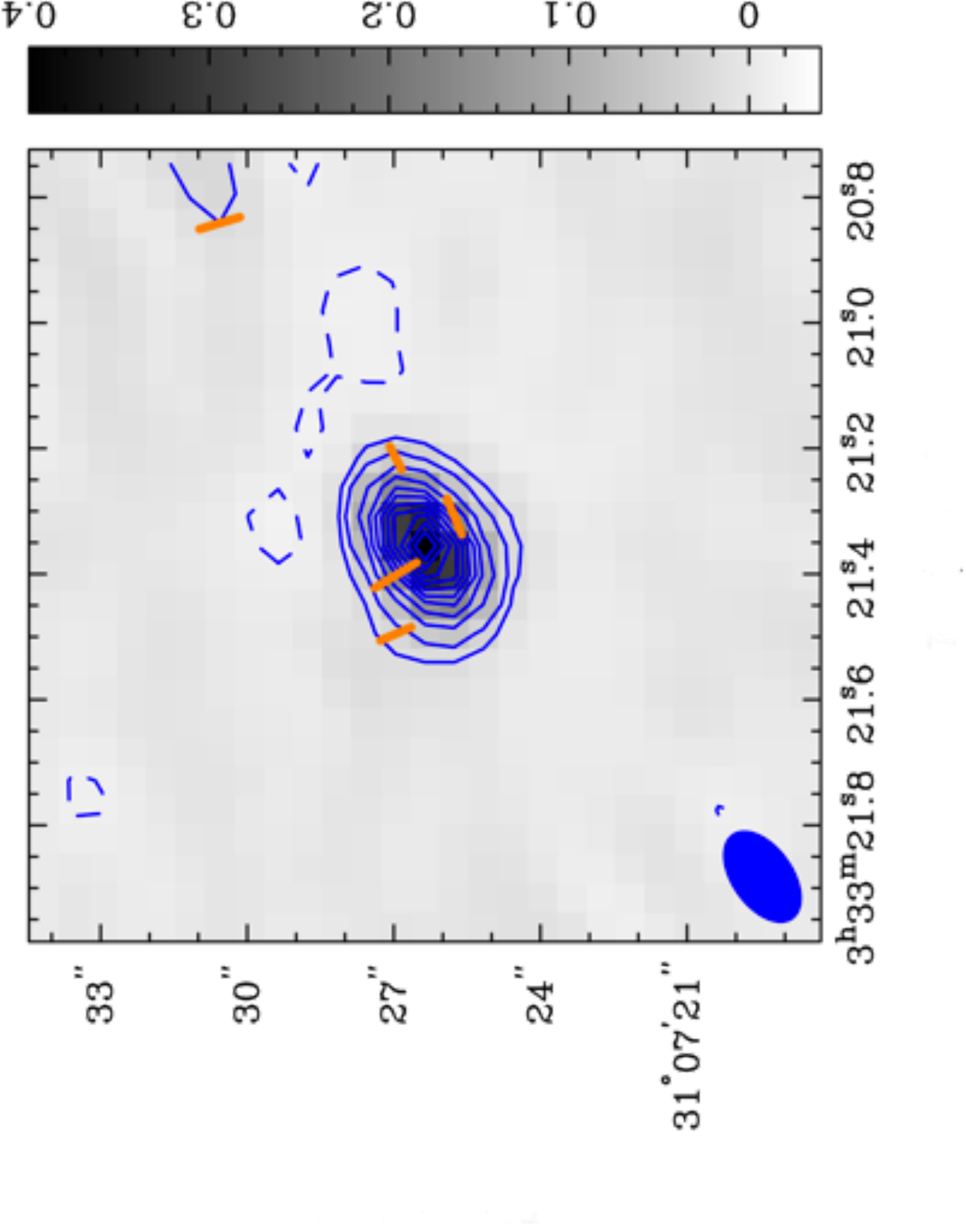}}{1} 
\end{annotate} &
\hspace{-0.7cm}\begin{annotate}{\includegraphics[height=4.8cm, angle=-90]{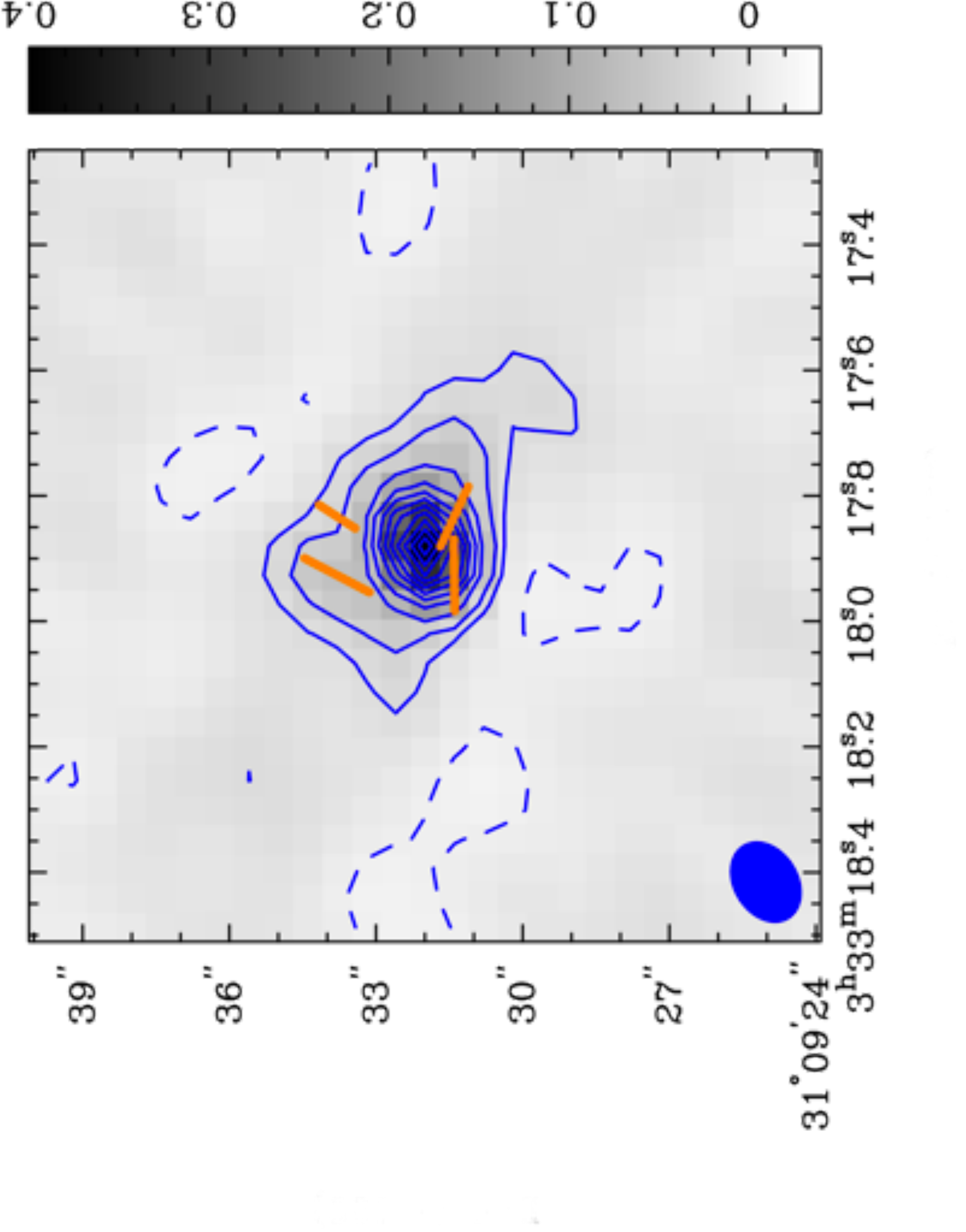}}{1} 
\end{annotate}&
\hspace{-0.7cm}\begin{annotate}{\includegraphics[height=4.8cm, angle=-90]{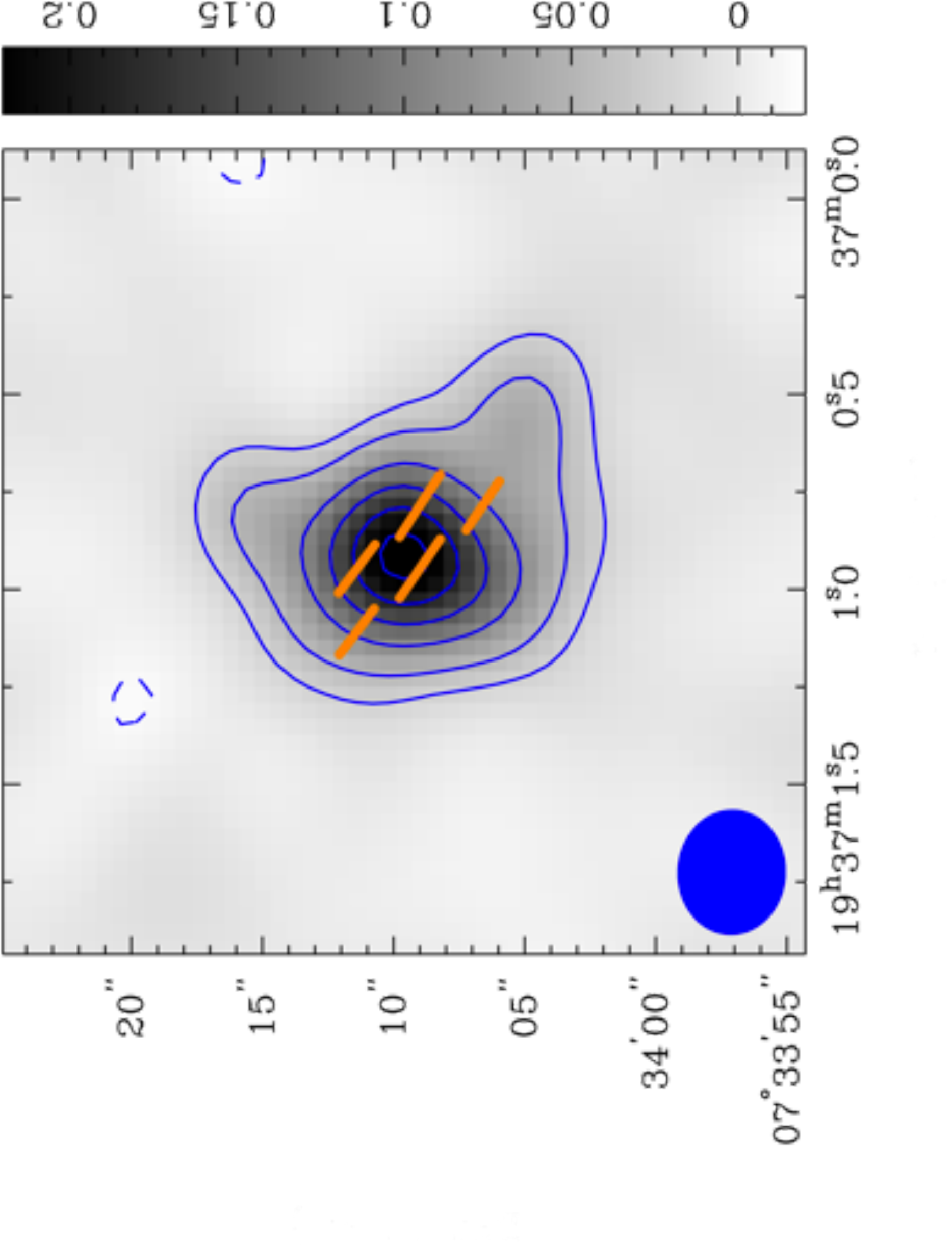}}{1} 
\end{annotate}&
\hspace{-0.7cm}\begin{annotate}{\includegraphics[height=4.8cm, angle=-90]{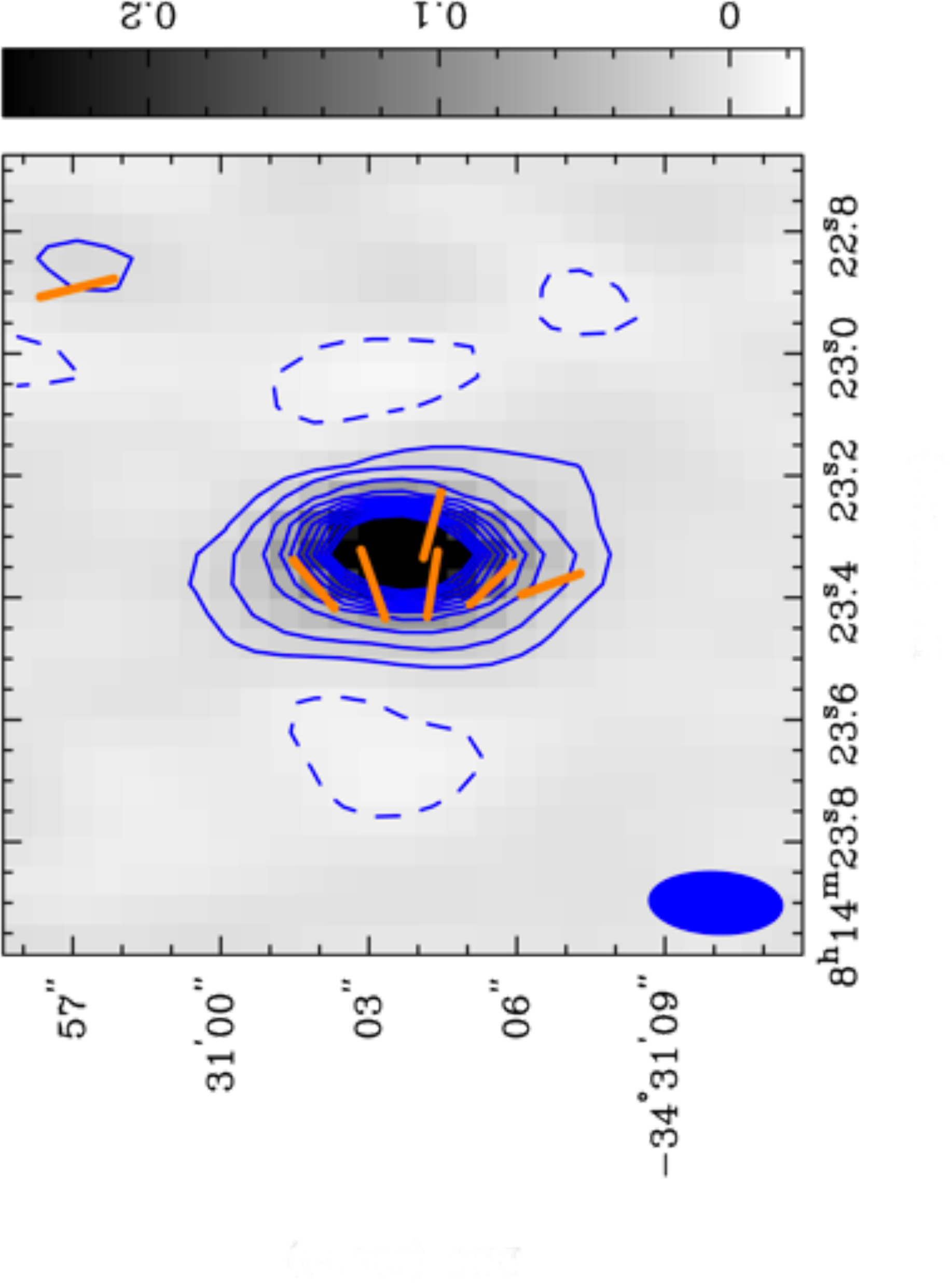}}{1} 
\end{annotate}\\
\vspace{-110pt}\hspace{0.9cm}{\large B1-b} & 
\vspace{-110pt}\hspace{0.8cm}{\large B1-c} & 
\vspace{-110pt}\hspace{0.8cm}{\large B335} &
\vspace{-110pt}\hspace{0.8cm}{\large BHR7} \\
\end{tabular}

\clearpage \vspace{-20pt}
\begin{tabular}{p{4.2cm}p{4.2cm}p{4.2cm}p{4.2cm}}
\hspace{-0.5cm}\begin{annotate}{\includegraphics[height=4.8cm, angle=-90]{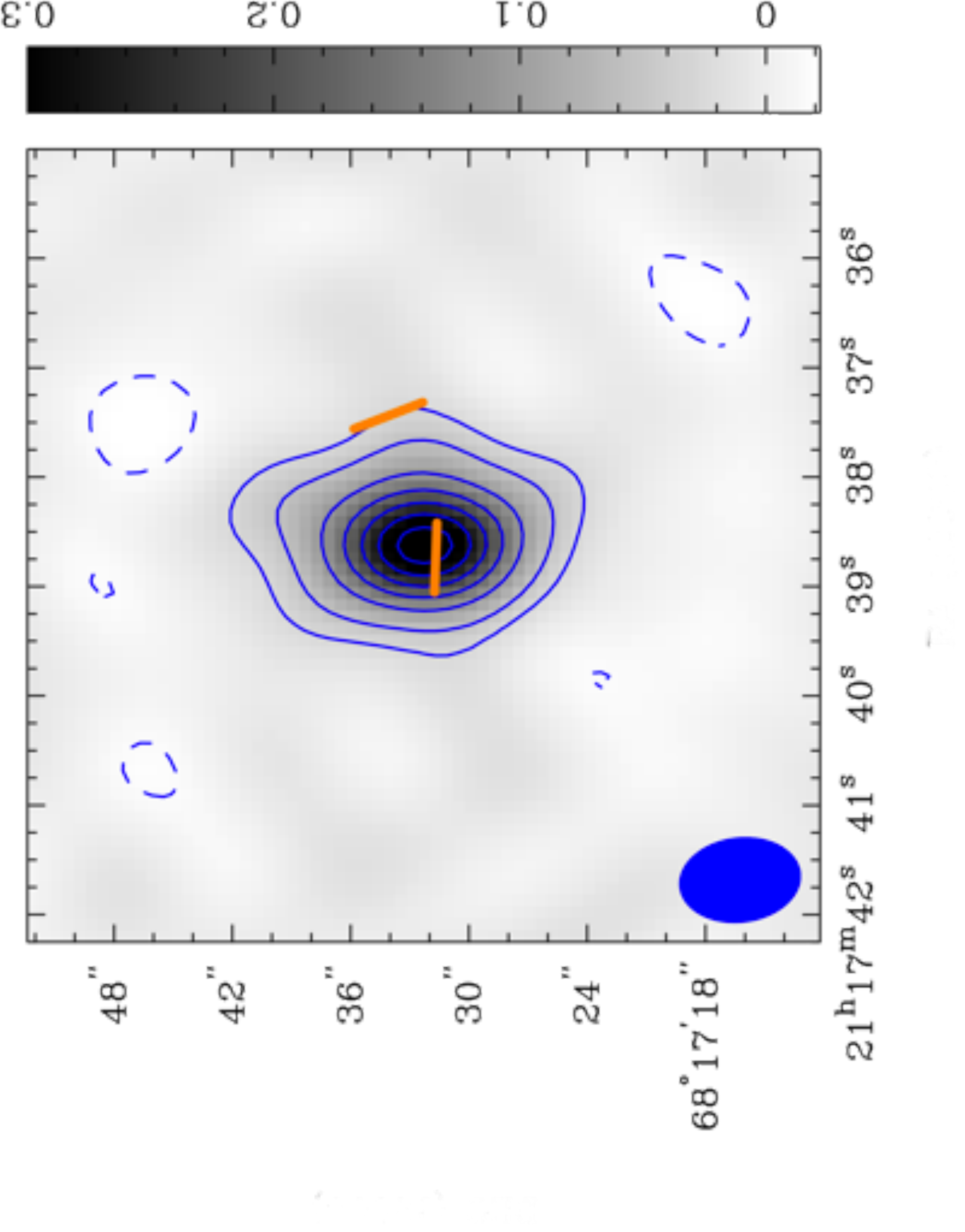}}{1} 
\end{annotate}&
\hspace{-0.7cm}\begin{annotate}{\includegraphics[height=4.8cm, angle=-90]{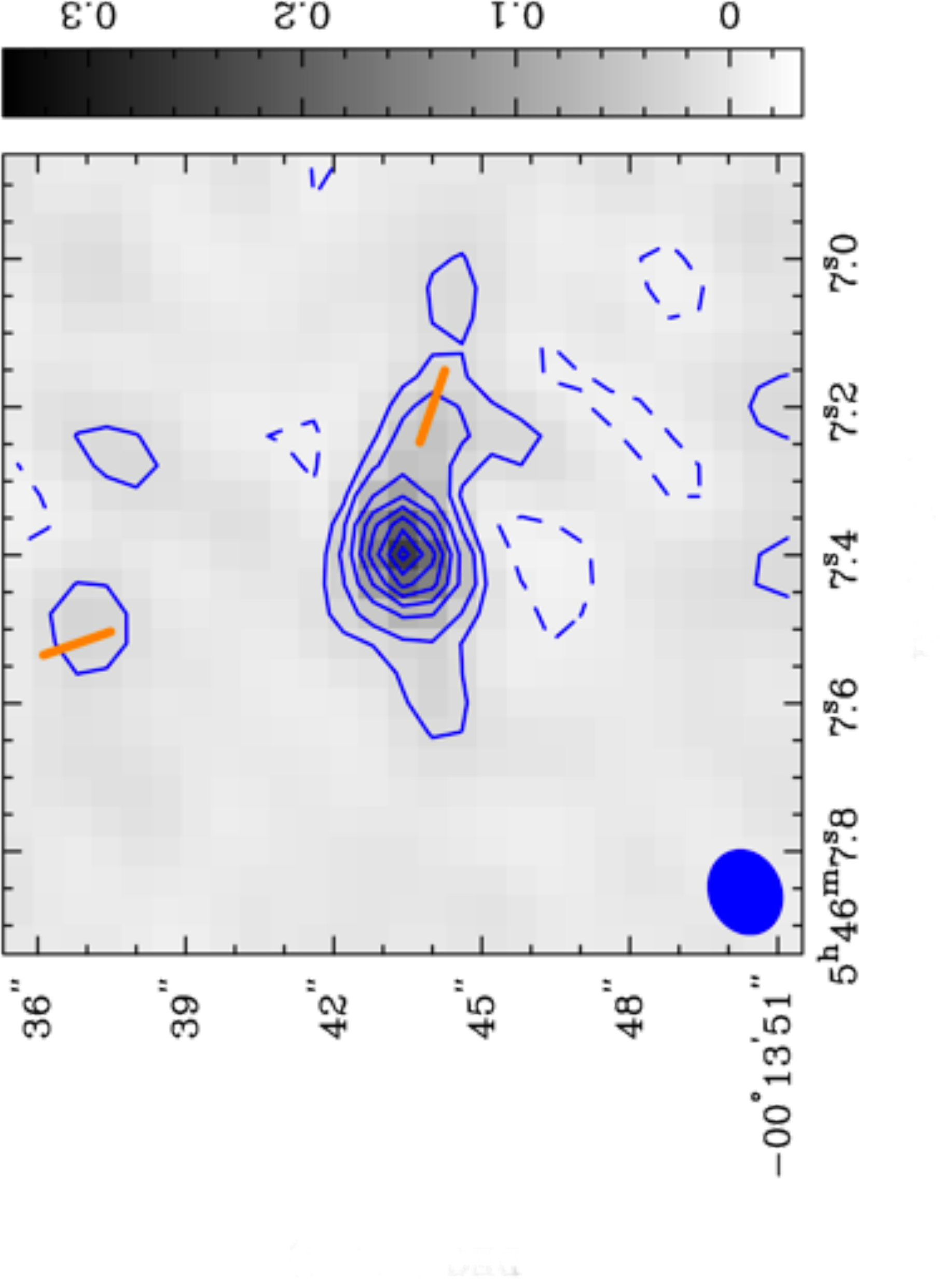}}{1} 
\end{annotate}&
\hspace{-0.7cm}\begin{annotate}{\includegraphics[height=4.8cm, angle=-90]{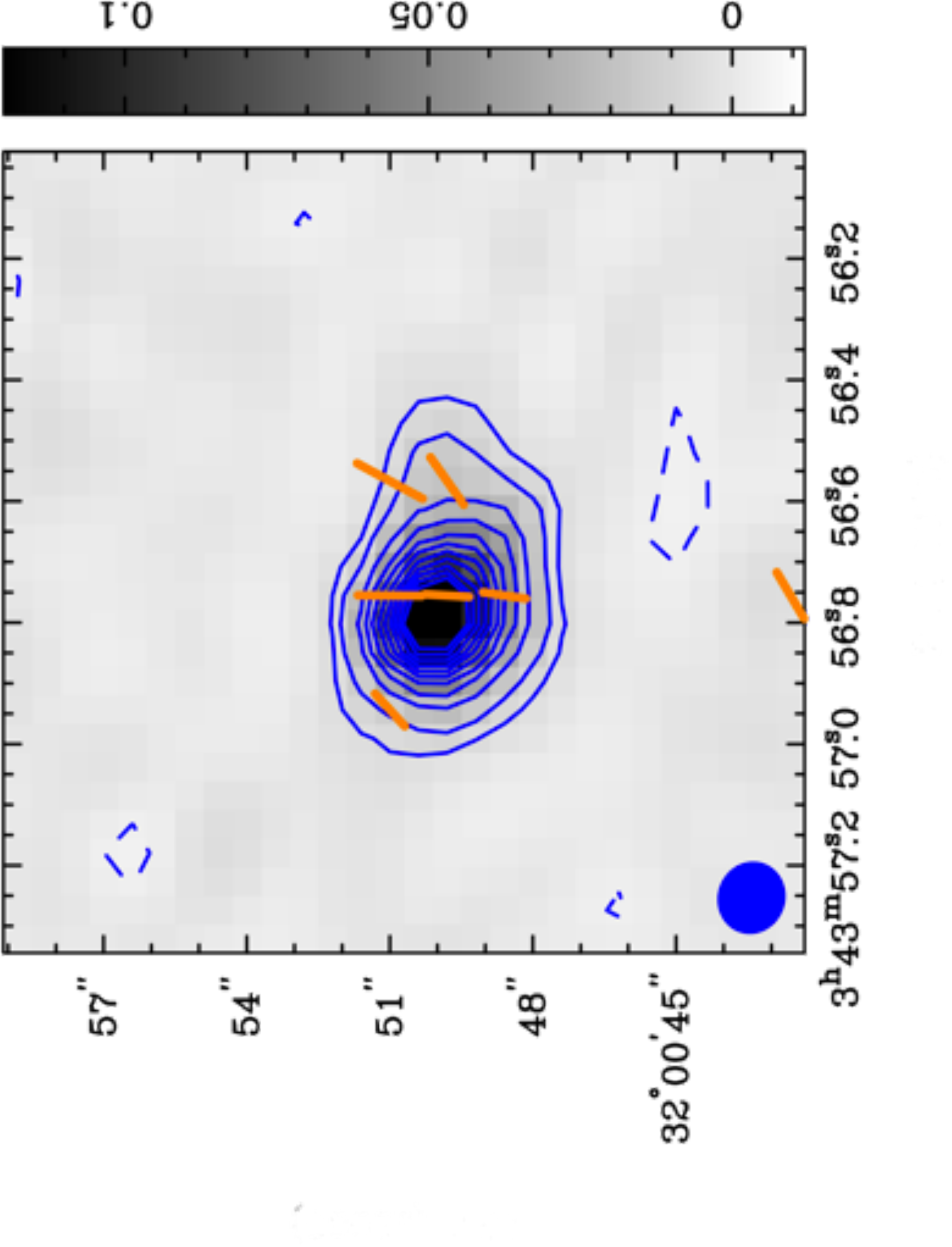}}{1} 
\end{annotate}&
\hspace{-0.7cm}\begin{annotate}{\includegraphics[height=4.8cm, angle=-90]{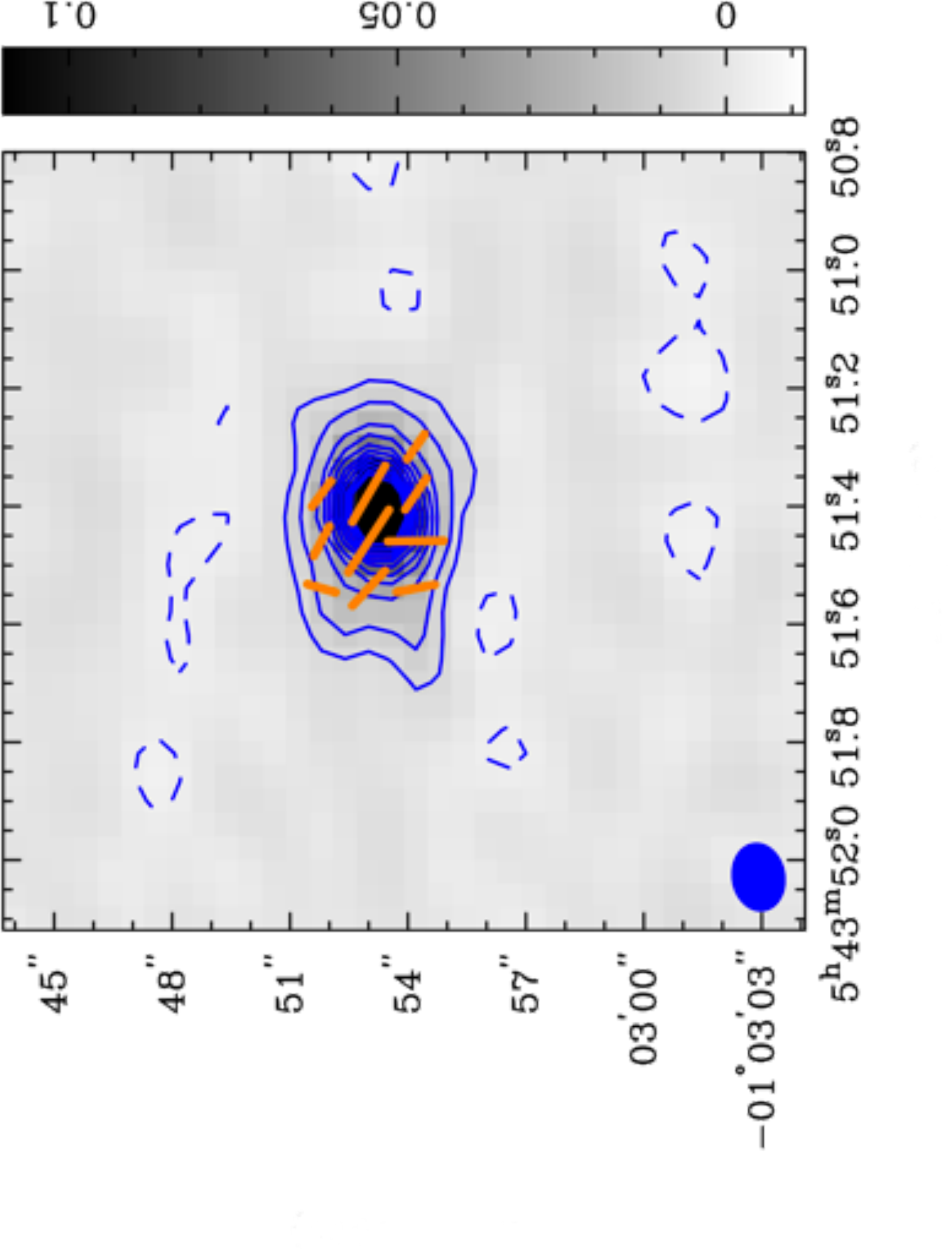}}{1} 
\end{annotate}\\
\vspace{-110pt}\hspace{0.9cm}{\large CB230} & 
\vspace{-110pt}\hspace{0.8cm}{\large HH25-MMS} &
\vspace{-110pt}\hspace{0.8cm}{\large HH211-mm} & 
\vspace{-110pt}\hspace{0.8cm}{\large HH212} \\
\end{tabular}

\clearpage \vspace{-20pt}
\begin{tabular}{p{4.2cm}p{4.2cm}p{4.2cm}p{4.2cm}}
\hspace{-0.5cm}\begin{annotate}{\includegraphics[height=4.8cm, angle=-90]{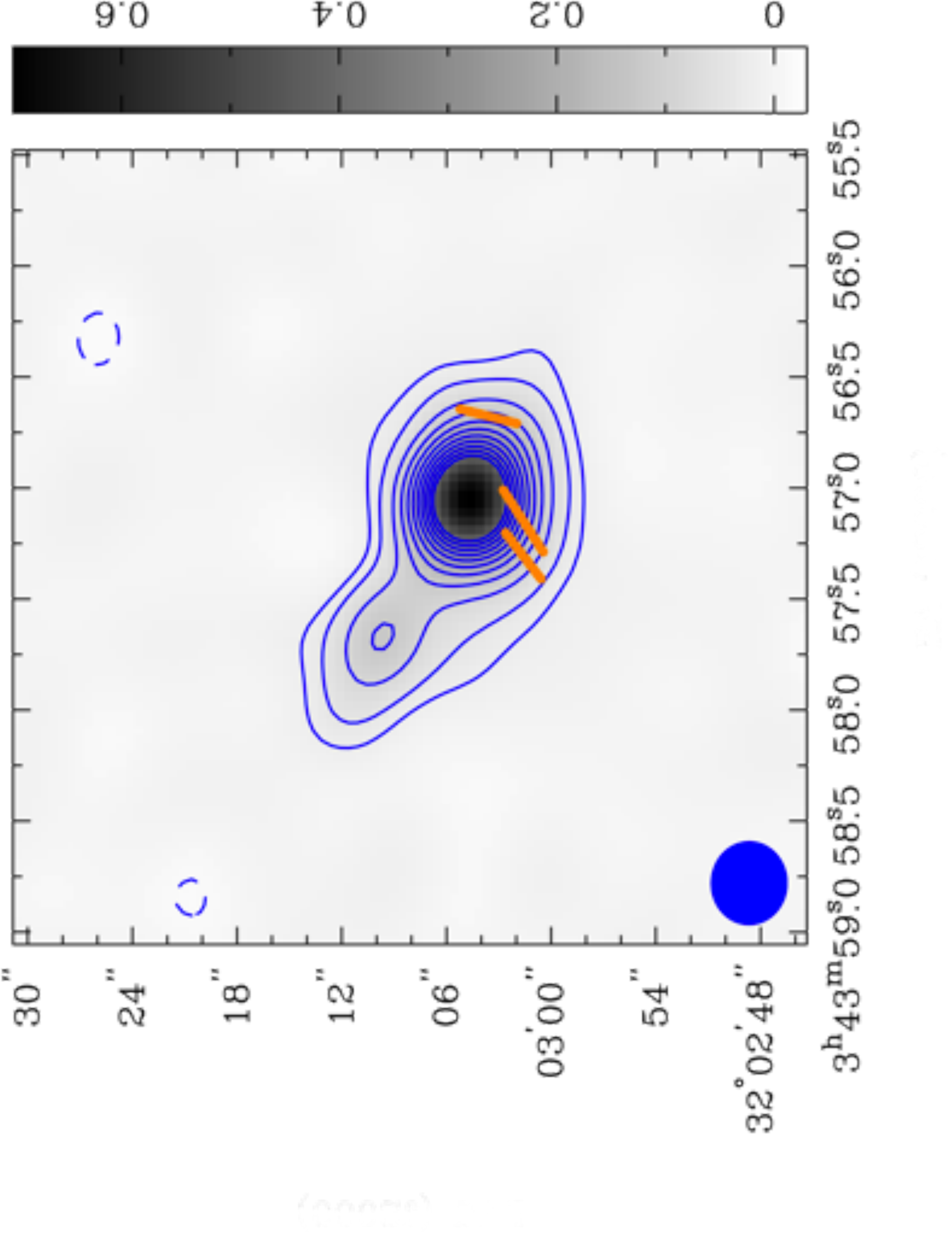}}{1} 
\end{annotate} &
\hspace{-0.7cm}\begin{annotate}{\includegraphics[height=4.8cm, angle=-90]{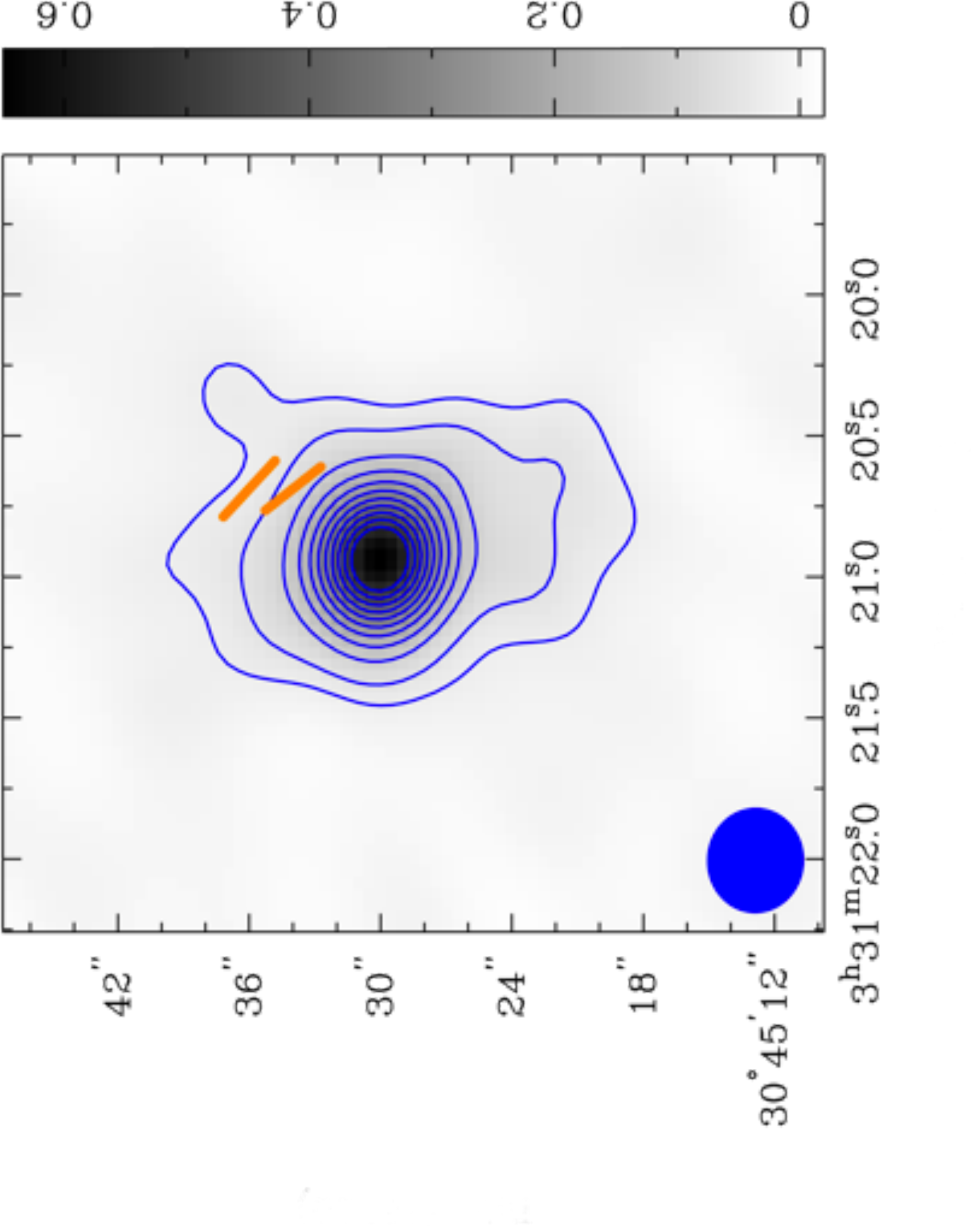}}{1} 
\end{annotate} &
\hspace{-0.7cm}\begin{annotate}{\includegraphics[height=4.8cm, angle=-90]{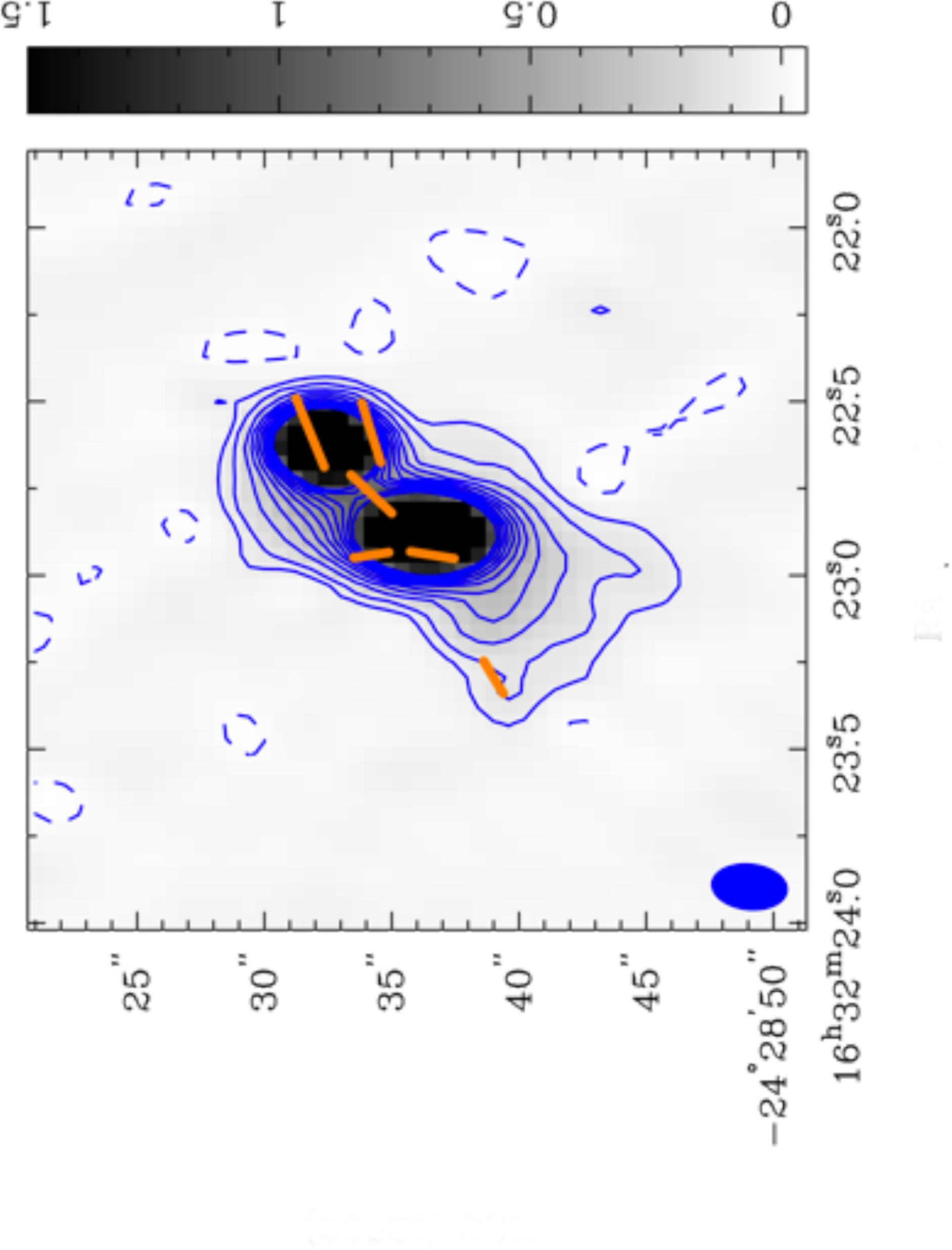}}{1} 
\end{annotate} &
\hspace{-0.7cm}\begin{annotate}{\includegraphics[height=4.8cm, angle=-90]{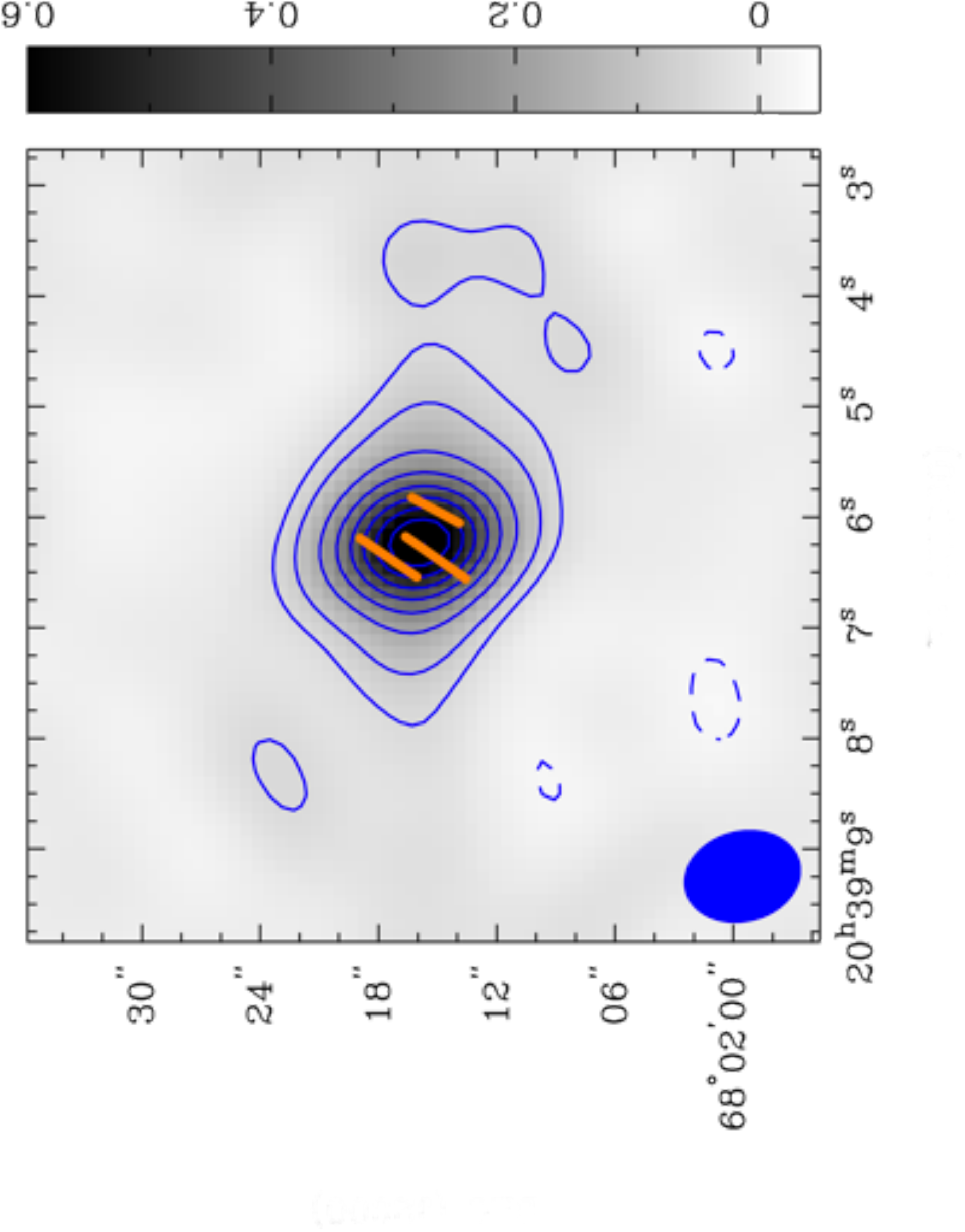}}{1} 
\end{annotate} \\
\vspace{-110pt}\hspace{0.9cm}{\large HH797} &
\vspace{-115pt}\hspace{0.8cm}{\large IRAS03282} & 
\vspace{-110pt}\hspace{0.8cm}{\large IRAS16293} & 
\vspace{-110pt}\hspace{0.8cm}{\large L1157} \\
\end{tabular}

\clearpage \vspace{-20pt}
\begin{tabular}{p{4.2cm}p{4.2cm}p{4.2cm}p{4.2cm}}
\hspace{-0.5cm}\begin{annotate}{\includegraphics[height=4.8cm, angle=-90]{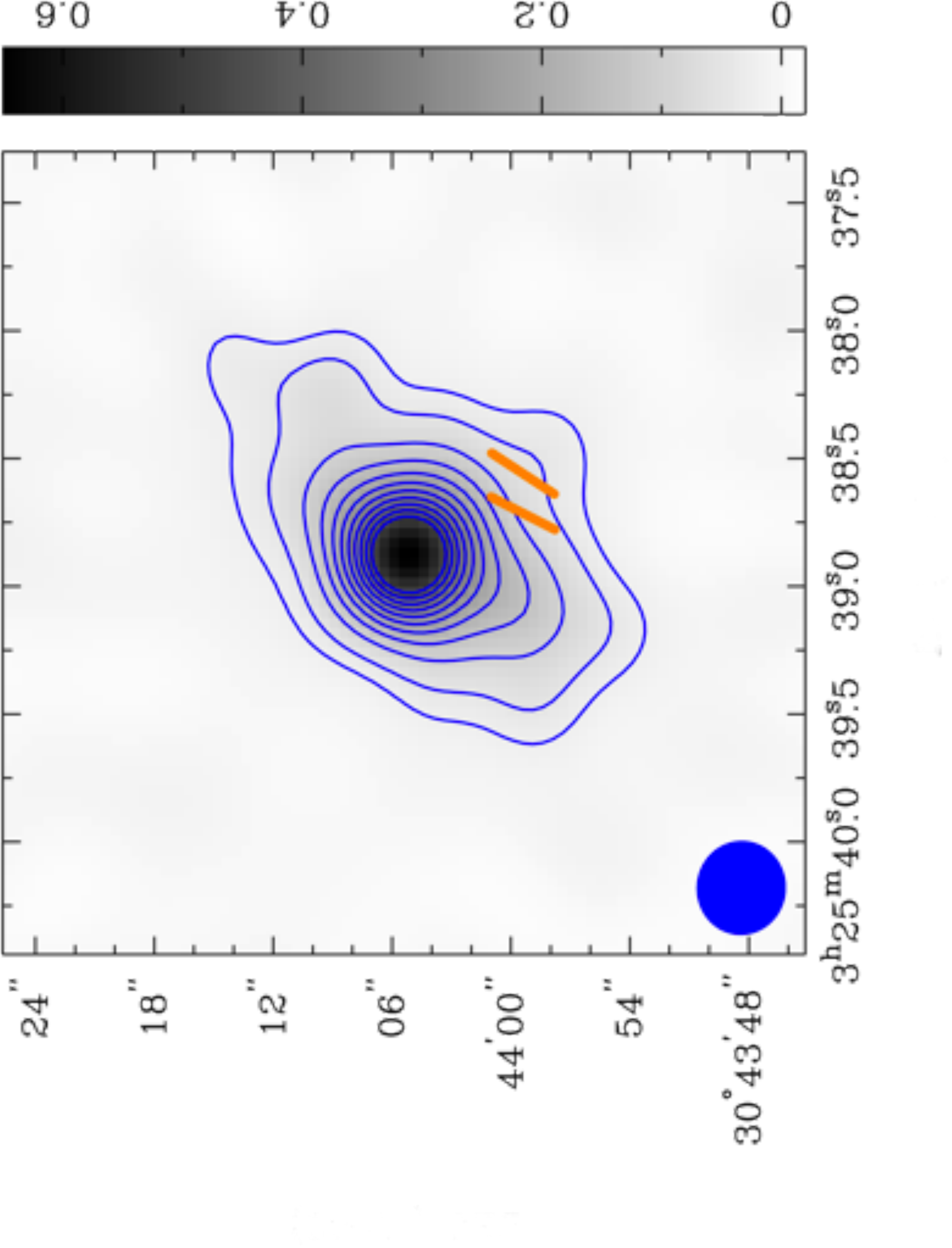}}{1} 
\end{annotate} &
\hspace{-0.7cm}\begin{annotate}{\includegraphics[height=4.8cm, angle=-90]{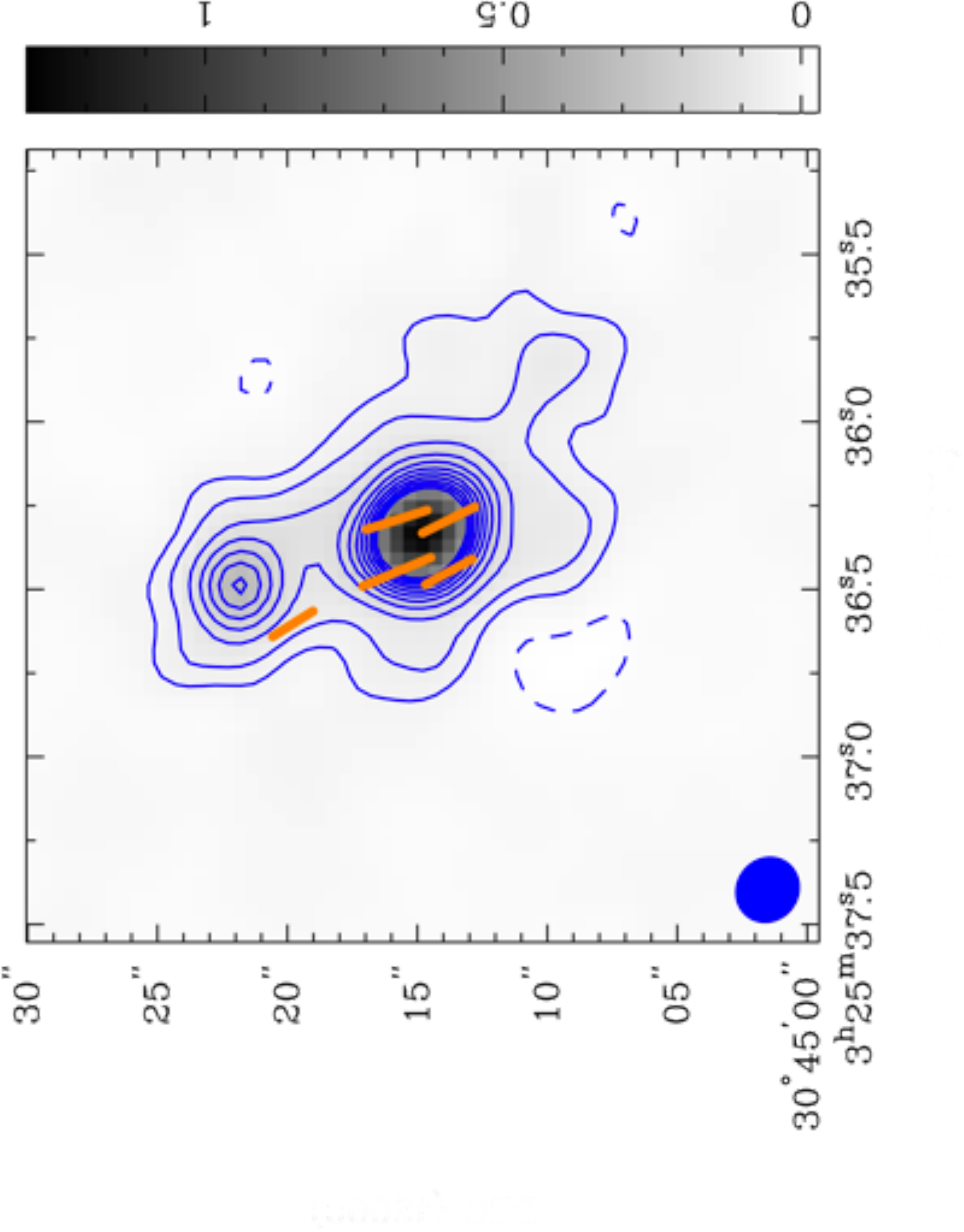}}{1} 
\end{annotate} &
\hspace{-0.7cm}\begin{annotate}{\includegraphics[height=4.8cm, angle=-90]{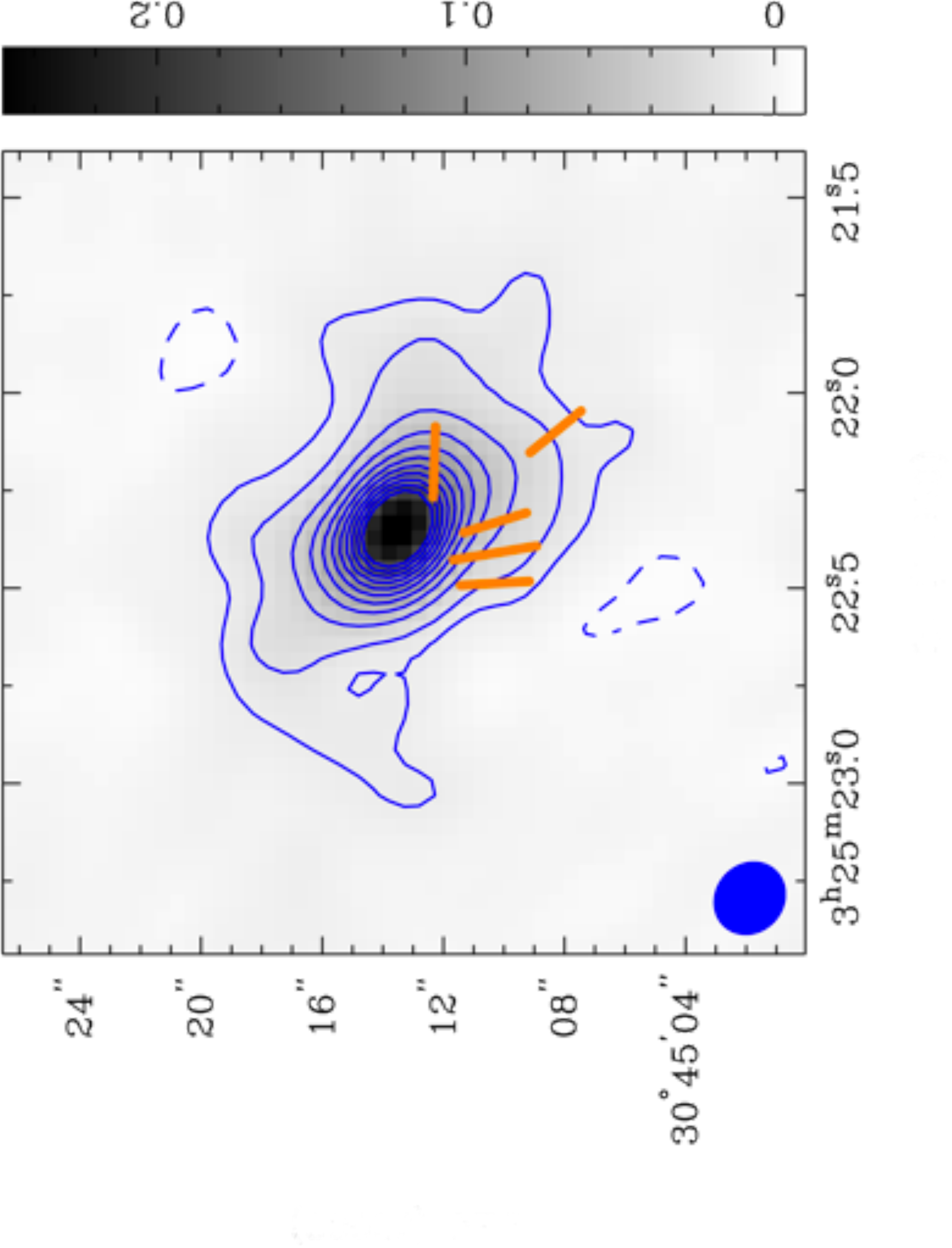}}{1} 
\end{annotate} &
\hspace{-0.7cm}\begin{annotate}{\includegraphics[height=4.8cm, angle=-90]{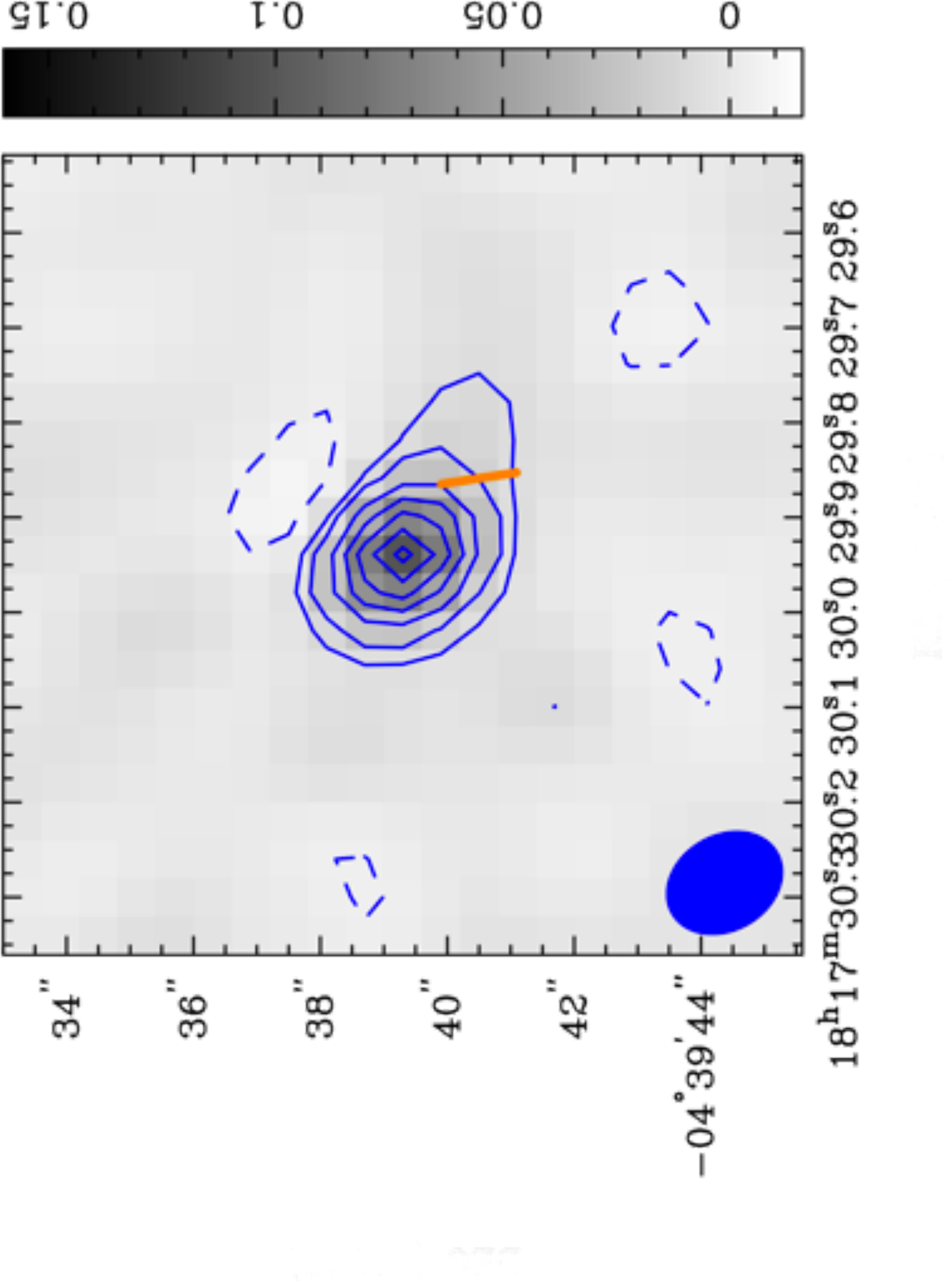}}{1} 
\end{annotate} \\
\vspace{-110pt}\hspace{0.9cm}{\large L1448C} &
\vspace{-110pt}\hspace{0.8cm}{\large L1448N-B} & 
\vspace{-110pt}\hspace{0.8cm}{\large L1448-2A} & 
\vspace{-110pt}\hspace{0.8cm}{\large L483-mm} \\
\end{tabular}

\clearpage \vspace{-20pt}
\begin{tabular}{p{4.2cm}p{4.2cm}p{4.2cm}p{4.2cm}}
\hspace{-0.5cm}\begin{annotate}{\includegraphics[height=4.8cm, angle=-90]{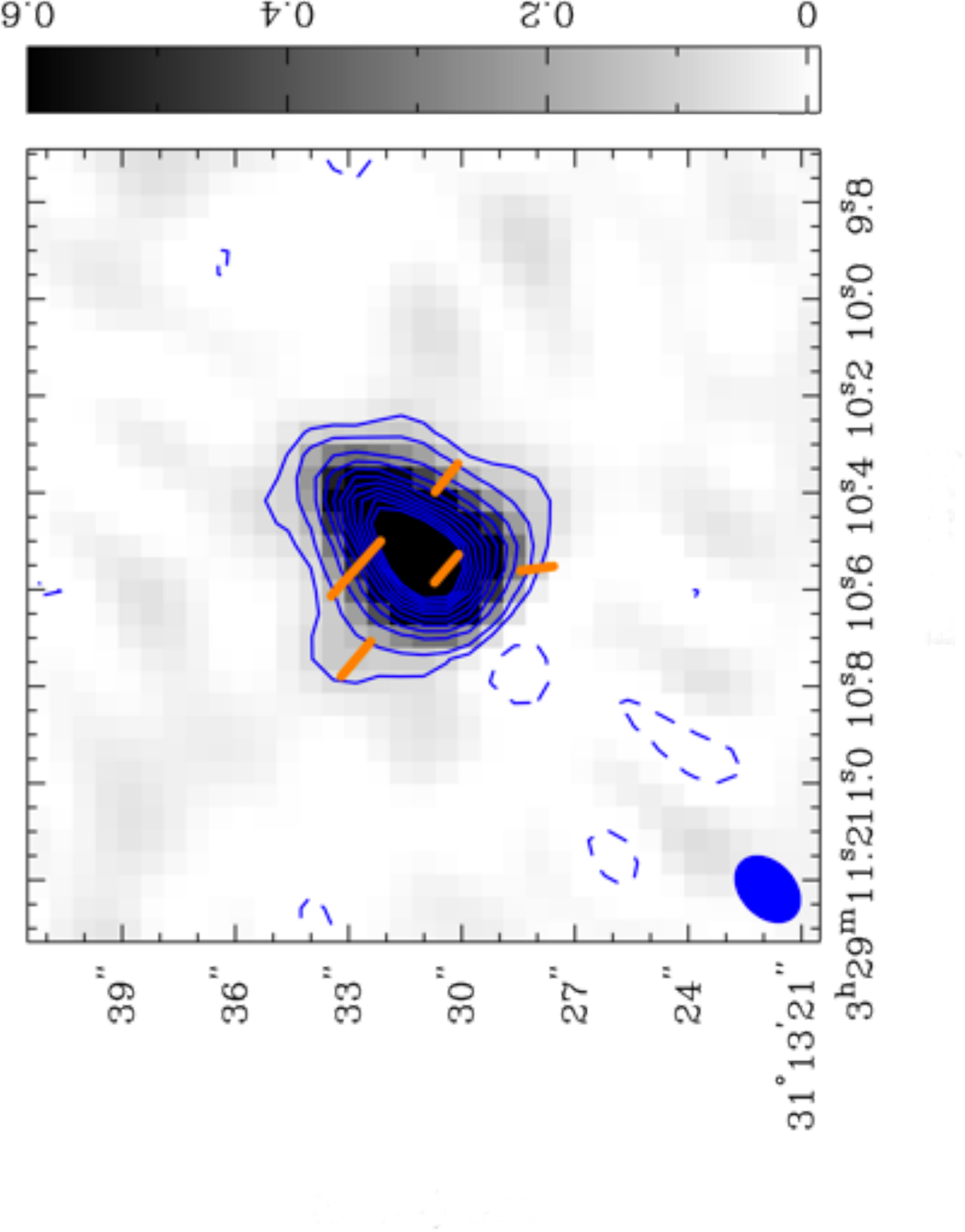}}{1} 
\end{annotate} &
\hspace{-0.7cm}\begin{annotate}{\includegraphics[height=4.8cm, angle=-90]{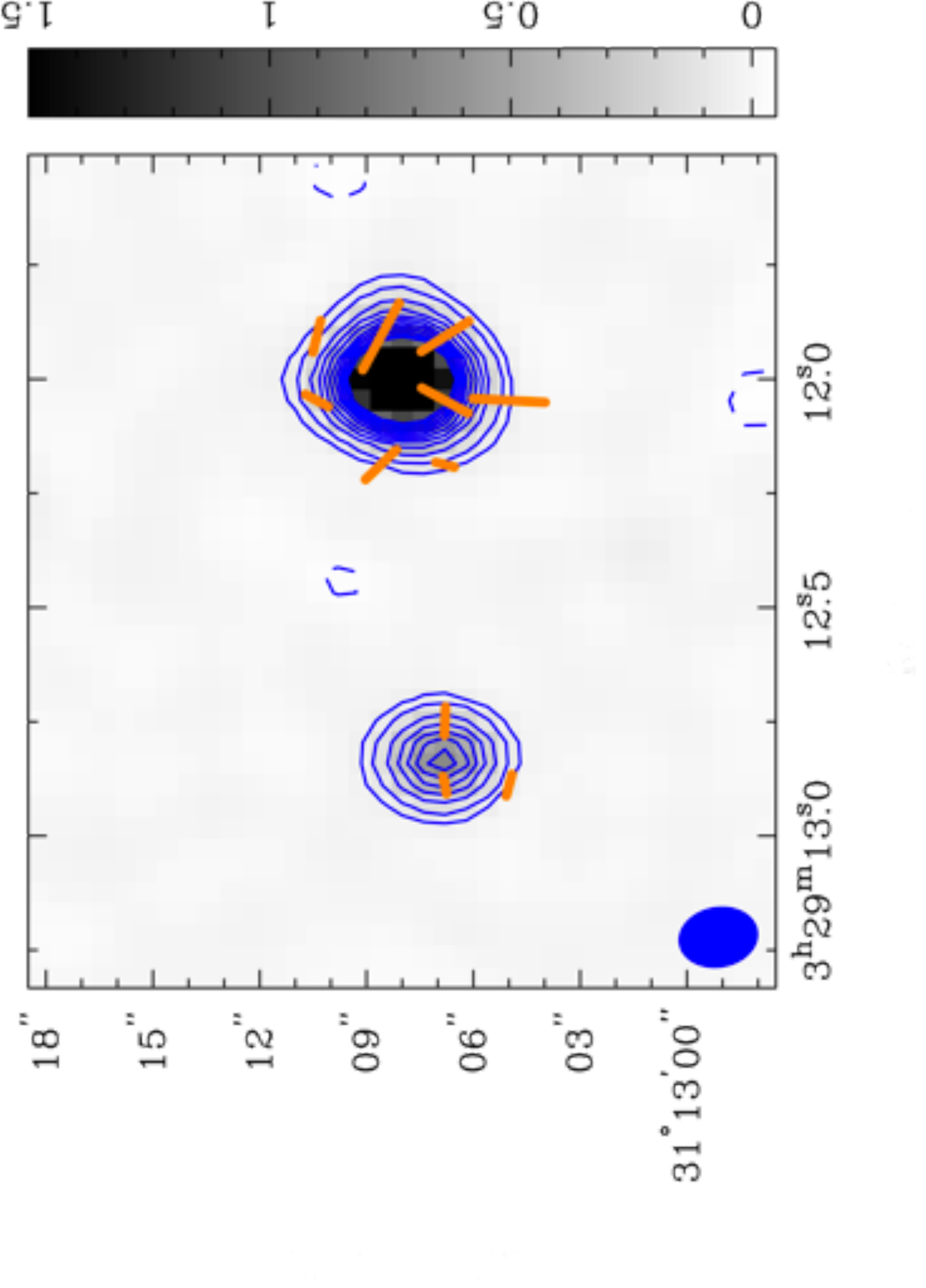}}{1} 
\end{annotate} &
\hspace{-0.7cm}\begin{annotate}{\includegraphics[height=4.8cm, angle=-90]{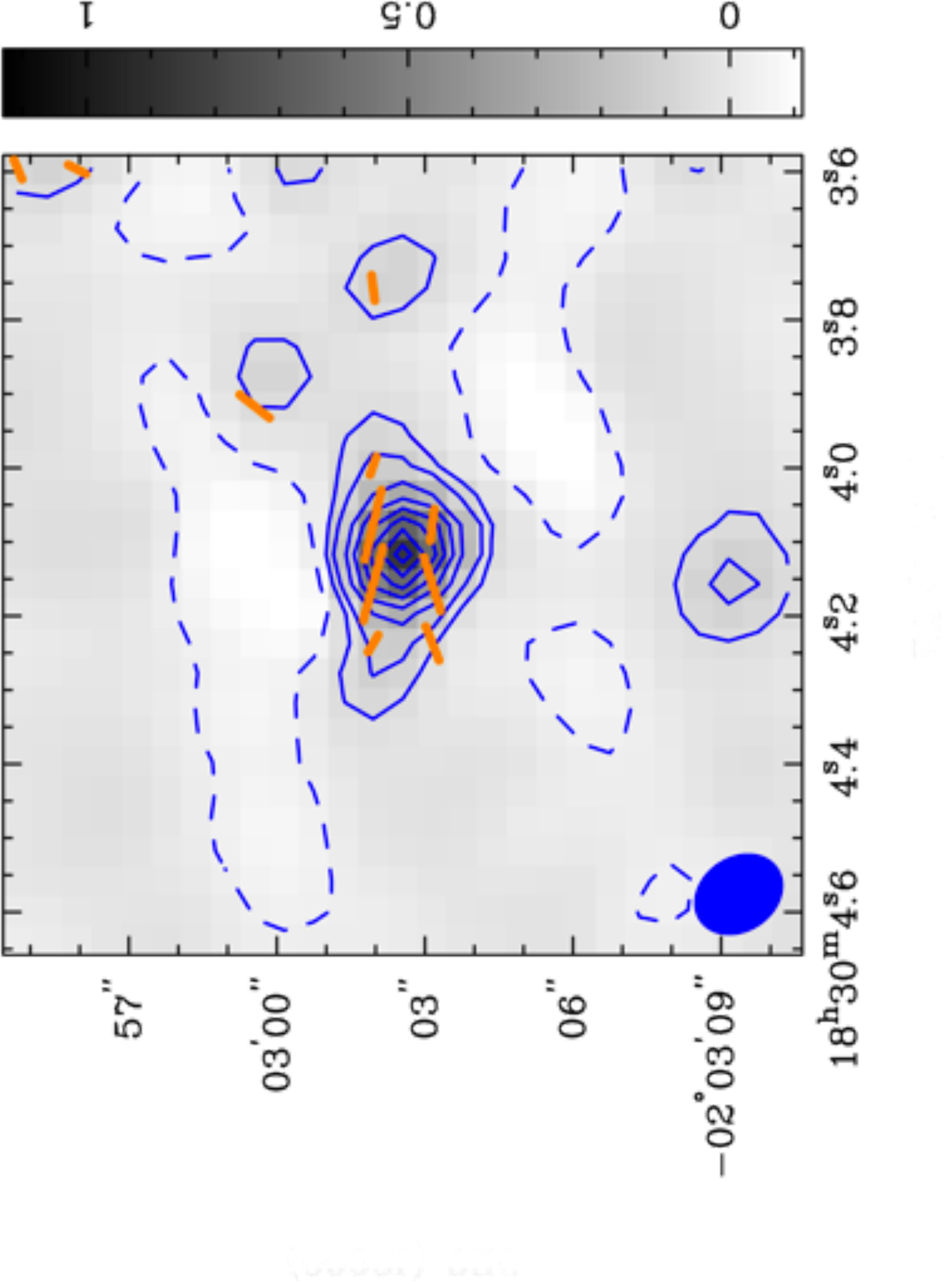}}{1} 
\end{annotate} &
\hspace{-0.7cm}\begin{annotate}{\includegraphics[height=4.8cm, angle=-90]{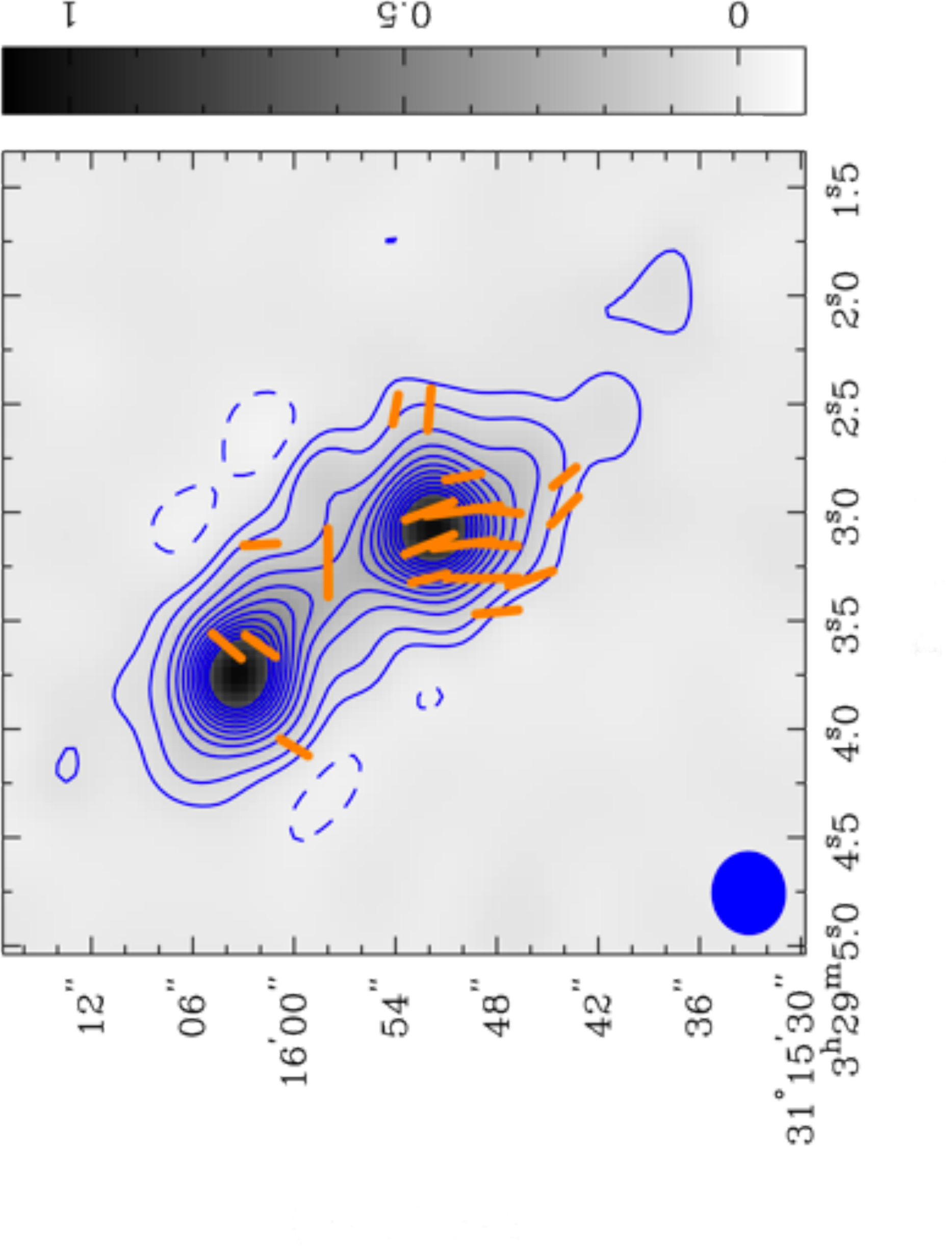}}{1} 
\end{annotate} \\
\vspace{-110pt}\hspace{0.9cm}{\large IRAS4A} & 
\vspace{-105pt}\hspace{0.8cm}{\large IRAS4B} & 
\vspace{-110pt}\hspace{0.8cm}{\large SSMM18}  &
\vspace{-110pt}\hspace{0.8cm}{\large SVS13-B} \\ 
 \end{tabular}
 \vspace{-15pt}
\caption{B-field vectors (derived from the polarization vectors assuming a 90$\degr$ rotation) overlaid as orange 
segments on the SMA 850 \mic\ Stokes I continuum maps. Color scales are in Jy/beam. 
Contours at -3, 5, 10, 20, 30, 40, 50, 60, 70, 80, 90, and 100 $\sigma$ appear in blue. 
The filled ellipses in the lower left corner indicate the synthesized beam of the SMA maps. }
\label{StokesI_Borientation_all}
\vspace{10pt} 
\end{figure*}
%%%%%%%%%%%% SMA maps with B overlaid %%%%%%%%%%%%%%

%%%%%%%%%%%% SMA maps with B overlaid %%%%%%%%%%%%%%
\begin{figure*}
\centering
\vspace{20pt}
\includegraphics[width=18cm]{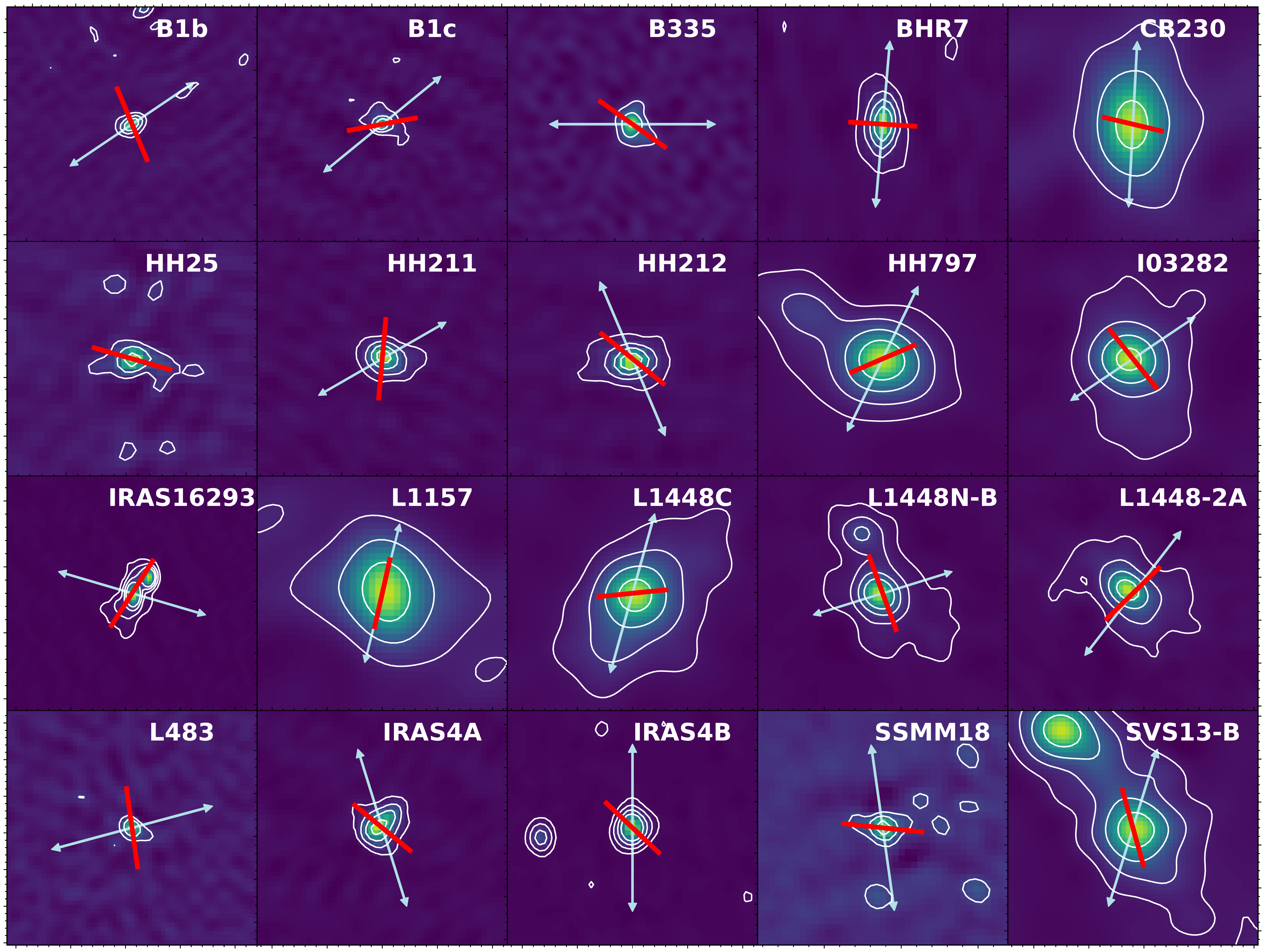}
\caption{Mean magnetic field orientation in the 20 Class 0 protostellar envelopes (red segments) overlaid on the dust emission maps
(contour and color maps). Contours are indicated for detections at 5, 20, 50, and 100 $\sigma$. The common physical scale of each map 
is 8000 $\times$ 8000 au. We indicate the outflow axis for each source with cyan arrows. For L1448C, we report the central 2-$\sigma$ 
detection \citep[see][and $\S$3.1.]{Galametz2018}.}
\label{StokesI_Borientation}
\vspace{10pt} 
\end{figure*}
%%%%%%%%%%%% SMA maps with B overlaid %%%%%%%%%%%%%%

\vspace{-2pt}
\begin{equation}
P_{i} = \sqrt[]{Q^2+U^2 - \sigma_{Q,U}^2},
\vspace{-2pt}
\end{equation}

\begin{equation}
p_{frac} = P_{i}~/~I,
\vspace{-2pt}
\end{equation}

\begin{equation}
\label{epPA}
P.A. = 0.5 \times\ arctan(U/Q),\end{equation}

\noindent with $\sigma_{Q,U}$ the average rms of the Q and U maps.
We applied a 5$\sigma$ cutoff on Stokes I and a 3$\sigma$ cutoff on 
Stokes Q and U in order to only discuss locations where polarized emission is robustly detected. 
The synthesized beams and rms of the various cleaned maps are provided in Table~\ref{SMAmapsrms}. 
Maps are produced with a pixel size of 0.6\arcsec. In the appendix (see Table~\ref{PixelSizeInfluence}), we
show that this choice does not affect the mean B-field orientation we derived.

The Stokes I dust continuum emission maps are shown in Fig.~\ref{StokesI_Borientation_all}.
A description of the morphology of this continuum emission as well as details of the source 
multiplicity are provided in Appendix \ref{SourceCharacteristics}.
The polarized intensity and polarization fraction maps are shown in Fig.~\ref{PolaMaps}. 
To our knowledge, we present the first detections of polarized dust emission at envelope scales 
toward BHR7 and HH25.

\section{Analysis}

\subsection{Mean magnetic field orientation}

The polarization angles were rotated by 90$\degr$ to obtain the magnetic field direction. 
The B vectors obtained for the 20 sources are overlaid on the Stokes I maps in Fig.~\ref{StokesI_Borientation_all}.
We note that the strength of SMA observations is twofold. First, their interferometric nature allows us to filter 
out the large-scale B~field permeating the surrounding host cloud, and we focus on the fields 
in the inner protostellar envelopes. Second, the modest spatial resolution of our observations allows us to cancel out 
the more complex topology of the field at small ($<$ 500 au) scales because of the 
intense gravitational pull of the infalling material and the launching of protostellar outflows \citep{Kwon2019}. 
Our observations are also expected to be less prone than high-angular resolution observations to selectively 
tracing the B~field in locations where the dust grain alignment efficiency may be highly inhomogeneous, 
for instance, along irradiated cavity walls that are located very close to the central protostellar objects 
\citep[see, e.g.,][]{LeGouellec2019}. Because most Class 0 disks only 
contribute at scales much smaller than the scales probed by our beam \citep[fewer than 25\% of 
the Class 0 disks extend beyond 60 au; see][]{Maury2019}, dust polarization due to self-scattering 
is unlikely to contribute to the polarization observed at envelope scales with the SMA.  \\

To trace the main direction of B at envelope scales, we extracted the mean
B-field orientation within the central 1000 au region of each source. To perform the calculation, we used the polarization
angle and polarization angle error maps produced with our data-reduction procedure within the {\it idl}/wmean function.
The weighted mean is calculated as

\begin{equation}
\mu = \frac{\Sigma x_i / \sigma_i^2}{\Sigma 1/\sigma_i^2}
,\end{equation}

\noindent with $\mu$ the mean position angle of the B~field, x$_i$ the individual position angles detected within the central 1000 au region,
and $\sigma_i$ their associated errors. We report the magnetic field position angles in Table~\ref{AnglesVelGrad} and overlay them 
(with red segments) on the Stokes I maps for the full sample in Fig.~\ref{StokesI_Borientation}. 
The errors provided in Table~\ref{AnglesVelGrad} are the external uncertainties {\it eu} based on the spread of the values obtained 
multiplying the internal uncertainty {\it iu} by the square-root of the reduced chi-squared, with

\begin{equation}
iu = \frac{1}{\sqrt{\Sigma 1/\sigma_i^2}} 
\end{equation}

and 

\begin{equation}
eu = iu~\sqrt{\frac{\chi^2}{N-1}}~~~~where~~~~\chi^2= \Sigma{\frac{(x_i - \mu)^2}{\sigma_i^2}}
.\end{equation}

\noindent This method is appropriate for calculating a mean B-field angle recovered at 1000 au scales in protostellar envelopes, 
from individual detections presenting a large dynamic range in signal-to-noise ratios, as are some of the polarization detections in the SMA map 
of each individual source, but also to propagate the individual errors and angle dispersions into an error on the mean value. 
While most protostars show small dispersions of their individual detections around the mean B~field (18 sources out of 20 
have dispersions $<$20$^{\circ}$), we note that two protostars (IRAS16293A and Per-B1c) 
present large angle dispersions around their mean position angle.  \\

To quantify the effect of the pixel size that we used during the map-making procedure on the mean B-field orientation, we rederived the polarization maps for 
pixels equal to 0.6\arcsec\ (our fiducial pixel scale), 0.7\arcsec\, , 0.8\arcsec\, , 0.9\arcsec\ , and 1.0\arcsec\ and reestimated the 
mean magnetic field position angles (see Table~\ref{PixelSizeInfluence}). We observe that the pixel size affects the position angles we derived only very little; the uncertainties on the orientation are mostly dominated by the dispersion of the position angles 
within the 1000 au central region. We note that for HH211, the interval of values obtained by changing the pixel size ranges from 
162$^{\circ}$ to 175$^{\circ}$: the error in position angle may be closer to 10$^{\circ}$ than the 5$^{\circ}$ we report in Table~\ref{AnglesVelGrad}
for this source. \\

Additionally, because other methods have been developed for calculating mean polarization angles, we propose in 
Appendix~\ref{Complement} a simple comparison between the mean angles obtained from our method and three other averaging methods that are used to 
analyze single-dish observations of dust polarized emission. The three additional methods rely on 
(i) a simple averaging (no weighting),
(ii) an averaging of the Stokes values before computing an angle \citep[e.g., used in the single-dish maps of clouds by][]{Li2006}, or 
(iii) summing individual Stokes fluxes to remove the variations around a mean value (considering that Stokes Q and U are positive and 
negative, their sum cancels most variations and should converge toward the most widespread value in the map).
This second test shows that the position angle of the mean B~field computed with the different methods presents only small variations in the mean
position angle, with a dispersion smaller than the error bars from our method reported in Table~\ref{AnglesVelGrad}. This indicates that our 
measurement are robust envelope-scale values. One exception is HH212, for which the interval of values 
obtained using the various methods ranges from 51 to 59$^{\circ}$. As for HH211, the error on position angle may be closer to 
10$^{\circ}$ than the 4$^{\circ}$ error reported in Table~\ref{AnglesVelGrad} for this source. \\

The magnetic field orientation of 12 sources of the sample is discussed in \citet{Galametz2018}. 
We add here additional notes concerning the choices we made in the current analysis.
For IRAS4B, the B~field is complex, with an average B-field direction in the eastern part of 149$\pm$45$^{\circ}$ 
and an average direction in the western part of 51$\pm$19$^{\circ}$. Both sides give a misalignment with the 
outflow of 30-50$^\circ$. In this analysis, we used a misalignment of 40$^\circ$ for this source. 
The B-field orientation used for IRAS03292 is the average value detected in the off-centered region where B is detected by the SMA 
\citep[see][]{Galametz2018}. In the case of L1448C, the only robust detections ($>$3-$\sigma$) are vectors located outside the central 
1000 au region, as shown in our Fig.~\ref{StokesI_Borientation_all}. For this analysis, we thus decided to report the 2-$\sigma$ detection of the polarized dust 
emission in the central region from \citet{Galametz2018}: its position angle is in agreement with the 1.3~mm B-field detections reported 
by \citet{Hull2014} (e.g. a P.A. of 95$^{\circ}$). We stress that the B-field topology in the outer envelope layers may be complex \citep{Cox2018} 
and differ from the main B-field reported here.
For L1448-2A, the hourglass shape is resolved with the SMA: we therefore used the central vector detected 
in the TADPOL survey \citep{Hull2014}. Finally, for IRAS16293-A, we estimated the mean magnetic field orientation in a smaller (3\arcsec\ in radius) 
aperture to avoid contamination by the companion source IRAS16293-B. The average value (173$\pm$40$^{\circ}$) is consistent with the NS 
orientation also found in \citet{Rao2009}, and the uncertainty is large. \\

We provide details of the magnetic field orientation of the additional eight sources in Appendix A. Most of the additional sources 
present misaligned configurations between the B and outflow position angles. For B1-bS, we note that we decided to use the average 
north-east value for our analysis because apparently it better traces the magnetic field at envelope scales connected with the field traced on larger scales. 
For L483, the eastern line segment shows a discrepancy with the global orientation of the western line segments: we did not 
include it in our analysis (see the discussion in Appendix A, which shows that this choice does not affect the correlation we find, however). \\

Finally, Table~\ref{AnglesVelGrad} also provides the outflow position angles retrieved from the literature (see references in the table). 
The outflow position angle uncertainties can vary from one source to another and depend on the outflow inclination and potential overlap of the red- and 
blueshifted components. We assumed a conservative 10$^{\circ}$ error on the outflow position angle for all sources.

%%%%%%% Position angle table %%%%%%%%%%%%
\begin{table*}
\centering
\caption{Position angles of the outflow and magnetic field and velocity gradient characteristics}
\begin{tabular}{ccccccccccc}
\hline
\hline
&&\\
Name                    & \multicolumn{2}{c}{Outflow}   && B$_{mean}$ && Line used& \multicolumn{4}{c}{Velocity gradients}       \\
&&\\
\cline{2-3}
\cline{5-5}
\cline{8-11}
&&\\
& P.A. $^a$ & Ref. && P.A. &&  & Gradient & P.A. & Ref.         \\
& ($\degr$) &&& ($\degr$) &&& (km~s$^{-1}$pc$^{-1}$) & ($\degr$) &      \\
&&\\
\hline
&&\\
Per B1-bS               
& 120 & [1]     &&      26$\pm$7    && N$_2$D$^+$   & 23$\pm$3 &  -8$\pm$8      & [26] \& T. w.   \\
Per B1c                 
& 125 & [2] &&  99$\pm$39  && N$_2$H$^+$ & 7.5 & 50 & [2] \& T. w.              \\
B335 
& 90 & [3]  &&  55$\pm$3    && N$_2$H$^+$   & $\sim$1.0   & - & [24]                    \\
BHR7-MMS                
& 174 & [4] &&  87$\pm$19        &      & N$_2$D$^+$ & 14$\pm$1.0 & -36$\pm$6 & [4] \& T. w.          \\
%&&&&&                                          & H$_2$CO       & 41$\pm$0.1 & -19$\pm$0.1 & [4] \& T. w.            \\
CB230                   
& 172 & [5] &&  85$\pm$4          && N$_2$H$^+$ & 3$\pm$0.1 & 98$\pm$1.3 & [21]  \\
HH25-MMS                
& - & [6]       &&      74$\pm$17   && - & - & - & -            \\ %outflow = 155, but not centered in SMM2
HH211-mm                
& 116 & [7] &&  174$\pm$5   && N$_2$H$^+$       & 7$\pm$0.03 & 26$\pm$0.3 & [21,23]       \\
HH212                   
&  23 & [8] &&  51$\pm$4   && NH$_3$    & 4.5   & 113   & [22]          \\
HH797                   
& 150 & [9] &&  110$\pm$7  && - & - & - & -             \\
IRAS03282               
& 120 & [10] && 43$\pm$6   && N$_2$H$^+$        & 9$\pm$0.01 & 114$\pm$0.01 & [21]  \\
IRAS16293-A             
& 75 \& 145     & [11] && 173$\pm$40 && CN      &   25          & -        &  [25]       \\
L1157                   
& 146 & [12] && 149$\pm$4  && N$_2$H$^+$ & 0.8$\pm$0.4 & 113$\pm$65 & [20]      \\
L1448C                  
& 162 & [13] && 95$\pm$4   && N$_2$H$^+$ & 13$\pm$1 & -179$\pm$1.0 & [20]       \\
L1448N-B                
& 105 & [14] && 23$\pm$4 && N$_2$H$^+$ & 13$\pm$1 & 100$\pm$1 & [20]    \\
L1448-2A                
& 134 & [5] && 139$\pm$9 $^b$ && N$_2$H$^+$ & 2$\pm$1 & -177$\pm$21 & [20]      \\
L483                    
& 105 & [15] && 8$\pm$11  && N$_2$H$^+$ & 9$\pm$0.03 & 45$\pm$0.2 & [21]        \\
NGC~1333 IRAS4A 
& 20 & [16] && 55$\pm$13 && N$_2$H$^+$ & 7$\pm$1 & 37$\pm$2 & [20]      \\
NGC~1333 IRAS4B 
& 0 & [17] && 51$\pm$19  && N$_2$H$^+$ & 3$\pm$1 & -71$\pm$14 & [20]    \\
Serp SMM18              
&  8 & [18] && 84$\pm$16  && N$_2$H$^+$ & 12$\pm$0.01 & 69$\pm$0.04 & T. w.      \\
SVS13-B                 
& 160 & [19] && 18$\pm$5   && N$_2$H$^+$ & 5$\pm$1 & 16$\pm$4 & [20]    \\
&&\\
\hline
\end{tabular}
\begin{list}{}{}
\item[$^a$] Position angles in the table are provided east of north. 
\item[$^b$] Taken from \citet{Hull2013}.
\item[References -]
T.~w. refers to this work,
[1] \citet{Gerin2015}, 
[2] \citet{Matthews2008}, 
[3] \citet{Hirano1988},
[4] \citet{Tobin2018},
[5] \citet{Hull2013},
[6] \citet{Dunham2014},
[7] \citet{Gueth1999},
[8] \citet{Lee2017},
[9] \citet{Tafalla2006},
[10] \citet{Hatchell2007},
[11] \citet{Rao2009},
[12] \citet{Bachiller2001},
[13] \citet{Dutrey1997},
[14] \citet{Kwon2006},
[15] \citet{Oya2018},
[16] \citet{Choi2006},
[17] \citet{Choi2001},
[18] \citet{Maury2019},
[19] \citet{Bachiller1998},
[20] \citet{Gaudel2020},
[21] \citet{Tobin2011},
[22] \citet{Wiseman2001},
[23] \citet{Tanner2011},
[24] \citet{Saito1999},
[25] Antonio Hern\'{a}ndez-G\`{o}mez's PhD thesis, private communication.
[26] \citet{Huang2013}.
\end{list}
\label{AnglesVelGrad}
\end{table*}
%%%%%%%%%%%%%%%%%%%%%%%%%

\subsection{Kinematic properties of the protostellar envelopes}

\noindent{\it \textup{The CALYPSO sources. }} For sources that are part of the CALYPSO sample (i.e., L1157, L1448C, L1448N-B, L1448-2A, 
IRAS4A, IRAS4B, and SVS13-B), \citet{Gaudel2020} recently presented observations of the 
dense gas kinematics using C$^{18}$O and N$_2$H$^+$ measurements. They derived specific 
angular momentum estimates throughout their collapsing protostellar envelopes from 50 au 
to 10000 au scales. In their analysis, velocity maps are produced from a combined 
Plateau de Bure (PdBI) + 30-m N$_2$H$^+$ dataset. A hyperfine structure line profile was used to determine the 
velocity of the molecular line emission in order to produce a velocity map. Then velocity gradients were
fit in a 40\arcsec\ $\times$ 40\arcsec\ region around the sources and determined by the least-squares minimization,
with v$_{grad}$ = v$_0$ + a$\Delta\alpha$ + b$\Delta\beta$, with $\Delta\alpha$ and $\Delta\beta$ 
the offsets with respect to the central source \citep{Goodman1993}.
Serp SMM18 is also part of the CALYPSO sample but is not included in 
the sample studied in \citet{Gaudel2020}. For this source, we determined the velocity gradient 
strength and position angle using the same 2D fitting technique. The velocity map is shown in 
the appendix in Fig.~\ref{VelocityRecalculated}. 

\vspace{5pt}
\noindent{\it \textup{CB230, HH211-mm, IRAS03282, and L483}. } For these sources, 2D velocity gradients and position angles on 
similar scales have been estimated in \citet{Tobin2011} using the N$_2$H$^+$ line and 
fitting a plane to the entire velocity field (with R $>$ 10\arcsec\ for L483 and R of 
about 20-25\arcsec\ for the other sources). We use their velocity gradients and position 
angles in the following analysis (see their Table 10). L1157 is also part of their sample, 
with a greater velocity gradient strength (from 3.5 to 10.5 km~s$^{-1}$~pc$^{-1}$) than was 
derived in \citet{Gaudel2020} (0.8 km~s$^{-1}$~pc$^{-1}$) and different position angles 
depending on the use of PdBI, Combined Array for Research in Millimeter-wave Astronomy (CARMA),
or Very Large Array (VLA) data. We kept the values from \citet{Gaudel2020} 
in this paper, but recall that the N$_2$H$^+$ velocity field is extremely complex 
for this source: the velocity gradient direction is close to the outflow axis, but the symmetry is
perturbed by a redshifted line emission southeast of the protostar, with potential contribution 
of the outflow cavity walls to the N$_2$H$^+$ emission for this source \citep{Tobin2011,Gaudel2020}.

\vspace{5pt}
\noindent{\it \textup{Per B1-bS.}  }We calculated the 2D velocity gradients from the N$_2$D$^+$ datacube obtained from N. Hirano
and presented in \citet{Huang2013}. Compared to its companion source B1-bN, the B1-bS line profiles are dominated by the 
V$_{LSR}$= 6.3 km~s$^{-1}$ component. We fit this hyperfine structure with the {\it hfs$\_$cube} procedure \citep{Estalella2017}, fitting the 
line when detected at a 5$\sigma$ level. The procedure returns a velocity map from which we
determined the gradient on a 20\arcsec\ $\times$ 20\arcsec\ region around the source to avoid contamination by the companion.
The map of the gas velocities in the envelope of B1-bS that we used to extract the main velocity gradient in this source 
is shown in  Fig.~\ref{VelocityRecalculated}. 
Because a smaller region was fit, the magnitude of the velocity gradient might only provide
an upper limit for this source: velocity gradients tend to increase when probed at smaller physical radii, 
as shown, for instance, by \citet{Gaudel2020} (their Table~2).

\vspace{5pt}
\textup{\textup{\noindent{\it \textup{Per B1-c.} }}}We calculated the 2D velocity gradients from the N$_2$H$^+$ velocity map 
obtained from B. Matthews and presented in \citet{Matthews2006}. As explained in their analysis, 
the spectral resolution was not sufficient to separate the hyperfine splitting components, therefore the moment map was 
taken over an isolated line component of N$_2$H$^+$. The velocity map of this source is presented in 
Fig.~\ref{VelocityRecalculated}. We fit the velocity gradient in a 20\arcsec\ $\times$ 20\arcsec\ region 
around the source to encompass the full N$_2$H$^+$ emission presented in \citet{Matthews2006} (their Fig.~8). 
As for Per B1-bS, the magnitude of the velocity gradient might provide an upper limit for this source 
because a smaller region was fit.
 
\vspace{5pt}
\noindent{\it \textup{BHR7-MMS.} } We calculated the 2D velocity gradients from both the H$_2$CO and N$_2$D$^+$ datacubes obtained from J. Tobin 
and presented in \citet{Tobin2018}. We used Gaussian line profiles to model the H$_2$CO emission and the {\it hfs$\_$cube} procedure \citep{Estalella2017},
fitting the hyperfine structure of the N$_2$D$^+$ line when detected at a 5$\sigma$ level. 
For H$_2$CO, as there is no robust detection beyond a radius of 7\arcsec\ (equivalent to 3000 au for BHR7),
we estimated the velocity gradient in a 14\arcsec\ $\times$ 14\arcsec\ region around the source and obtained a gradient of 40 km~s$^{-1}$~pc$^{-1}$. 
We note, however, that in the cold envelope, there might not be enough CO in the gas phase to form H$_2$CO: 
a significant part of the H$_2$CO emission could thus come from warmer gas belonging to the outflow, as 
suggested by the position angle (-19$^{\circ}$, thus aligned with the outflow) of the velocity gradient we derive.
In this respect, the N$_2$D$^+$ emission might better trace the gas kinematics in the envelope. The velocity map 
derived from the N$_2$D$^+$ datacube is presented in Fig.~\ref{VelocityRecalculated}. Using these data 
over a 30\arcsec\ $\times$ 30\arcsec\ region (equivalent to the $\pm$20\arcsec\ regions used for the Perseus sources),
we obtain a weaker but still strong velocity gradient of $\sim$15 km~s$^{-1}$~pc$^{-1}$. 
We chose to use this measurement in the remaining analysis.

\vspace{5pt}
\noindent{\it \textup{B335. }} B335 is a particular case; very little rotation is observed in the source. \citet{Menten1984} only reported a velocity shift from NH$_3$ 
observations of a few 10$^{-2}$ km~s$^{-1}$ over half an arcmin scale, which was confirmed by the weak velocity gradient estimated 
by \citet{Caselli2002}. Based on \citet{Saito1999} (their Fig 5), we derive a velocity gradient of 1 km~s$^{-1}$~pc$^{-1}$ over the 
interval $\pm$60\arcsec\ (equivalent to the $\pm$20\arcsec\ regions used in Perseus) in the NS direction (i.e., perpendicular to the outflow). 
A recent study by \citet{Watson2020}, based on the reflection nebulosity of a nearby star, suggests that the distance to B335 
could be 165~pc (we use 100~pc in this study). This greater distance, although it does not affect the mean B-field position angle derived for the source much, would lead to a velocity gradient about a factor of 2 lower than we used in this analysis.
We note that there is still much uncertainty in the position angle of the velocity gradient for this source; the C$^{18}$O velocity gradient is tilted by 18$\degr$ 
with respect to the east-west outflow direction \citep{Yen2015,Yen2015b}. We did not analyze the position angle of the velocity gradient
for this source.

\vspace{5pt}
\noindent{\it \textup{HH212 and IRAS16293-A.} } The remaining velocity gradients were directly taken from the literature or were obtained from private communication. 
For HH212, \citet{Wiseman2001} determined a velocity gradient of about 4--5 km~s$^{-1}$pc$^{-1}$ over a 25--30\arcsec\ region.
For IRAS16293, the gradient calculated from the H$^{13}$CO$^+$ velocity map of \citet{Rao2009} is huge (430 km~s$^{-1}$~pc$^{-1}$) 
and covers the whole IRAS16293-A/IRAS16293-B system. ALMA and SMA are both observing this large velocity gradient that might be 
partly contaminated by the various outflows emerging from this complex system. When CN observations are 
used (Antonio Hern\'{a}ndez-G\'{o}mez's PhD thesis\footnote{\url{https://tel.archives-ouvertes.fr/tel-02492210/}}, private communication), 
then the velocity gradient observed perpendicular to the main EW outflow decreases to 25 km~s$^{-1}$~pc$^{-1}$. We use this value in the paper.

%%%%%%%%%%%% Misalignment_VelGrad %%%%%%%%%%%%%%
\vspace{40pt}
\begin{figure*}
\centering
\begin{tabular}{cc}
\hspace{-15pt}\includegraphics[width=13cm]{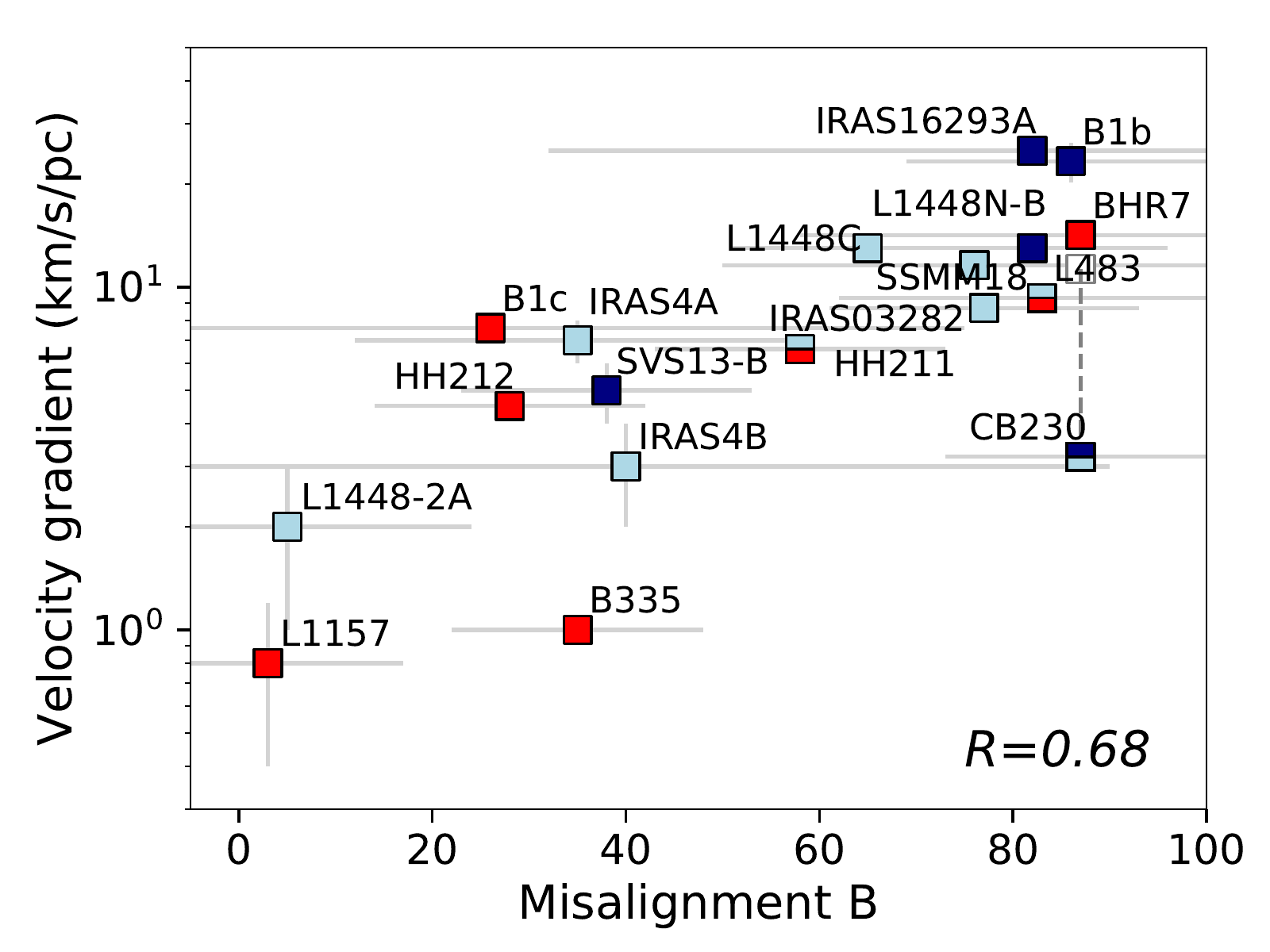} \\&
\end{tabular}
\vspace{-10pt}
\caption{Projected angle between the mean magnetic field within the 1000 au central region and the outflow direction as a function of the velocity 
gradient of the source estimated from line measurements. Sources are color-coded as a function of their fragmentation below 5000 au scales 
(with red, light blue, and dark blue for sources with a detection of a single, double, and 3-4 dust peaks). 
For certain sources, two colors are used: For L483, the ALMA 1.2 mm map from \citet{Oya2017} revealed a continuum source detected 
at a 5$\sigma$ level in the south-west region that might suggest fragmentation. 
For HH211, \citet{Lee2009} detected a companion source with the SMA in the southwestern extension of the source,
but VLA and ALMA observations have questioned the binary nature of the source \citep{Tobin2016,Lee2019}. 
In both cases, the nature of the additional source needs to be better investigated to be confirmed or refuted as a companion. 
Finally, the companion to CB230-A appears to host two near-IR objects, 
thus could be part of a triple system \citep{Massi2008}. For this source, we also indicate 
both the velocity gradient derived by \citet{Tobin2011} from a 1D fit perpendicular to the outflow direction 
(empty symbol) and a 2D fit of the total velocity field (filled symbol) because the two 
measurements lead to different values of the velocity gradient.
}
\label{MisalignmentVelGrad}
\end{figure*}
%%%%%%%%%%%% Misalignment_VelGrad %%%%%%%%%%%%%%

\vspace{5pt}
\noindent{\it \textup{HH25-MMS and HH797.} } Finally, we were unable to find observations for HH25 that would allow us to estimate a velocity gradient. Although it is located in the 
window of observations, the H$^{13}$CO$^+$ line is unfortunately not detected in this object by SMA. For HH797, the complex 
velocity pattern derived from the C$^{18}$O datacube from \citet{Palau2014} did not allow us to estimate a clean velocity 
gradient strength or direction. We decided to drop these two sources for the remaining analysis. 

\vspace{5pt}
\noindent 
The velocity gradient strengths (in km~s$^{-1}$~pc$^{-1}$) and position angles used in this analysis are summarized in Table~\ref{AnglesVelGrad}. 
The bulk of the velocity gradients resides in the [0--20] km~s$^{-1}$~pc$^{-1}$ range. This range is consistent with that derived for a sample
of 17 nearby protostellar systems by \citet{Tobin2011} (their Fig.~26 right). 
We note that velocity gradients aligned in the equatorial plane are commonly interpreted as 
envelope rotation. Recent analyses have revealed a more complex interpretation, 
with sources showing shifts or even reversal of the gas velocity gradients within envelopes. In some sources, the observed 
gradients might even originate from ongoing infall or be linked with turbulence \citep{Gaudel2020}. 
In all cases, however, large velocity gradients trace more dynamical envelopes with higher kinetic energy.

%%%%%%%%%%%%%%%%%%%%%%%%%%%%%%%%%%%%%%%%
%       Results   %
%%%%%%%%%%%%%%%%%%%%%%%%%%%%%%%%%%%%%%%%

%%%%%%%%%%%%  %%%%%%%%%%%%%%
\vspace{40pt}
\begin{figure*}
\centering
\begin{tabular}{cc}
\hspace{-10pt}\includegraphics[width=9.3cm]{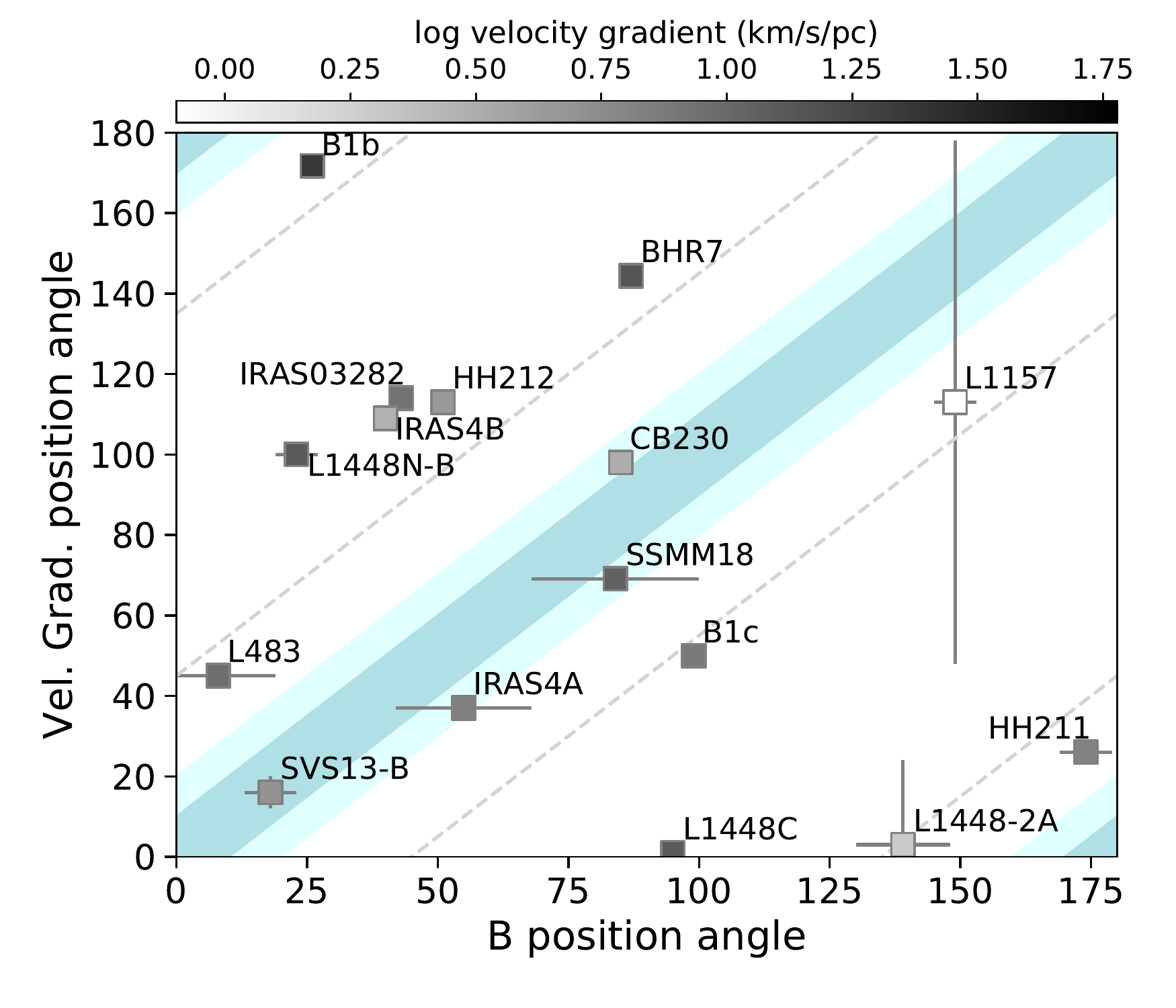} &
\hspace{-10pt}\includegraphics[width=9.3cm]{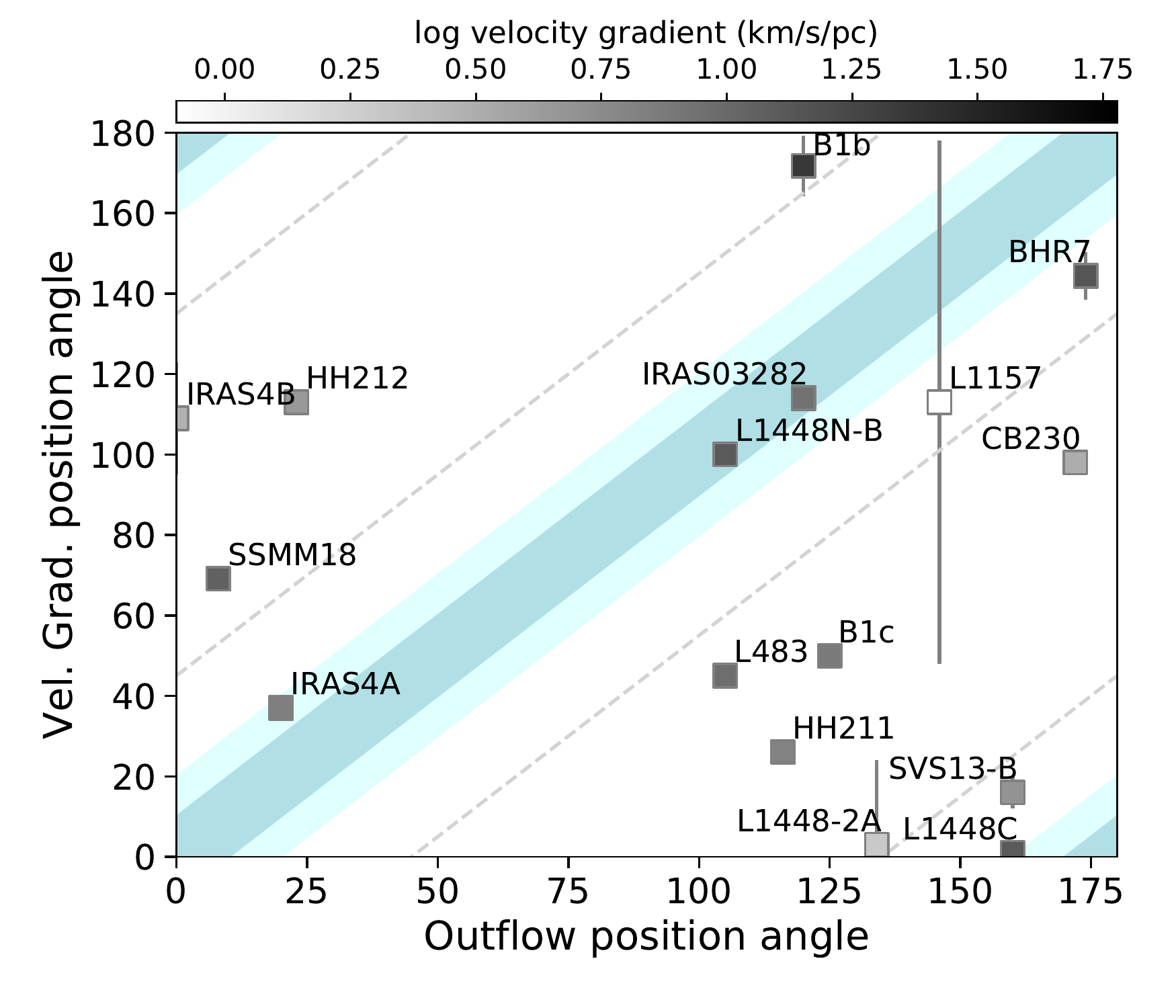} \\
\hspace{-10pt}\includegraphics[width=8.3cm]{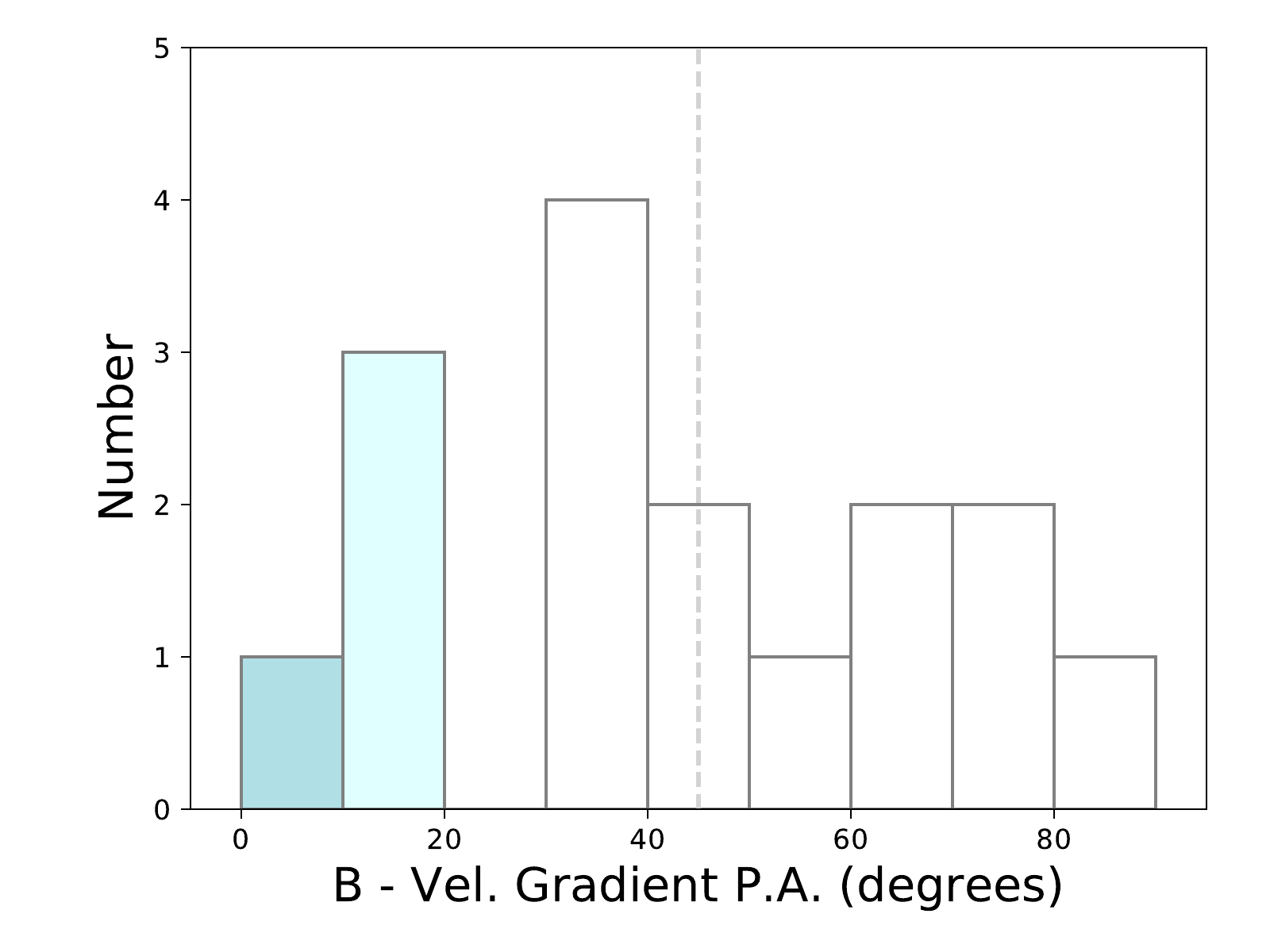} &
\hspace{-10pt}\includegraphics[width=8.3cm]{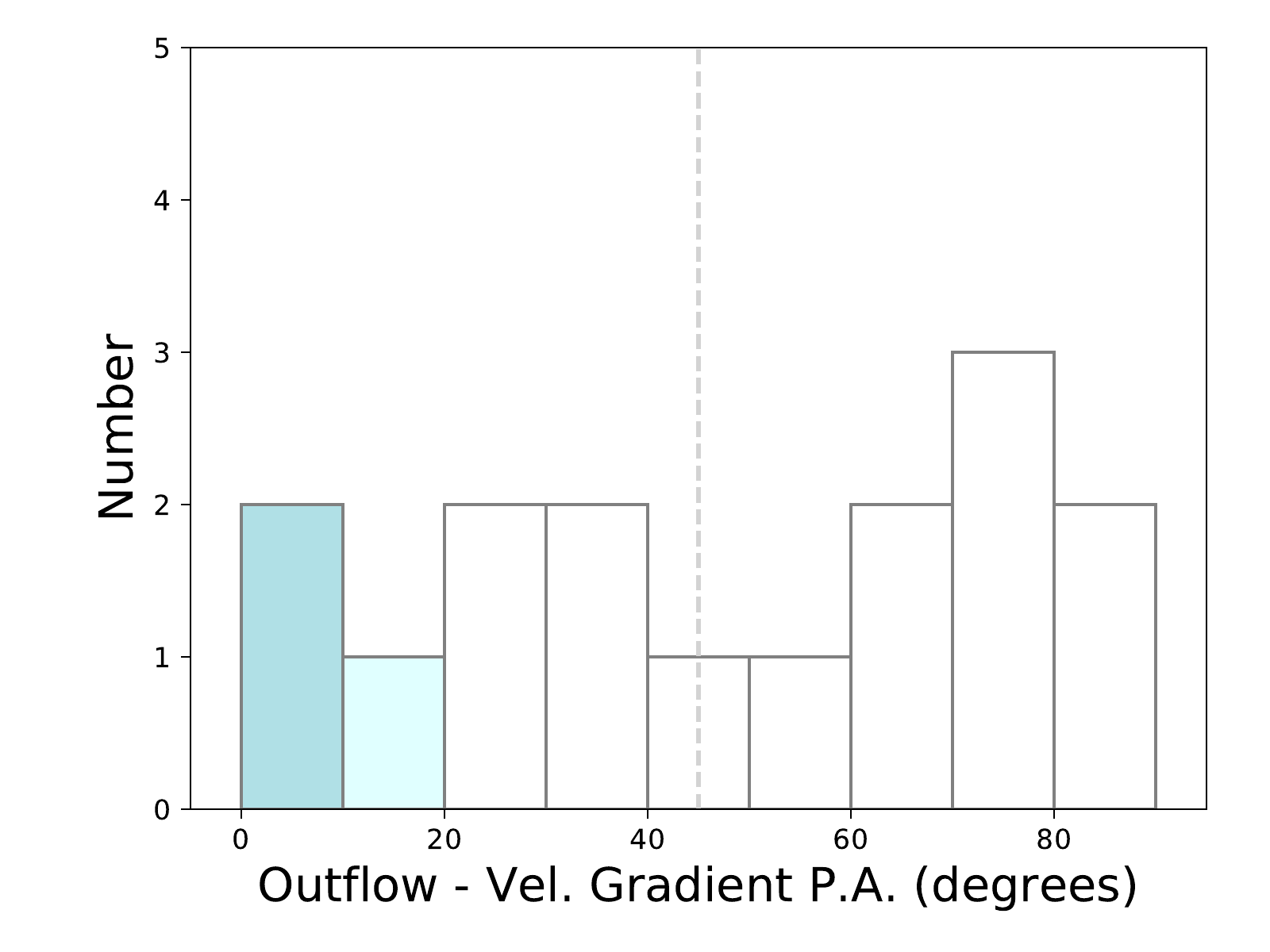} \\
\end{tabular}
\caption{{\it Top left:} Velocity gradient position angle as a function of the mean magnetic field position angle. 
The light and darker blue stripes indicate a projected angle between the two smaller than 20 and 10$\degr$ 
, respectively. The dashed line indicates a difference of 45$\degr$ between the two position angles. Protostars are 
color-coded as a function of their velocity gradient, with increasing gradients as the color darkens.
{\it Top right:} Relation between the velocity gradient position angle and the outflow direction. We use the same 
convention for the stripes and lines. We note that the blue regions indicate sources whose velocity gradient might 
partly be tracing the outflow motion rather than the envelope motions. 
{\it Bottom:} Corresponding histograms of the misalignment of (left) the B~field and (right) the outflow
axis with respect to the velocity gradient position angle. Colors and lines delineate the same angle offsets as in the top panels.}
\label{PAcomparison}
\end{figure*}
%%%%%%%%%%%% Comparison B PA versus Vel Grad PA %%%%%%%%%%%%%%

%%%%%%%%%%%%%%%%%%%%%%%%%%%%%%%%%%%%%%%%
%       Velocity gradients   %
%%%%%%%%%%%%%%%%%%%%%%%%%%%%%%%%%%%%%%%%

\section{Discussion}

\subsection{Misaligned B~fields associated with the small angular momentum of the protostellar gas}

 \citet{Galametz2018} qualitatively noted a higher occurrence of misaligned B-field lines in sources in which large velocity gradients were detected in the equatorial plane
at scales of thousands of au. The measurements of the envelope kinematics combined with B position angle measurements allow us to quantify the relation.
In Fig.~\ref{MisalignmentVelGrad} we show the projected angle between the B-field orientation and the outflow axis as a function of the 
velocity gradient (in km~s$^{-1}$~pc$^{-1}$) that we used as a proxy to probe the gas dynamics in the surrounding envelope. 
Errors on the misalignment angles (x-axis) are the addition of the B-field orientation error quoted in Table~\ref{AnglesVelGrad} 
and that of the outflow position angle error.
The colors used in the plot are discussed in \S~\ref{frag}. 

We observe a reasonably good (R=0.68) positive correlation between the misalignment of B with respect to the outflow and the strength of the velocity 
gradient traced at envelope scales. This quantitatively demonstrates that a relation exists between the orientation of the magnetic field and the 
kinematic energy in envelopes. This is consistent with the result presented in \citet{Yen2015} and based on a sample of 17 Class 0 and I protostars where no source 
with large specific angular momenta were found with a strongly aligned configuration. The correlation observed in Fig.~\ref{MisalignmentVelGrad} 
may be interpreted in various ways. One interpretation might be that the misalignments of B at envelope scales are driven by the strong 
rotational or infall motions of the envelope, while another interpretation might be that an aligned B~field could have favored the smaller 
velocity gradients we observe. \\

If the misalignment of B with respect to the outflow at envelope scales were driven by the envelope kinematics 
(i.e., if the initially aligned B-field lines were to be twisted by the infalling or rotating matter), we would expect a relation of the position 
angle of B and that of the velocity gradient. We plot the velocity gradient position angle as a function of the mean 
magnetic field position angle in Fig.~\ref{PAcomparison} (left). 
To facilitate visualization, the light and darker stripes indicate where the projected angle between the two directions 
is smaller than 20 and 10$\degr$ , respectively. The dashed line indicates a difference of 45$\degr$ between the two position angles. 
We color-code sources as a function of the velocity gradient strength. The bottom left panel shows the 
corresponding histogram of the angular difference between the position angle of B and the velocity gradient position angle. 

We observe that regardless of the height (so potentially effect) of the velocity gradient, we do not observe an alignment of the magnetic field 
direction and the velocity gradient position angle. This suggests that the 
B-field lines do not preferentially follow the direction of the matter infall and collapse dynamics. 
On the contrary, the sources seem to be scattered across the plot, as highlighted by the relatively flat corresponding histogram (Fig.~\ref{PAcomparison} bottom left). 
This could indicate that the correlation observed in Fig.~\ref{MisalignmentVelGrad} has a more complex explanation than matter infall or rotation 
that causes the misaligned magnetic field lines. This could favor the second interpretation, namely that in sources showing an aligned
configuration of B, magnetic braking could be more efficient at removing angular momentum, leading to the smaller velocity gradient we observe.

Reinforcing the interpretation that the correlation may be due to more efficient magnetic braking at removing angular momentum 
in sources initially in an aligned configuration of B further, we stress that most sources with low values of their envelope gas velocity gradient 
 also have an envelope B~field that is well aligned with the field observed in their surrounding environment.
The large-scale magnetic field lines probed around L1157 by Planck or probed at intermediate scales with SHARP \citep{Stephens2013,Chapman2013} 
are consistent with the SMA B orientation. We note that results from optical polarimetry are also consistent, with a small angle offset $<$20$^{\circ}$ \citep{Sharma2020}. 
For B335, the orientation of the near-IR polarization vectors seems to also fit with the orientation of the submillimeter polarization vectors \citep{Bertrang2014} 
and with the east-west direction found with JCMT-POL by \citet{Yen2019}. Finally, the B-field lines
in NGC~1333 IRAS4A, HH211, or L1448-2A appear to be extremely well ordered at scales traced by the SHARP or SCUBA instruments \citep[see]{Hull2014} down to the SMA scales.
B1-c does not follow this trend, however, with a large-scale B~field oriented with a position angle of 35$\degr$ \citep{Matthews2002}, that is, nearly perpendicular 
to the outflow direction, compared to the SMA 100$\degr$ orientation.

The nature of the gas kinematics recovered at these envelope scales (i.e., whether the angular momentum originates from rotation, 
infall, or even from turbulent motions inherited from the initial conditions and/or turbulence), is unknown. Its amount appears to be intimately 
linked to the magnetic configuration in protostellar envelopes, however.

%%%%%%%%%%%%%%%%%%%%%%%%%%%%%%%%%%%%%%%%
%       Fragmentation   %
%%%%%%%%%%%%%%%%%%%%%%%%%%%%%%%%%%%%%%%%
\subsection{Indication of higher multiplicity in systems with misaligned magnetic fields ?}
\label{frag}

In order to investigate the effect of the magnetic field on the envelope fragmentation, we color-code the sources 
of Fig.~\ref{MisalignmentVelGrad} depending on whether they are fragmented below 5000 au scales. 
We indicate in particular whether the source hosts a single, double, triple, or quadruple dust peak (using submillimeter 
direct imaging). Details of the fragmentation for each individual source are provided in Appendix A. 
We stress that very close multiplicity (below 100 au scales) is not considered in our analysis.

We observe that sources standing as single objects (in red) mostly appear to reside in the bottom left 
corner of the plot, that is, with relatively small velocity gradients of their surrounding 
envelopes and aligned magnetic field orientation with respect to the outflow axis. 
Uncertainties also remain about the nature of potential companions detected in HH211 and L483, hence these sources appear 
with two colors in Fig.~\ref{MisalignmentVelGrad}.
In order to assess whether the two populations (single versus multiple) belong to the same distribution, 
we applied a 2D Kolmogorov-Smirnov test. We used the two python scripts {\it ndtest.py}\footnote{https://github.com/syrte/ndtest/blob/master/ndtest.py}
 and {\it KS2D.py}\footnote{https://github.com/Gabinou/2DKS/blob/master/KS2D.py} to estimate the K-S statistics and p-values.
These 2D testings are based on statistical methods developed by \citet{Peacock1983} and \citet{Fasano1987}. Excluding the sources for which 
the exact multiplicity nature is unsure (i.e., HH211 and L483), both methods return the same low p-value of 0.13, indicating that the single and multiple 
source populations likely do not belong to the same population. More sources would, however, be necessary to reinforce this statistical test.
We also note that the two sources SVS13-B and IRAS4B have large-scale companions (SVS13-A and IRAS4B2 located at 4200 au and 3200 au 
 from SVS13-B and IRAS4B1, respectively), but they do not seem to be fragmented below 3000 au scales, to our knowledge. These sources have 
moderate velocity gradients and B is misaligned, so they are both located in the bottom left quadrant of Fig.~\ref{MisalignmentVelGrad}. 
When our separation criterion is whether fragmentation is observed above and below 3000 au scales (compared to 5000 au scales as before), 
the p-value of the previous K-S test drops to 0.04, which is also consistent with our conclusion that a dichotomy exists between the single 
versus multiple source population and that the magnetic field alignment might affect the way in which the envelope fragments.

Observational studies have suggested that the magnetic field might affect the fragmentation rate at molecular clouds or filaments scales 
\citep[e.g., ][]{Teixeira2016,Koch2018}. Our analysis at protostellar envelope scales appears to support the theoretical predictions
that the magnetic field orientation in the envelope also plays a role in favoring or inhibiting the fragmentation processes of a dense 
protostellar core into multiple systems. We further discuss the predictions from MHD simulations and their relation to our 
results in the following section.

%%%%%%%%%%%%%%%%%%%%%%%%%%%%%%%%%%%%%%%%
%       Simulations   %
%%%%%%%%%%%%%%%%%%%%%%%%%%%%%%%%%%%%%%%%

\subsection{Consistency with predictions from MHD simulations}

In ideal MHD models, B~fields are shown to strongly regulate the transport of angular momentum and hence modify the 
final properties of stars and disks \citep[see][and references therein]{HennebelleInutsuka2019,Wurster2018,Zhao2020}. 
The inclusion of more realistic non-ideal MHD effects changed this view, with finer effects that may play a crucial role 
regarding the ability of B to interact with gas kinematics, such as ionization and dust properties \citep{Zhao2018}.
Some studies have shown that B is in particular less efficient at transporting angular momentum when it is initially misaligned 
with the rotation of the system \citep{Joos2012}. 

That a link might exist between the initial core-scale magnetic field orientation with respect to the rotation axis that drives the 
small-scale outflow launching and fragmentation has been 
predicted by models of protostellar formation. Through the collapse of a dense rotating infalling core and in the extreme case of an envelope rotation axis 
initially perpendicular to the magnetic field, \citet{PriceBate2007} have shown, for instance, that the magnetic tension appears to play an even more significant 
role in helping fragmentation. Inserting magnetic fields also appears to be a necessary condition to reproduce the fragmentation rates now observed in massive cores with 
ALMA, as suggested by \citet{Fontani2016}. The observations provide information of the main field direction at envelope scales: in an aligned configuration,
thus more organized magnetic field configuration, the magnetic pressure will be more efficient at stabilizing the envelope, reducing the rotation-induced
fragmentation at comparable scales (i.e., below 5000 au scales). If fragmentation were favored by less efficient magnetic braking processes, it could also be 
enhanced by `turbulent' fragmentation, that is, fragmentation linked to the increase of the kinematic energy of envelopes in misaligned configurations. 
Following our observational results, a study of non-ideal MHD models of protostellar collapse that searched for the roots of the correlation we report was 
initiated and is currently carried out by the team. \\

The question is whether a potential effect of the misaligned B-field orientation on 
the final protostellar disk sizes might be observed. If less angular momentum is transported from the envelope to the disk scales in 
the case of B-aligned configuration, as suggested by our study at envelope scales, 
smaller rotation-supported disks are expected. Observationally, the small size of disks in some Class 0 protostars 
has been interpreted as a consequence of an efficient magnetic braking that potentially
disrupts the disk formation \citep[][]{Maury2010,Yen2015,SeguraCox2018}. Magnetohydrodynamics simulations from \citet{Hirano2020} 
recently confirmed that in the later accretion phase, the smallest disk radius and mass are produced in alignment-configuration cases.
Using the CALYPSO sample, \citet{Maury2019} have shown that fewer than 25\% of the 26 Class 0 protostars may harbor large 
protostellar disks resolved at radii $>$60 au. Their results also favor a magnetized scenario for the disk formation. 
Unfortunately, most sources with large disk-like structures (L1527, SMM4, MM22, and GF9-2) 
are not included in the current sample. The disk sizes are mostly unresolved at $\sim$50 au scales in the 
remaining sources of our sample that overlap with the CALYPSO sample. This prevents us from studying the effect of the B misalignment 
on the disk formation itself. Such studies are currently hampered by the need of very high spatial resolution 
to study these disks with ALMA, for instance. Recent ALMA results from \citet{Cox2018} have analyzed, for instance, the 
polarization angle dispersion and tried to connect the signature or randomness of the B~field with the disk or non-disk nature of their sources.
The extremely complex morphology of magnetic fields in Class 0 protostars observed with ALMA at 50 au typical disk scales 
may suggest a less dynamically relevant B~field at these scales, as expected from non-ideal MHD models \citep[e.g., with ambipolar 
diffusion leading to a weaker coupling of the magnetic field lines with the circumstellar gas;][]{Mellon2009,Tsukamoto2015}. We stress, however, that 
the characterization of B in disks remains problematic, as several mechanisms can contribute to the production of polarized 
dust emission at these scales \citep{Kataoka2015,Cox2018}.

\subsection{Assessing the various caveats}

% Outflow contamination  %
\subsubsection{Outflow contamination}
\label{OutflowContamination}

Assessing the preferential direction and strength of the moving gas in protostellar envelopes is complicated by the presence 
of outflows (sometimes more than one, e.g., in the case of IRAS16293-A) 
that can contribute to the observed gradients, although our choice of mostly using a dense or cold gas tracer 
such as N$_2$H$^+$ should largely limit the contamination. In Fig.~\ref{PAcomparison} (right) 
we plot the velocity gradient position angle as a function of the outflow direction. 
The dashed line delineates a difference of 45$\degr$ between the two position 
angles. The light and darker stripes indicate when the projected angle between 
the two is smaller than 20 and 10$\degr$ , respectively, which indicates regions with sources 
whose velocity gradient might be more related to the outflow dynamics than to 
the envelope kinematics we analyzed here. 
The bottom right panel of Fig.~\ref{PAcomparison} provides the histogram of the misalignement of the outflow and the gradient. The dispersion 
in the outflow - gradient misalignment is consistent with the results from \citet{Tobin2011} (their Fig.~27). 
For only 3 sources is the velocity gradient roughly aligned with the outflow direction, that is, with $\Delta$(P.A.) $<$ 30$\degr$ 
(see Fig.~\ref{PAcomparison} right).

% Projection effects  %
\subsubsection{Projection effects}
\label{proj}

\begin{figure}
\centering
\includegraphics[width=8cm]{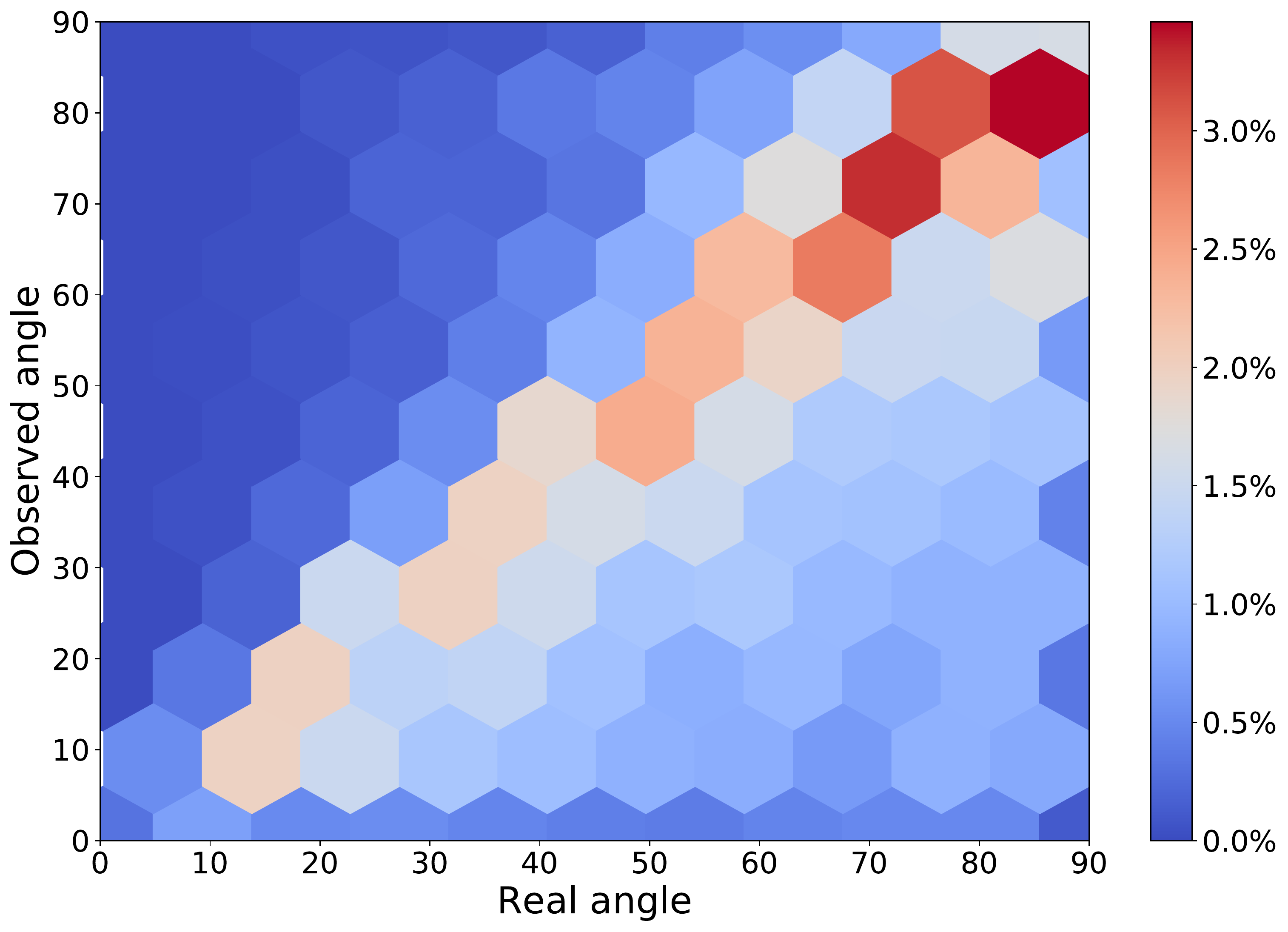} 
\caption{Monte Carlo test for 10000 randomly generated vector pairs projected onto a 3D plane
mimicking the plane of the sky. This plot shows the density distribution of the angles obtained by projecting
these vector pairs onto the plane compared to their real angle separation in 3D. The color bar indicates the 
probabilities of each projected vs. true angle pair.}
\label{PlotMonteCarlo}
\end{figure}

The position and misalignment angles quoted in this analysis are projected in the plane 
of the sky and might differ from the real angles between the various directions analyzed 
(velocity gradients, B, and outflows). 
In order to assess the uncertainties linked with projection effects, we performed a Monte Carlo
test. We developed a python script that randomly generates a 3D plane. This plane was our plane of the sky. 
We then generated 10000 random pairs of vectors in this 3D space. For each pair, vectors were projected 
onto the 3D plane (using the orthogonal basis of the plane) to estimate the two `projected vectors' and thus `observed angle'
between the two. The `real' and `observed' angles were then compared. 
Figure~\ref{PlotMonteCarlo} shows the density distribution
of the observed versus real angles. The 2D density plot is generated with 
the {\it hexbin} python function (Seaborn package). For simplicity, we modified 
the color bar to indicate the probabilities of each observed-real angle pair. 
As expected, the projected angles are most of the time equal to (i.e., on the diagonal) 
or lower (i.e., on the bottom right area) than the corresponding real angle. 
Qualitatively, the bright diagonal highlights that we did not make a fundamental 
mistake by taking the observed angle as a proxy for the real angle.

To quantify the effects, we separated the `observed' angles into 18 bins of 5$\degr$ each and provide 
the mean `real angle' in Table~\ref{TableMonteCarlo}. The standard deviations for each 
observed angle bins are provided as uncertainties. We note that the `true angle' 
distributions corresponding to each bin are not Gaussian. We observe that the largest discrepancies between 
the `real' and 'observed' angles appear for `observed' angles below 40$\degr$, with large error bars. 
In Fig.~\ref{MisalignmentVelGrad}, these projection effects could therefore shift the sources 
with a B misalignment lower than 40$\degr$ to the right and realign these sources along 
a more global linear correlation because the projection effects do not strongly affect sources 
beyond a 40$\degr$ misalignment.
Synthetic observations derived from MHD simulations are currently developed by the 
team to complement these tests on projection effects (Valdivia et al. in prep.).

%%%%%%%%%%%% Monte Carlo results - Plot %%%%%%%%%%%%%%
%%%%%%%%%%%% Monte Carlo results - Plot %%%%%%%%%%%%%%

%%%%%%% Monte Carlo results - Table %%%%%%%%%%%%%%%%%%%%%%%%
\begin{table}
\centering
\caption{Mean `true angle' per `observed angle' bins}
\begin{tabular}{cc}
\hline
\hline
\vspace{-5pt}
&\\
Observed angle          &       Mean `true angle' \\
(\degr) & (\degr) \\
\vspace{-5pt}
&\\
\hline
&\\
\vspace{1pt}[0~-~5]      & 38.8~$\pm$~25.4 \\\relax
\vspace{1pt}[5~-~10]     & 36.2~$\pm$~25.3 \\\relax
\vspace{1pt}[10~-~15]    & 41.3~$\pm$~24.3 \\\relax
\vspace{1pt}[15~-~20]    &  38.7~$\pm$~22.5 \\\relax
\vspace{1pt}[20~-~25]    &  41.4~$\pm$~21.6 \\\relax
\vspace{1pt}[25~-~30]    &  46.0~$\pm$~19.7 \\\relax
\vspace{1pt}[30~-~35]    &  48.3~$\pm$~18.5 \\\relax
\vspace{1pt}[35~-~40]    &  50.0~$\pm$~18.8 \\\relax
\vspace{1pt}[40~-~45]    &  52.4~$\pm$~16.1 \\\relax
\vspace{1pt}[45~-~50]    &  54.6~$\pm$~16.3 \\\relax
\vspace{1pt}[50~-~55]    &  60.5~$\pm$~15.0 \\\relax
\vspace{1pt}[55~-~60]    &  65.5~$\pm$~14.6 \\\relax
\vspace{1pt}[60~-~65]    &  65.4~$\pm$~14.5 \\\relax
\vspace{1pt}[65~-~70]    &  69.6~$\pm$~14.0 \\\relax
\vspace{1pt}[70~-~75]    &  73.4~$\pm$~14.2 \\\relax
\vspace{1pt}[75~-~80]    &  76.6~$\pm$~14.4 \\\relax
\vspace{1pt}[80~-~85]    &  79.1~$\pm$~15.1 \\\relax
\vspace{1pt}[85~-~90]    &  80.2~$\pm$~14.5 \\
\vspace{-5pt}
&\\
\hline
\end{tabular}
\label{TableMonteCarlo}
\vspace{-10pt}
\end{table}
%%%%%% Monte Carlo results - Table  %%%%%%%%%%%%

\subsubsection{Dependence on tracers} 

One of the caveats of this analysis is also that it depends on the molecular line tracers chosen. Both C$^{18}$O and 
N$_{2}$H$^{+}$ datacubes are available for the CALYPSO sources. We decided to select N$_{2}$H$^{+}$ as a tracer 
of the envelope kinematics. Several analyses have shown it to be a robust tracer of the outer envelopes, thus the scales 
we trace with the SMA, rather than of the central regions where this molecule is usually depleted and whose kinematics gas
are then usually traced through C$^{18}$O or H$_2$D$^+$ \citep{Bergin2002,Anderl2016,Gaudel2020,Maret2020}. 
We used the NH$_3$ molecule for HH212: the joined analysis of N$_2$H$^+$ and NH$_3$ by \citet{Tobin2011} has confirmed 
that both lines trace similar physical conditions at the scales we study here. 
We also selected N$_{2}$D$^{+}$ and CN as complementary tracers of cold dense gas. Thus, unless some strongly 
different chemical effects are at work in the different protostellar envelopes, all sources have been observed with homogeneous gas tracers. 
Variations in the opacity from one source to another could bias the physical scales at which the velocity gradients are probed, 
with the observed N$_{2}$H$^{+}$ arising from projected shells or more external regions than what we assumed. 
N$_{2}$H$^{+}$ or N$_2$D$^+$ could also be absent near the protostar, although this should not have a strong effect 
on the scales probed in this analysis. A more extended analysis using other tracers of the gas dynamics would help test the robustness 
of our conclusions at various envelope radii and/or for gas that is subject to different local conditions.

%\subsubsection{Distance uncertainties} 
%One also has to remember than by gathering sources from different star-forming clouds, thus at various distances, 
%our SMA survey is not probing exactly the magnetic field morphology at comparable physical scales in each source. 
%The observations of sources with various facilities (single-dish, interferometers) and thus resolutions that the B-field 
%morphology, in particular its orientation, can change from the host cloud to the core and disk scales 
%\citep[][among others]{Brauer2016,Zhang2014,Maury2018}. On top of these considerations, the determination of the 
%exact distance to a single star-forming cloud is also difficult. Gaia results for instance led to strong revisions within the 
%past 10 years \citep{Ortiz-Leon2018,Ortiz-Leon2018_2}. Current efforts are on-going in order to derive distances in a 
%more homogeneous manner \citep{Zucker2019,Zucker2020}. 

%%%%%%%%%%%%%%%%%%%%%%%%%%%%%%%%%%%
%       Summary                 %
%%%%%%%%%%%%%%%%%%%%%%%%%%%%%%%%%%%

\section{Summary}
 
We have carried out dust polarization observations at 0.87 mm for 20 Class 0 protostars with the SMA. We 
analyzed the misalignment of the magnetic field orientation (derived over the central 1000 au scales) with the outflow orientation
in order to compare this misalignment with the gas kinematics. The dynamics of the gas was traced through measurements of velocity 
gradients at envelope scales using molecular line emission. 
Our analysis provides for the first time a quantification of the striking 
correlation that links the misalignment of the magnetic field orientation at envelope scales and the angular momentum 
of the gas reservoir that is directly involved in the star formation process. Comparing the trend with the presence of 
multiple stellar systems, we also show that 
sources that tend to stand as single objects mostly reside in environments with a weak velocity gradient and/or a rather aligned B-field orientation
compared to the outflow axis. Altogether, the observations tend to show a coherent picture for the role of the magnetic 
field in forming stars and their protoplanetary disks: they suggest that strong B~fields in an aligned configuration may be more efficient 
in regulating both the gas kinematics and the level of fragmentation during the early embedded phases of star and disk formation.
Our findings could be in line with theoretical expectation from the most recent magnetized models of star formation, which predict reduced 
angular momentum at smaller scales due to magnetic braking, although a thorough exploration of the physical causes behind the observed 
correlations should be explored in numerical models. Our results provide a strong observational confirmation of the cornerstone role of
B in regulating the formation of stellar systems and settling the primordial conditions from which the future disk, star, and planets will form.

%%%%%%%%%%%%%%%%%%%%%%%%%%%%%%%%%%%
% Acknowledgements      %
%%%%%%%%%%%%%%%%%%%%%%%%%%%%%%%%%%%

\section*{Acknowledgments}
We thank the anonymous referee for his/her constructive suggestions 
that improve our methodology descriptions and overall quality of the manuscript.
This project has received funding from the European Research Council (ERC) 
under the European Union Horizon 2020 research and innovation programme 
(MagneticYSOs project, grant agreement N$\degr$ 679937, PI: Maury). 
J.M.G. is supported by the grant AYA2017-84390-C2-R (AEI/FEDER, UE).
We thank John Tobin for providing us with the BHR7 H$_2$CO and N$_2$D$^+$ datacubes 
\citep[presented in][]{Tobin2018}. We also thank Naomi Hirano and Brenda Matthews 
for providing us with the N$_2$H$^+$ data cubes for B1b and B1c respectively (data 
presented in \citet{Huang2013} and \citet{Matthews2008}). We finally thank Aina Palau 
for providing us with the C$^{18}$O data cube for HH797 \citep[see][]{Palau2014}. 
This publication is based on data of the Submillimeter Array. The SMA is a joint 
project between the Smithsonian Astrophysical Observatory and the Academia 
Sinica Institute of Astronomy and Astrophysics, and is funded by the Smithsonian 
Institution and the Academia Sinica.

%%%%%%%%%%%%%%%%%%%%%%%%%%%%%%%%%%%
%       Bibliography            %
%%%%%%%%%%%%%%%%%%%%%%%%%%%%%%%%%%%
\vspace{-10pt}
\bibliographystyle{aa}
\bibliography{/Users/mgalamet/Documents/Work/Papers/mybiblio_MY.bib}

%%%%%%%%%%%%%%%%%%%%%%%%%%%%%%%%%%%
%       Appendix                                                        
%%%%%%%%%%%%%%%%%%%%%%%%%%%%%%%%%%%
\appendix

\section{Individual source characteristics}
\label{SourceCharacteristics}

\noindent{\bf B1-bS}

\vspace{2pt}
\noindent{\it Continuum morphology at 850 \mic\ .}
When the large-scale envelope of Per B1-b is elongated in the north-south 
direction and encompasses the two sources B1-bN and B1-bS \citep{Chen2013},
our SMA observations only focus on B1-bS. They reveal a core elongated in the northwest-southeast direction at 1000-2000 au scales, a morphology very similar to that observed at 3mm with the Nobeyama Array \citep{Hirano1999} or 850 \mic\ with ALMA \citep{Gerin2017}. 

\vspace{2pt}
\noindent{\it Fragmentation. }
The two submillimeter sources in B1-b were first discovered by \citet{Hirano1999}. The companion of B1-bS is called B1-bN 
and is located $\sim$16\arcsec\ north. It was studied with the VLA by \citet{Tobin2016}. 
The observations allowed them to show that the two sources have similar temperatures, but that B1-bS is twice as luminous as B1-bN. 
Both sources seem to be extremely different in terms of their richness in complex organic molecules (COM), as revealed by recent ALMA observations \citep[][]{Marcelino2018}. The ALMA results do not reveal further fragmentation in B1-bS below 200 au scales. 
The source Per-emb-41, located only 13\arcsec\ in the southwest of B1-bS, appears to be more evolved \citep{Tobin2016}.
f
\vspace{2pt}
\noindent{\it Magnetic field orientation. }
On the large scales observed by SCUBA \citep{Matthews2002} or the B-fields In STar-forming Region Observations (BISTRO) survey \citep{Coude2019}, the B1-b system already shows strong variations 
in its polarization position angle, with a mean value of the magnetic field direction position angle toward B1-bS of about 30$\degr$. This orientation
is consistent with the average B-field orientation obtained with the SMA in the northeast region (26$\degr$). The western vectors are oriented at 111$\degr$, that is, perpendicular to the eastern vector and along the outflow direction. We decided to use the average northeast value for this analysis because it appears to trace the magnetic field better at envelope scales connected with the magnetic field traced on larger scales. \\

\noindent{\bf B1-c}

\vspace{2pt}
\noindent{\it Continuum morphology at 850 \mic.\ } The large-scale 2.7 and 3.3 mm continuum emission observed by \citet{Matthews2006} presents extensions mostly along the outflow. In contrast, the 850 \mic\ SMA continuum is flattened in the direction perpendicular to the source outflow. This is consistent with the SMA 1.3 mm dust continuum image presented in \citet{Chen2013}. The northern and southwestern plumes are also observed with ALMA at 870 \mic\ \citep[see][]{Cox2018}.

\vspace{2pt}
\noindent{\it Fragmentation. } 
The source is a single source at the SMA scales (this work),  at the VLA scales probed by the VLA/ALMA Nascent Disk and Multiplicity (VANDAM) 
survey \citep{Tobin2016}, and  at the ALMA scales probed by \citet{Cox2018}. 

\vspace{2pt}
\noindent{\it Magnetic field orientation. }          
At the large scales observed by SCUBA \citep{Matthews2002}, B1-c has a
polarization angle at 35$\degr$, that is, a magnetic field nearly perpendicular to the outflow
direction \citep[estimated at -55$\degr$;][]{Matthews2006}. \citet{Matthews2008} and the results from the BISTRO survey \citep{Coude2019} revealed a more complex pattern.  
Our SMA results are consistent with these more recent results: the SMA northern vectors are oriented at an angle of 152$\degr$ and the central vectors are oriented in an east-west orientation, with an average in position angle of 70$\degr$. The average in the central 1000 au region is 99$\degr$. Our results suggest that the hourglass configuration of the field lines observed with ALMA by \citet{Cox2018} already starts to be resolved at SMA scales. \\

\noindent{\bf BHR7-MMS}

\vspace{2pt}
\noindent{\it Continuum morphology at 850 \mic.\ } The SMA 850 \mic\ continuum emission is elongated in the north-south direction. By comparison, the SMA 1.3 mm continuum emission presented in \citet{Tobin2018} is much more compact and marginally extended in the east-west direction, but was observed with the Very Extended Configuration, which has a resolution that is three to four times higher than that of the current analysis.

\vspace{2pt}
\noindent{\it Fragmentation. } 
BHR7-MMS is an isolated dark cloud. The source does not seem to be fragmented at the intermediate scales probed
with the SMA \citep[][and this work]{Tobin2018}. Recently taken but not yet published ALMA observations of the source should
reveal the inner morphology of the source in the coming years. 

\vspace{2pt}
\noindent{\it Magnetic field orientation. }
To our knowledge, this is the first time that polarized dust emission is used to probe the magnetic field direction in this source. 
We observe that B is oriented east-west, perpendicular to the direction of the outflow. The northern and southern vectors are 
slighted tilted compared to the orientation of the central detection, which might be the signature of an hourglass morphology. \\

\noindent{\bf HH25-MMS}

\vspace{2pt}
\noindent{\it Continuum morphology at 850 \mic.\ } 
HH25-MMS is part of a string of embedded young stellar objects \citep[the HH24-26 complex;][]{Bontemps1995,Gibb_Davis1998}. SMA 870 \mic\ observations by \citet{Chen2013} have shown that three distinct sources compose the system (SMM1, SMM2, and SMM3) and are aligned in the north-south direction. If SMM1 were the driving source of the HH25 outflow, the SMA polarization data focus on the SMM2 source. Its SMA continuum appears to be extended in the east-west direction.
% Lbol = 6Ls - Tbol = 34K

\vspace{2pt}
\noindent{\it Fragmentation. }
As mentioned previously, SMM2 is part of a three-source system, with separations of 13 and 11\arcsec\ with SMM1 and SMM3, respectively. On larger scales, HH25-MMS also has a Class I companion, HH26IR, located 1.5\arcmin\ to the southwest. 

\vspace{2pt}
\noindent{\it Magnetic field orientation. }
Elongated in the north-south direction, the HH25-MMS system is separated into three distinct sources (SMM1, SMM2, and SMM3) resolved by \citet{Chen2013}. The SMA observation focuses on SMM2. This is the first known detection of polarized dust emission toward this source. However, we do not detect polarization toward the very central region but in its western extension. The magnetic field at this position is tilted in the 74$\degr$ direction, that is, mainly perpendicular to the projected line connecting the three sources. In this analysis, we did not compare the B-field orientation with the known outflow of the region (revealed by a VLA 3.6 cm survey) because there is strong evidence that the outflow originates from SMM1 and not from SMM2 \citep{Bontemps1995,Chen2013}.\\

\noindent{\bf HH211-mm}

\vspace{2pt}
\noindent{\it Continuum morphology at 850 \mic.\ } As observed at 230 GHz by \citet{Gueth1999}, the HH211-mm submm continuum emission is compact. Also resolved with the SMA, the continuum is slightly elongated in the southwest direction, that is, perpendicular to the outflow axis.

\vspace{2pt}
\noindent{\it Fragmentation. }
Higher resolution SMA observations also performed at 850 \mic\ have revealed that HH211-mm hosts two sources separated by 0.3\arcsec\  \citep{Lee2009}. 
The first, SMM1 is the protostar from which the collimated outflow originates, and SMM2, in its southwest, is responsible for the southwestern 
extension we observe in our analysis. Modeling the jet wiggle, \citet{Lee2010} also suggested that the HH211-SMM1 source itself might be a proto-binary source 
with a separation of $\sim$5 au. Using ALMA data, \citet{Lee2019} have however recently questioned the nature of SMM2 as a secondary source.
The companion source is not detected either as part of the VANDAM Perseus survey \citep{Tobin2016}.

\vspace{2pt}
\noindent{\it Magnetic field orientation. }
The central field lines traced by the SMA observations have a north-south direction, that is, they are roughly perpendicular to the outflow direction. This average 175$\degr$ orientation is consistent with the results from the TADPOL survey from \citet{Hull2014} and the SCUBA-POL results from \citet{Matthews2009}. This is also consistent with the central, northeast, and northwest field lines traced at a 0.6\arcsec\ resolution with the SMA by \citet{Lee2014} down to the ALMA scales presented in \citet{Lee2018}. Farther away from the center, the B-field lines seem to realign in the direction of the outflow axis (see the eastern and western vectors). \\

\noindent{\bf HH212}

\vspace{2pt}
\noindent{\it Continuum morphology at 850 \mic.\ }
The SMA observation reveals a disk-like shape in the direction perpendicular to the outflow, consistent with the continuum emission detected in \citet{Wiseman2001} and observed at 0.9 mm with ALMA \citep{Codella2014}. 

\vspace{2pt}
\noindent{\it Fragmentation. }
SMA observations at higher resolution than those presented in this paper have suggested that one, if not several, faint sources are located 
around the main body of the HH212-mms source \citep{Lee2008,Chen2013}. However, more recent ALMA observations by \citet{Codella2014} 
taken at a similar resolution (0\farcs5 resolution) did neither resolve nor detect these sources. These fainter sources detected with the SMA 
are probably detections within HH212 flattened envelope. The VANDAM team, using VLA observations, derived a Toomre {\it Q}
parameter for this source that is consistent with a marginally unstable disk, but did not detect multiplicity in the source \citep{Tobin2020}.
We therefore consider the source as a single source until further analysis confirms or refutes the presence of additional dust peaks.

\vspace{2pt}
\noindent{\it Magnetic field orientation. }
The field lines in the central plane are oriented in a northeast-southwest direction, close to the outflow axis of the source. 
The southern vectors seem to be following the outflow cavity walls.
We note that B traced with ALMA, in contrast, is perpendicular to the outflow direction \citep{Lee2018}. 
The authors suggested that this orientation might be due 
to dust polarization arising from self-scattering at these small scales. \\

\noindent{\bf L483}

\vspace{2pt}
\noindent{\it Continuum morphology at 850 \mic.\ }
Our SMA observations reveal a dust continuum elongated in the direction
perpendicular to the outflow direction. The southwestern extension is consistent with the extension observed by \citet{Park2000} at 3.4 mm.
%Lbol ? 10Ls

\vspace{2pt}
\noindent{\it Fragmentation. } 
L483 appears to be an isolated dense core down to the 2\arcsec\ $\times$ 1\farcs5 scales traced by the SMA observations, as has been suggested by \citet{Jorgensen2004} and \citet{Chen2013}. Recent 1.2mm ALMA observation resolves the southwestern extension, with a separate continuum source detected at a 3$\sigma$ level \citep{Oya2017}. Its nature  was not discussed, however. 

\vspace{2pt}
\noindent{\it Magnetic field orientation. }
Polarized emission is not detected toward the center of the source, but is dectected in the southwestern extension (two consistent detections). Its direction is perpendicular to the southeast-northwest outflow direction of the source and also perpendicular to the mean B-field orientation at core scales (93$\degr$) traced at 350 \mic\ using SHARP by \citet{Chapman2013}.
We also have a detection in the eastern part of the source, although it is associated with weaker continuum emission (hence its high associated 
polarization fraction), with a B position angle of 90$\degr$. We did not take this vector into account in our calculation of the magnetic field 
orientation because of its discrepancy with the other two detections. If the three vectors were into account, the weighted mean B position angle would be 50$\degr$, 
leading to a misalignment  of 55$\degr$ between the magnetic field and the outflow orientation. This would place L483 in the center of the correlation observed in Fig.~\ref{MisalignmentVelGrad} and reinforce the general correlation we observe. \\

\noindent{\bf Serpens South MM18}

\vspace{2pt}
\noindent{\it Continuum morphology at 850 \mic.}
The SMA continuum emission is extended in the western and southern part of the source. This extension is consistent with that found from the PdBI by \citet{Maury2019}.

\vspace{2pt}
\noindent{\it Fragmentation. } 
\citet{Maury2011} found that SerpS-MM18 is separated into two sources:  
MM18a is the primary protostar, followed by a weaker secondary source MM18b
10\arcsec\ in its southwest. This result has been confirmed by the observations of the Serpens South complex by \citet{Plunkett2018}. 

\vspace{2pt}
\noindent{\it Magnetic field orientation. }
H- and Ks-band polarization measurements have shown that the large-scale magnetic field is globally well ordered perpendicular to the main Serpens South filament, thus in an east-west direction \citep{Sugitani2011}. When we zoom in on SSMM18, the magnetic field lines are also oriented in this EW direction, with lines perpendicular to the outflow axis. The divergence of the eastern vectors could be a signature of an hourglass morphology. \\

\section{Effect of the pixel size on the mean B-field orientation}

%%%%%%% Position angle table %%%%%%%%%%%%
\begin{table}[!ht]
\centering
\caption{Effect of the pixel size on the mean B-field orientation}
\begin{tabular}{cccccc}
\hline
\hline
&&\\
Name                    & \multicolumn{5}{c}{B$_{mean}$ P.A.}   \\

&& \\
\cline{2-6} \\
                        & 0.6\arcsec\   & 0.7\arcsec\   & 0.8\arcsec\   & 0.9\arcsec\     & 1.0\arcsec\ \\
&& \\                           
\hline
&&\\
Per B1-bS               & 26$^{\circ}$          &       21$^{\circ}$    & 26$^{\circ}$            & 21$^{\circ}$          & 22$^{\circ}$          \\
Per B1c                         & 99$^{\circ}$          &       95$^{\circ}$    & 99$^{\circ}$            & 97$^{\circ}$          & 92$^{\circ}$          \\
BHR7-MMS                & 87$^{\circ}$          &       90$^{\circ}$    & 93$^{\circ}$            & 91$^{\circ}$          & 92$^{\circ}$          \\
HH25-MMS                & 74$^{\circ}$                  &       83$^{\circ}$    & 85$^{\circ}$            & 86$^{\circ}$          & -                             \\
HH211-mm                & 174$^{\circ}$                 &       167$^{\circ}$   & 175$^{\circ}$           & 162$^{\circ}$         & 170$^{\circ}$         \\
HH212                   & 51$^{\circ}$          &       59$^{\circ}$    & 56$^{\circ}$            & 57$^{\circ}$          & 47$^{\circ}$          \\
L483                    & 8$^{\circ}$           &       4$^{\circ}$             & 23$^{\circ}$            & 17$^{\circ}$          & 16$^{\circ}$          \\
Serp SMM18              & 84$^{\circ}$          &       83$^{\circ}$    & 84$^{\circ}$            & 85$^{\circ}$          & 83$^{\circ}$          \\
&&\\
\hline
\end{tabular}
\label{PixelSizeInfluence}
\end{table}
%%%%%%%%%%%%%%%%%%%%%%%%%

\section{Complementary tests on the mean B-field orientation estimates}
\label{Complement}

Other methods were developed to calculate mean polarization angles, in particular, to analyze
single-dish observations of dust polarized emission. We provide here the B-field position angle 
obtained from three other averaging methods. 
Table~\ref{ComplementB} gathers the various estimates derived.\\

Column (1) provides the (arithmetic, not weighted) mean B-field orientation within the central 1000 au region. 
Column (2) provides the B-field orientation estimated using the technique described in \citet{Li2006}. 
First, we computed $\mathrm{Q' = Q/P_i}$ and $\mathrm{U' = U/P_i}$ for each pixel in which the polarization is detected
in the central 1000 au region, then averaged all values of
$\mathrm{Q'}$ and $\mathrm{U'}$. We then derived the position angle average using $\mathrm{Q'}$, $\mathrm{U'}$
and Eq.~\ref{epPA}. 
Finally, Column (3) provides the B-field position angle obtained when all the Q and U fluxes of each pixel in which the polarization is detected
in the central 1000 au region are summed
separately, and the position angle average was then derived from the two sums. 
These methods, although adapted to compute a mean B orientation in single-dish observations with many independent detections 
with high S/N over a wide range of physical local conditions, may not be well adapted to interferometric data that contain only a few detections 
with large variations in their associated errors and that emanate from rather homogeneous local conditions and large variations in their associated errors.

The reference values we used in this analysis are reported in Table~\ref{AnglesVelGrad}.
The values of the B position angles estimated with the different averaging methods are extremely consistent 
with each other (with a standard deviation of 5$^{\circ}$ between the test and the nominal values), which is expected because the area over which the calculation was performed is small. These tests show that the main component 
of B in the protostellar envelopes we discussed here are, considering the intrinsic limitation of 
the data in hands, robust envelope-scale values, and our handling of the polarization data is therefore meaningful.

%%%%%%% Position angle table %%%%%%%%%%%%
\begin{table}[!ht]
\centering
\caption{Results from alternative methods for the B-field position angles}
\begin{tabular}{cccc}
\hline
\hline
&&\\
Name                    & \multicolumn{3}{c}{B$_{mean}$ P.A.}   \\
&& \\
\cline{2-4} \\
                & Mean  &       \citet{Li2006} &  From $\Sigma$U/$\Sigma$Q \\
                & (1) & (2) & (3) \\
&& \\                           
\hline
&&\\
Per B1-bS                       & 26.1$^{\circ}$        & 23.6$^{\circ}$                & 25.9$^{\circ}$                          \\
Per B1c                                 & 99.6$^{\circ}$        & 92.0$^{\circ}$         &  90.3$^{\circ}$\\   
B335                            & 54.7$^{\circ}$        & 55.0$^{\circ}$         &  54.7$^{\circ}$\\                             
BHR7-MMS                        & 84.0$^{\circ}$        & 84.8$^{\circ}$         &  86.4$^{\circ}$\\   
CB230                                   & 84.7$^{\circ}$        & 85.7$^{\circ}$         &  84.6$^{\circ}$\\     
HH25-MMS                        & 77.4$^{\circ}$        & 78.8$^{\circ}$         &  75.6$^{\circ}$\\   
HH211-mm                        & 173.3$^{\circ}$       & 173.9$^{\circ}$         &  173.4$^{\circ}$\\   
HH212                           & 51.1$^{\circ}$        & 59.8$^{\circ}$         &  56.6$^{\circ}$\\   
HH797                                   & 110.4$^{\circ}$       & 110.5$^{\circ}$         &  110.4$^{\circ}$\\    
IRAS03282                       & 42.2$^{\circ}$        & 44.7$^{\circ}$         &  42.4$^{\circ}$\\     
IRAS16293-A                     & 139.3$^{\circ}$       & 174.0$^{\circ}$         &  173.3$^{\circ}$\\    
L1157                                   & 149.4$^{\circ}$       & 148.2$^{\circ}$         &  149.0$^{\circ}$\\    
L1448C                                  & 95.0$^{\circ}$        & 94.9$^{\circ}$         &  95.0$^{\circ}$\\     
L1448N-B                        & 22.8$^{\circ}$        & 23.1$^{\circ}$         &  22.8$^{\circ}$\\                     
L483-mm                         & 7.9$^{\circ}$                 & 7.3$^{\circ}$                 &  7.9$^{\circ}$\\   
NGC~1333 IRAS4A                 & 53.4$^{\circ}$        & 57.3$^{\circ}$         &  54.0$^{\circ}$\\     
NGC~1333 IRAS4B         & 46.9$^{\circ}$        & 48.9$^{\circ}$        &  49.4$^{\circ}$\\       
SSMM18                          & 84.7$^{\circ}$        &  79.9$^{\circ}$         & 84.0$^{\circ}$ \\   
SVS13-B                                 & 20.0$^{\circ}$        & 20.8$^{\circ}$         &  19.1$^{\circ}$\\             

&&\\
\hline
\end{tabular}
\label{ComplementB}
\end{table}
%%%%%%%%%%%%%%%%%%%%%%%%%

\clearpage
\section{Polarization intensity and fraction maps}

%%%%%%%%%%%% Stokes Q, U, poli, pfrac maps %%%%%%%%%%%%%%
\begin{figure*}[!h]
\vspace{10pt}
\begin{tabular}{p{5.7cm}p{5.7cm}p{5.7cm}}
\hspace{2.5cm}{\Large \bf B1-b} & \hspace{2.5cm}{\Large \bf B1-c} & \hspace{2cm}{\Large \bf BHR7-MMS}\\
\hspace{-10pt}\vspace{-15pt}\includegraphics[width=5cm, angle=-90]{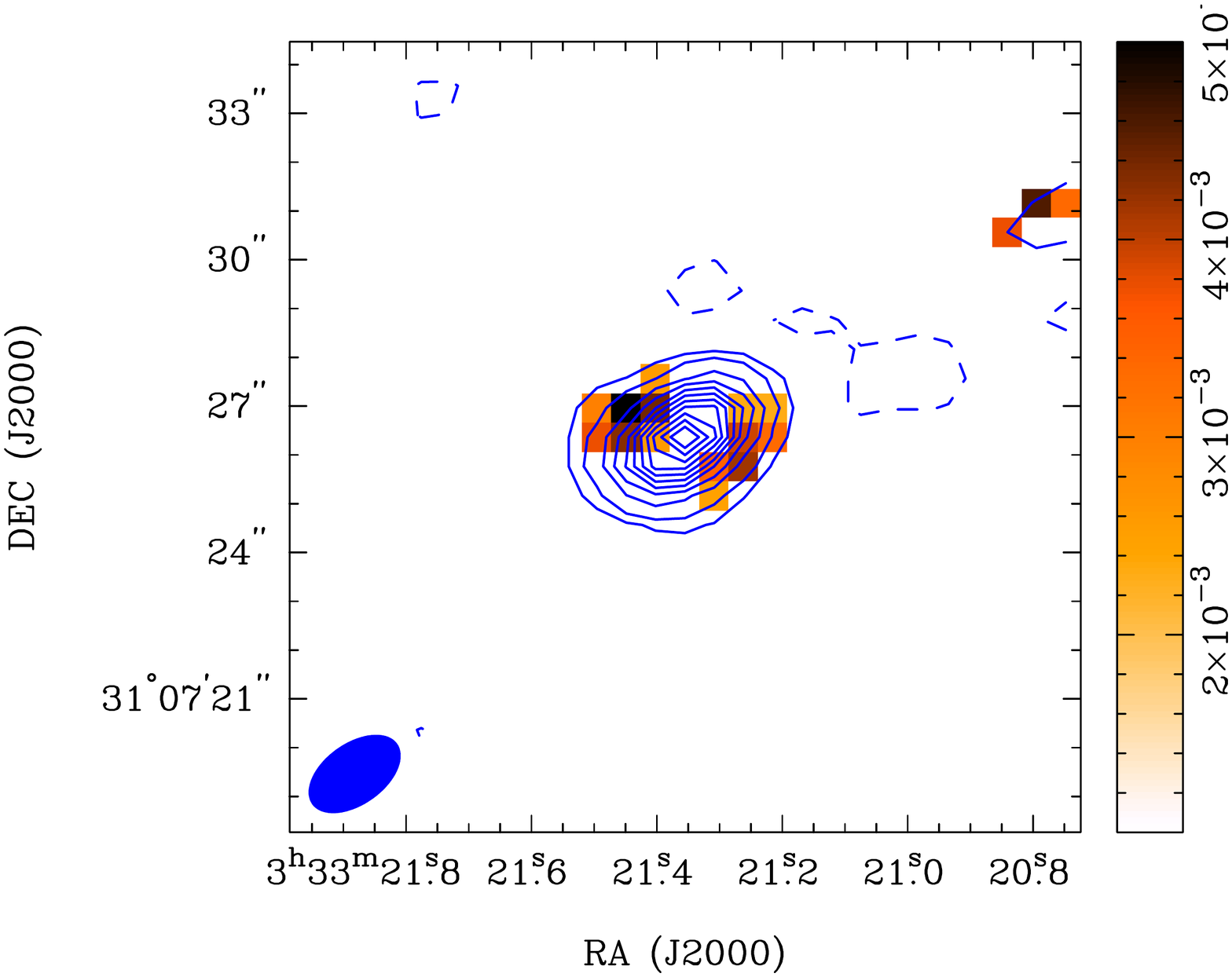} &
\hspace{-10pt}\vspace{-15pt}\includegraphics[width=5cm, angle=-90]{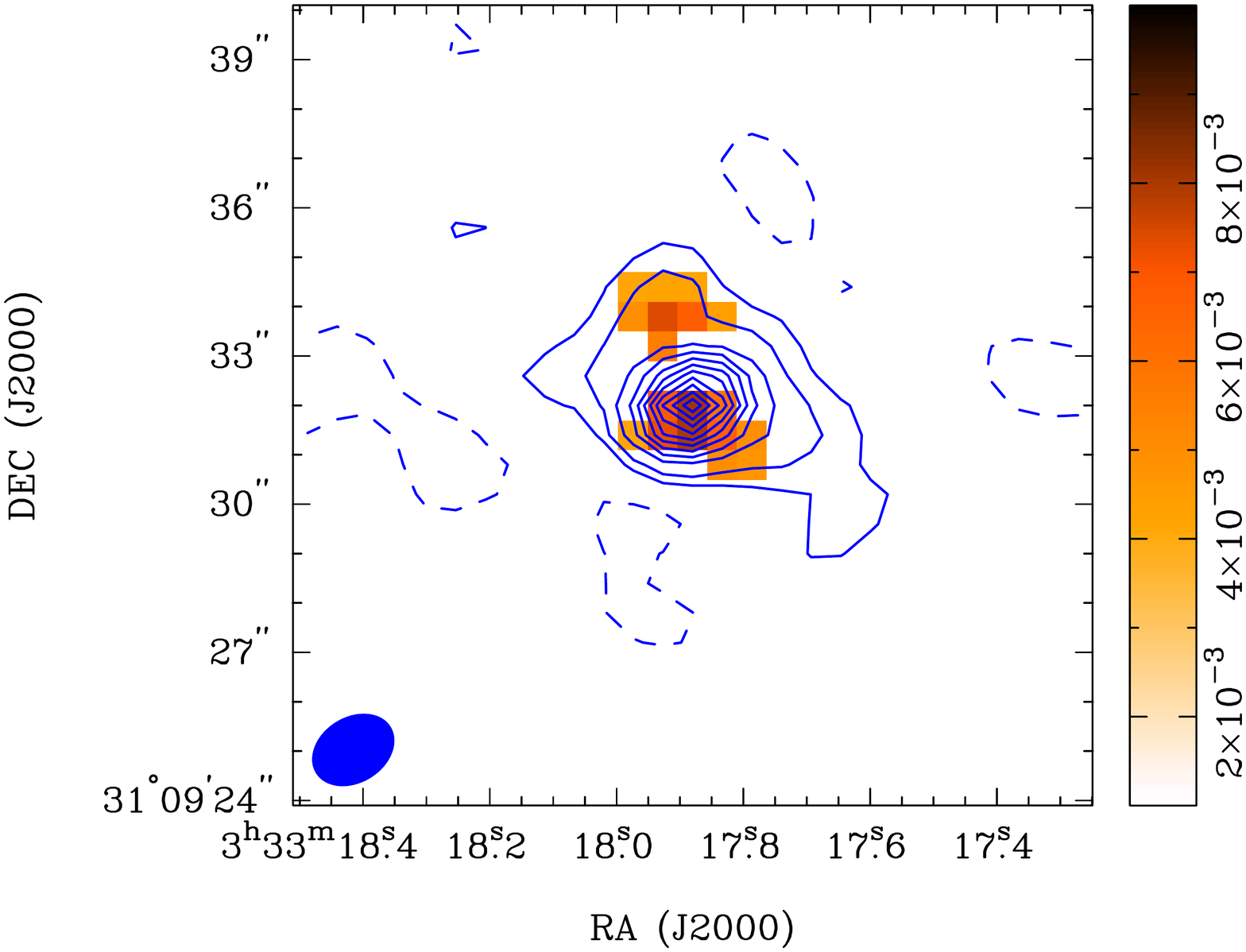} &
\hspace{-10pt}\vspace{-15pt}\includegraphics[width=5cm, angle=-90]{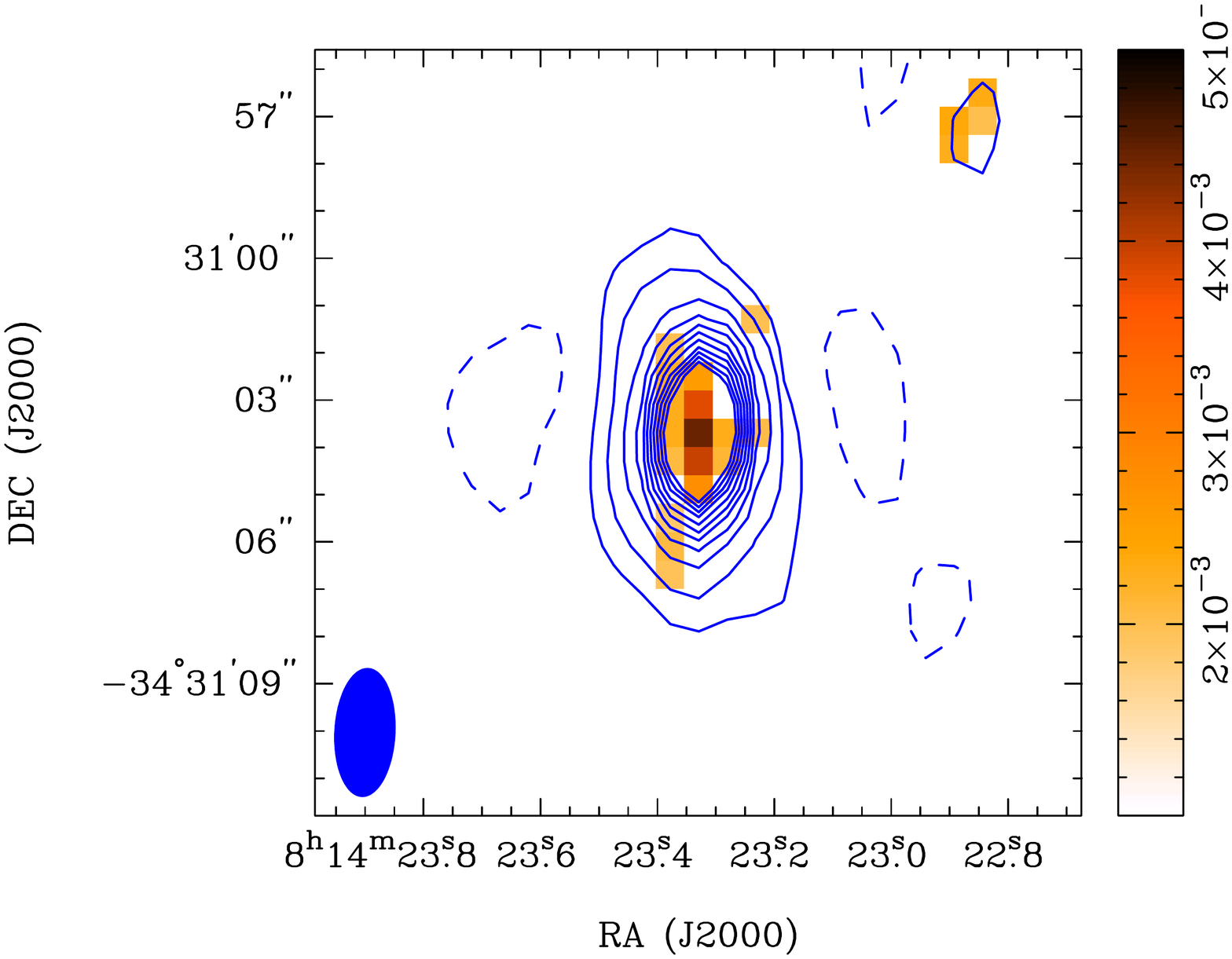} \\
\hspace{-10pt}\vspace{-15pt}\includegraphics[width=5cm, angle=-90]{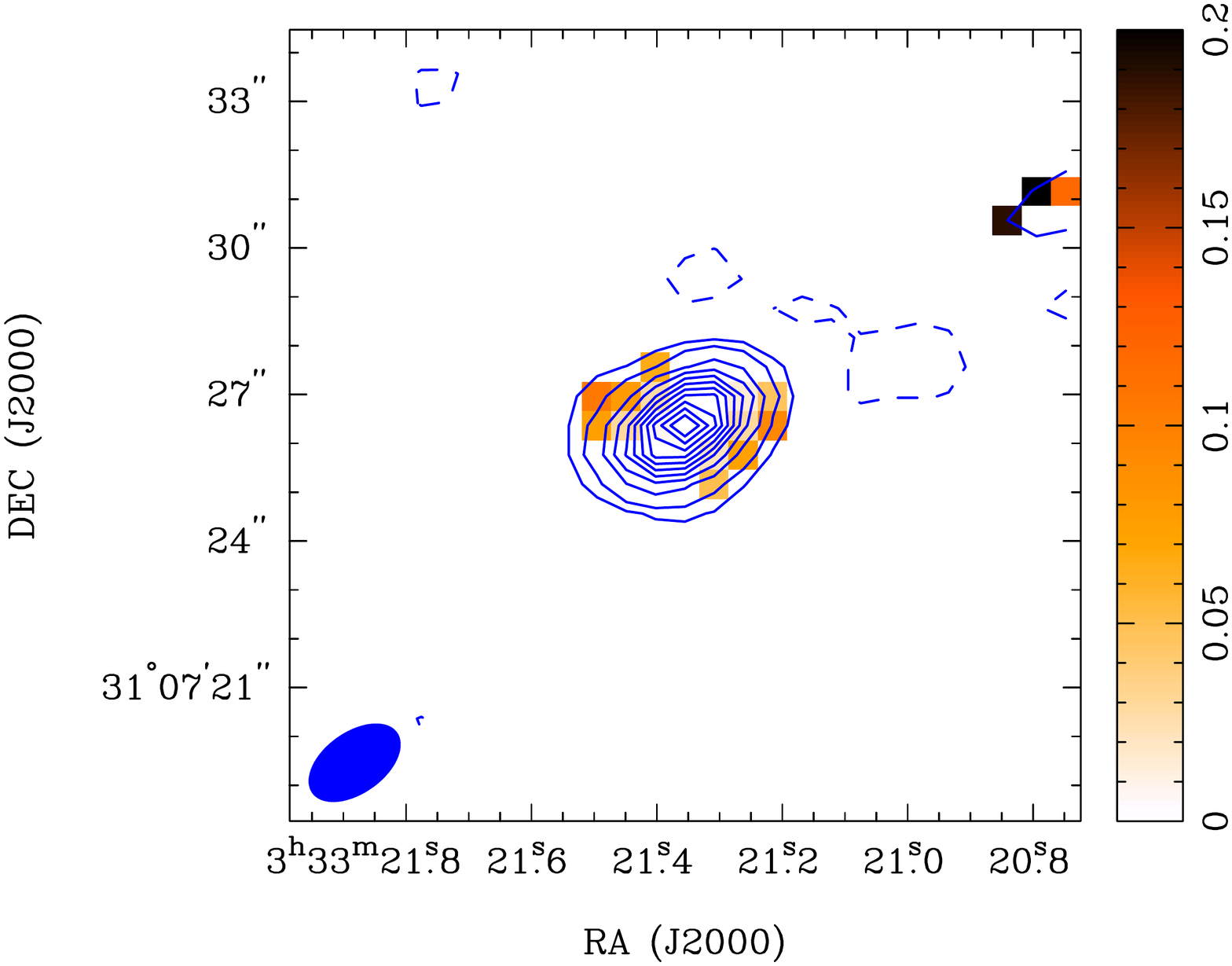} &
\hspace{-10pt}\vspace{-15pt}\includegraphics[width=5cm, angle=-90]{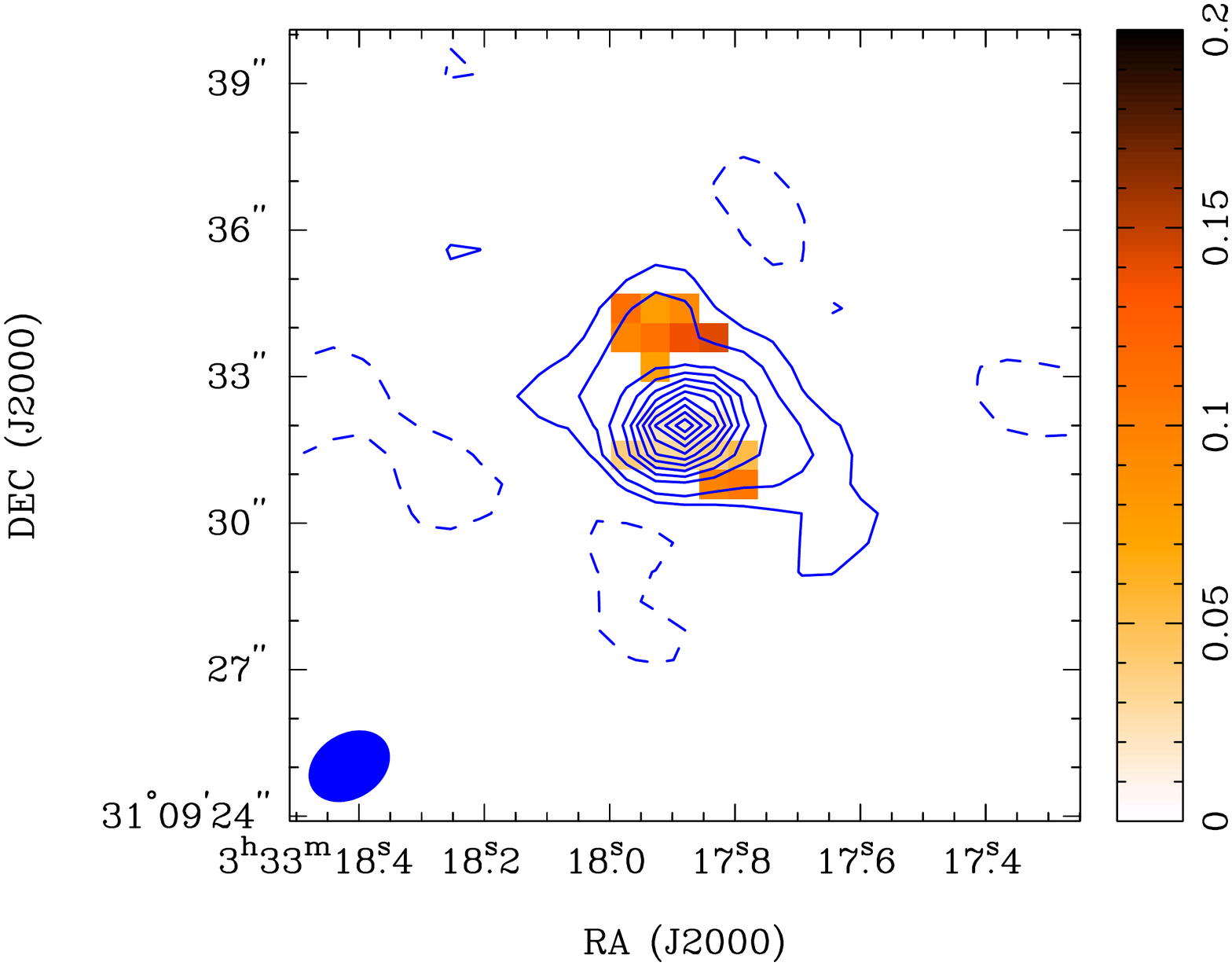} &
\hspace{-10pt}\vspace{-15pt}\includegraphics[width=5cm, angle=-90]{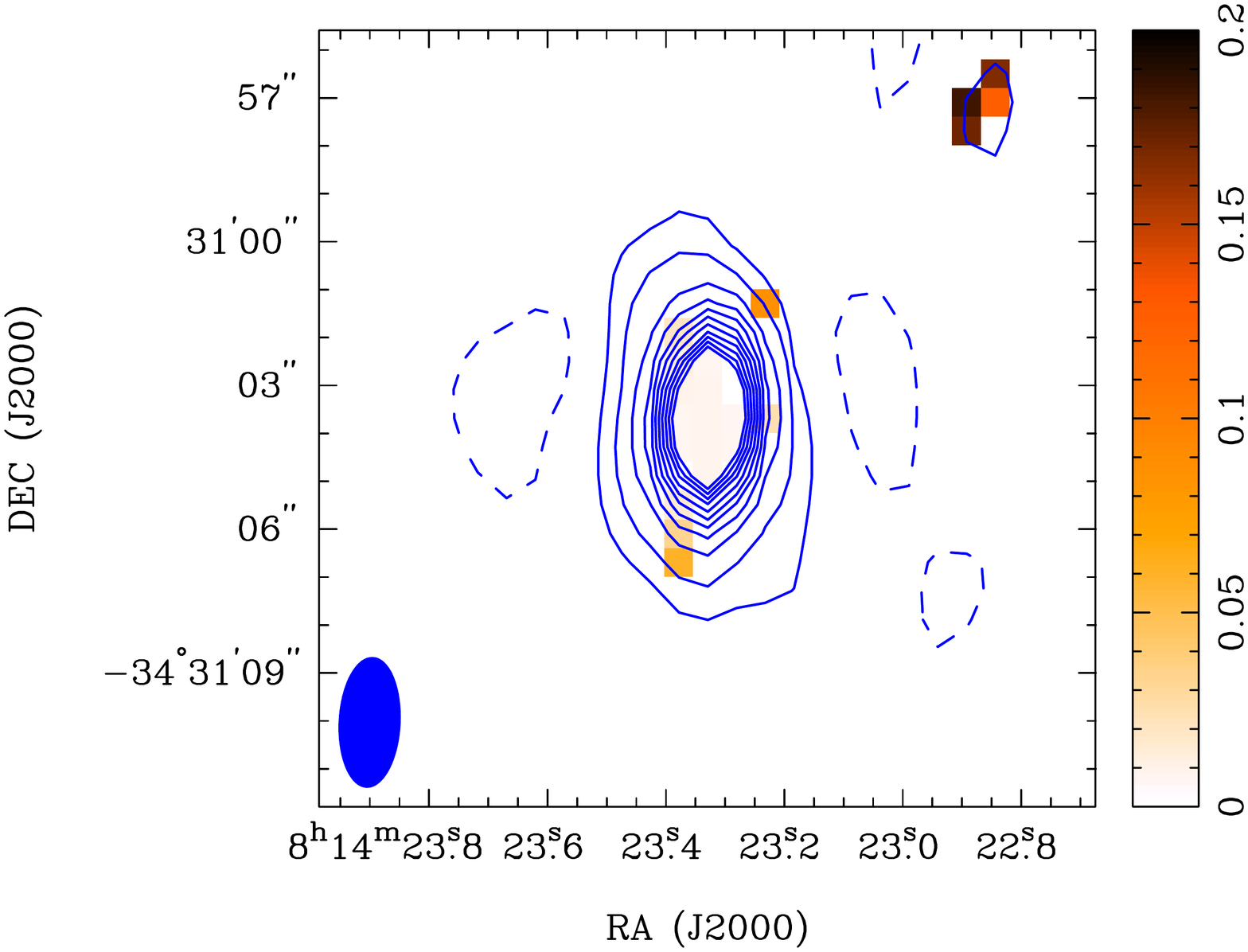} \\
&\\
&\\
&\\
\hspace{2.5cm}{\Large \bf HH25-MMS} & \hspace{2cm}{\Large \bf HH211-mm} & \hspace{2.3cm}{\Large \bf HH212}\\
\hspace{-10pt}\vspace{-15pt}\includegraphics[width=5cm, angle=-90]{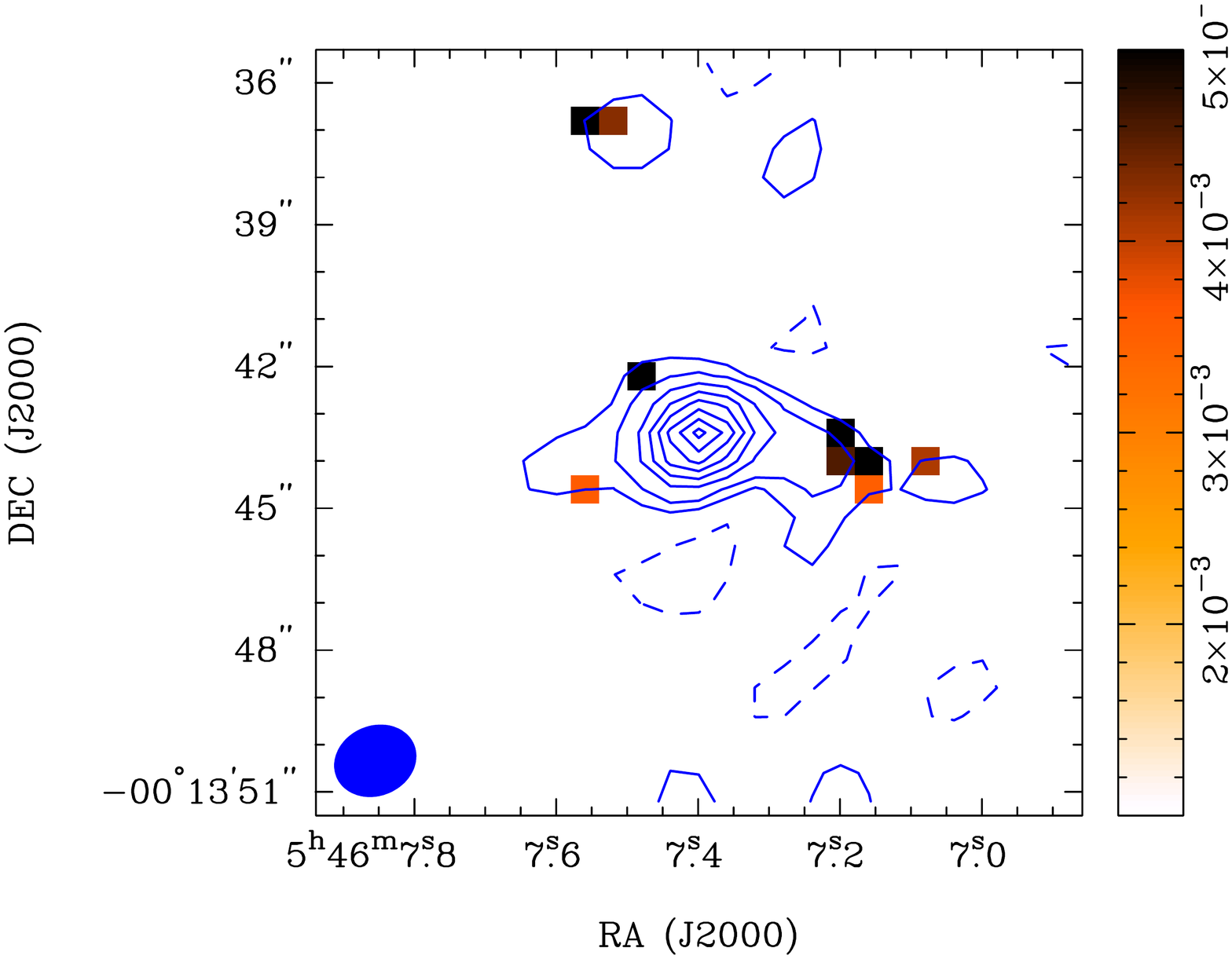} &
\hspace{-10pt}\vspace{-15pt}\includegraphics[width=5cm, angle=-90]{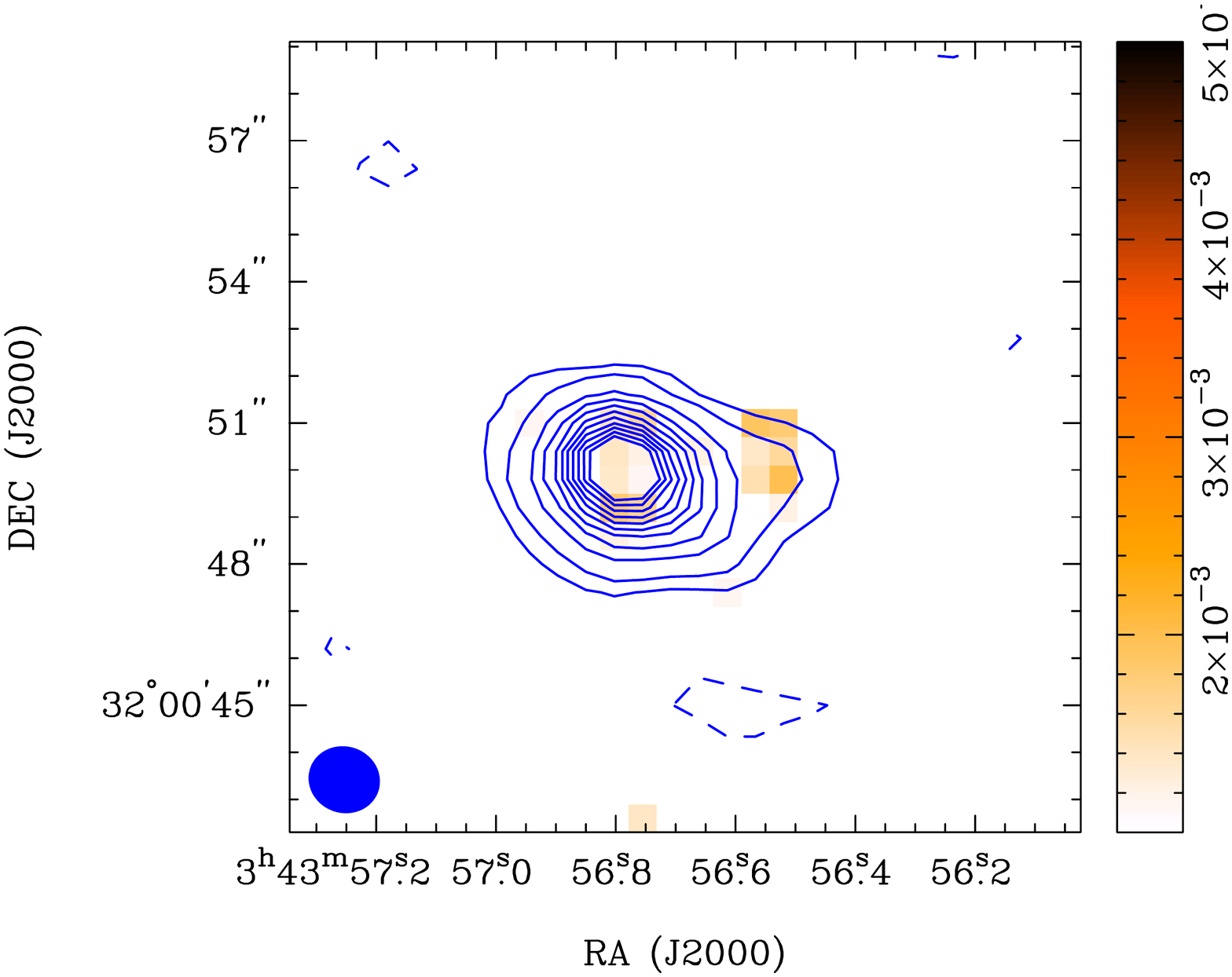} &
\hspace{-10pt}\vspace{-15pt}\includegraphics[width=5cm, angle=-90]{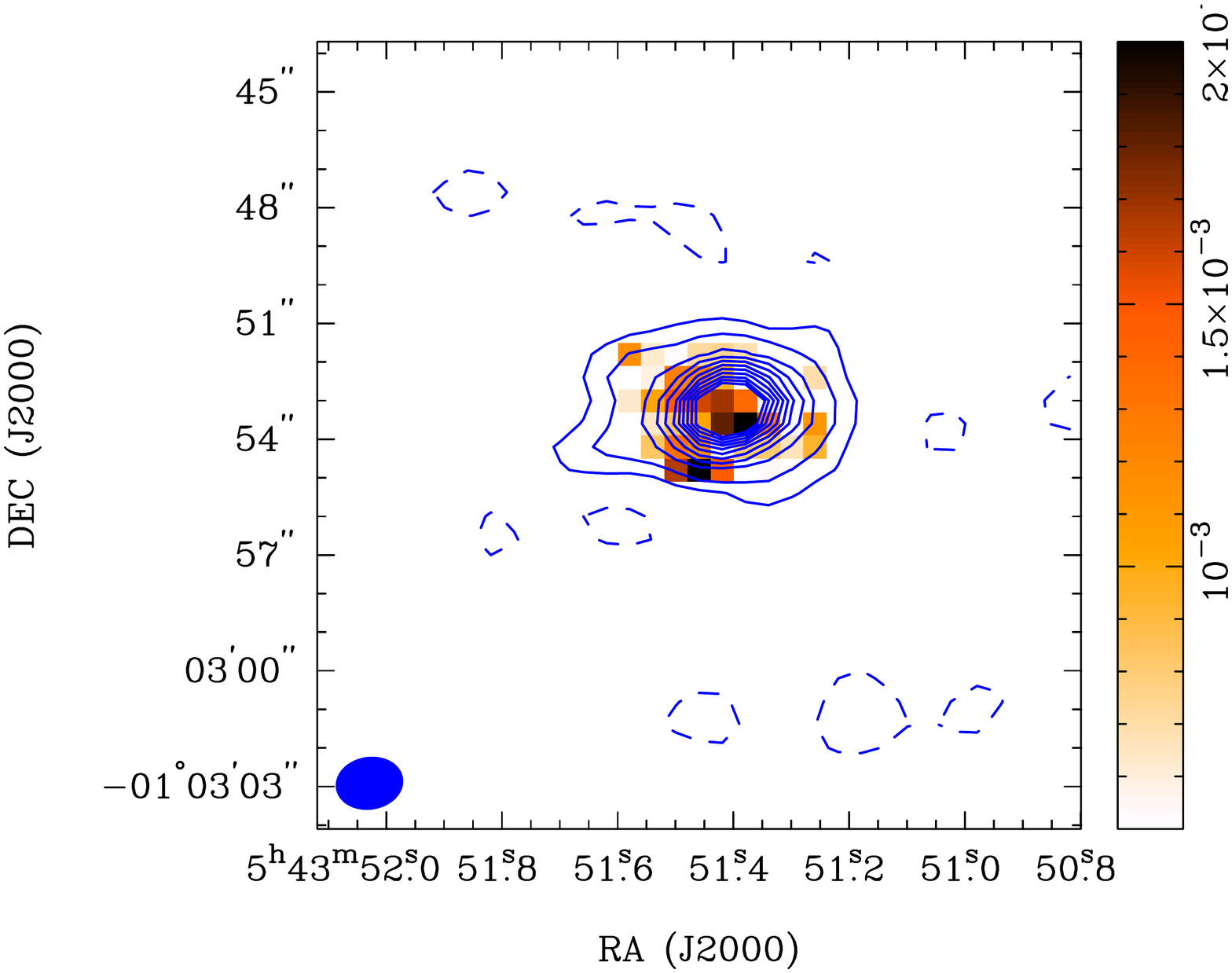} \\
\hspace{-10pt}\vspace{-15pt}\includegraphics[width=5cm, angle=-90]{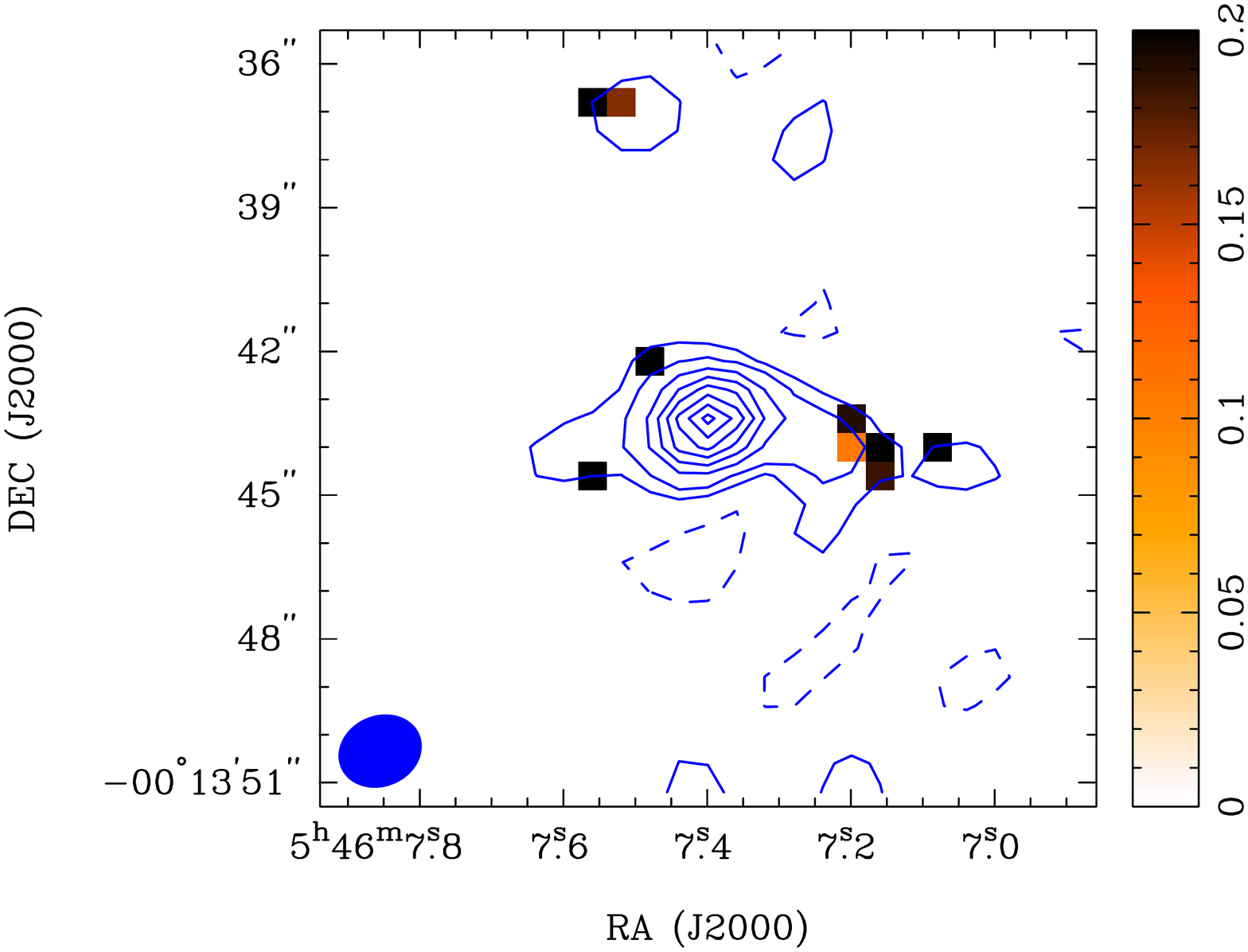} &
\hspace{-10pt}\vspace{-15pt}\includegraphics[width=5cm, angle=-90]{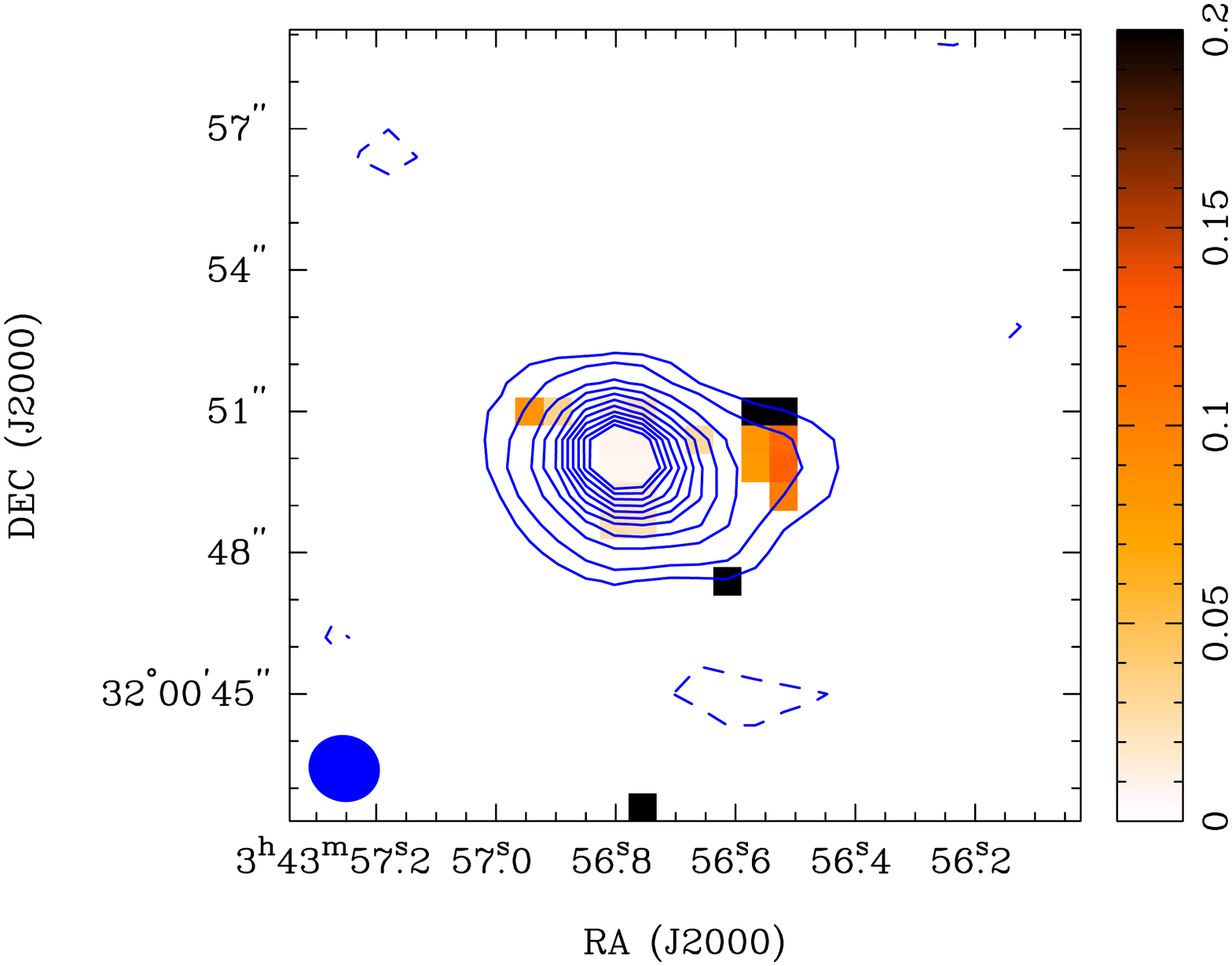} &
\hspace{-10pt}\vspace{-15pt}\includegraphics[width=5cm, angle=-90]{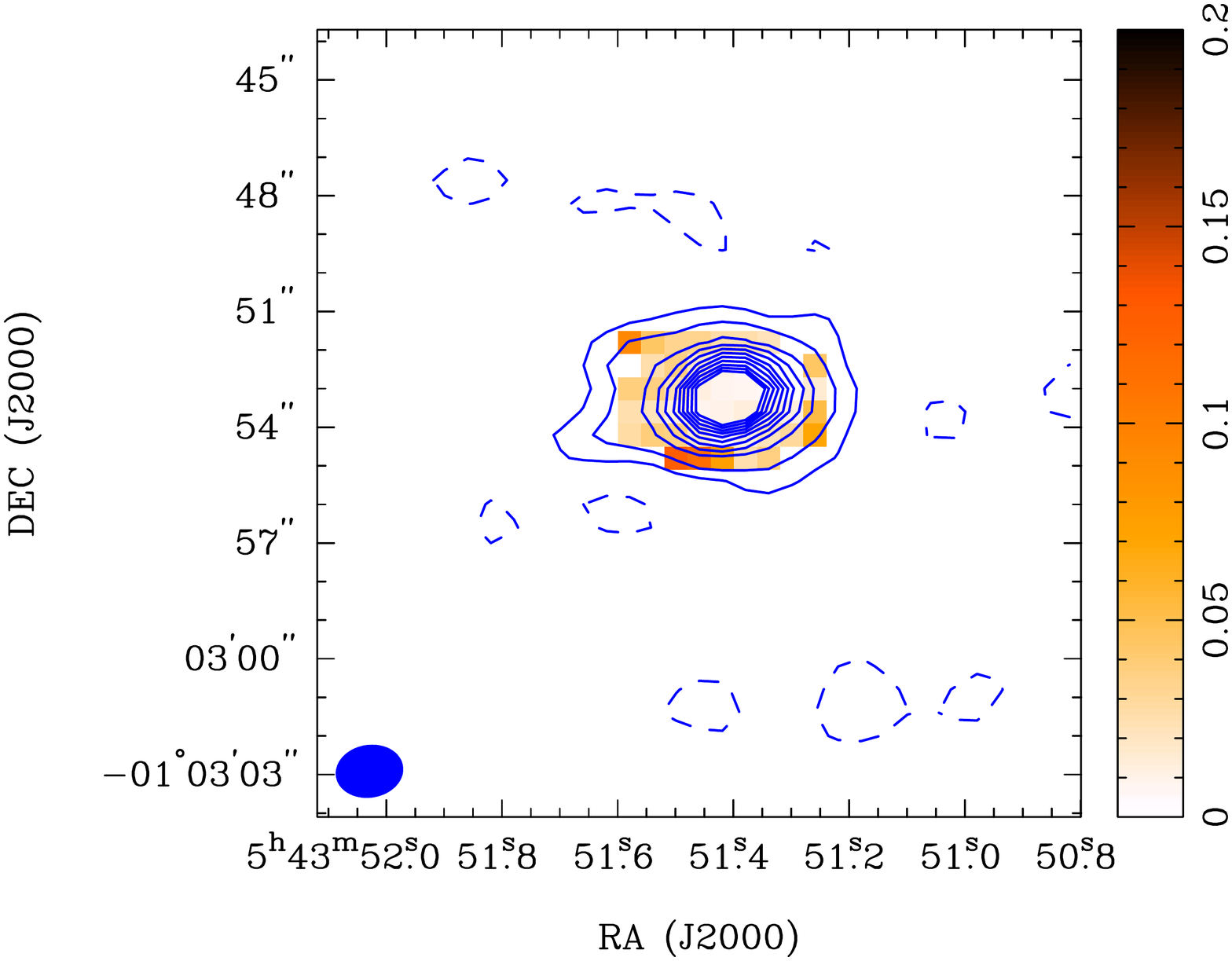} \\
& \\
\end{tabular}
\vspace{10pt}
\caption{
For each source, the {\it first row} shows the polarization intensity maps of the sample derived from the Stokes Q and U maps 
obtained with the SMA at 850 \mic\ The color scale is in Jy/beam. 
The Stokes I contours at -3, 5, 10, 20, 30, 40, 50, 60, 70, 80, 90, and 100 $\sigma$ are overlaid in blue. 
The filled ellipses in the lower left corner indicate the synthesized beam of the SMA maps. 
Their sizes are reported in Table~\ref{SMAmapsrms}. 
{\it Second row:} Polarization fraction map. }
\label{PolaMaps}
\vspace{10pt}
\end{figure*}
\addtocounter {figure}{-1}
\newpage
\begin{figure*}
\vspace{10pt}
\begin{tabular}{p{5.7cm}p{5.7cm}p{5.7cm}}
\hspace{2.5cm}{\Large \bf L483} & \hspace{2cm}{\Large \bf SSMM18} &\\
\hspace{-10pt}\vspace{-15pt}\includegraphics[width=5cm, angle=-90]{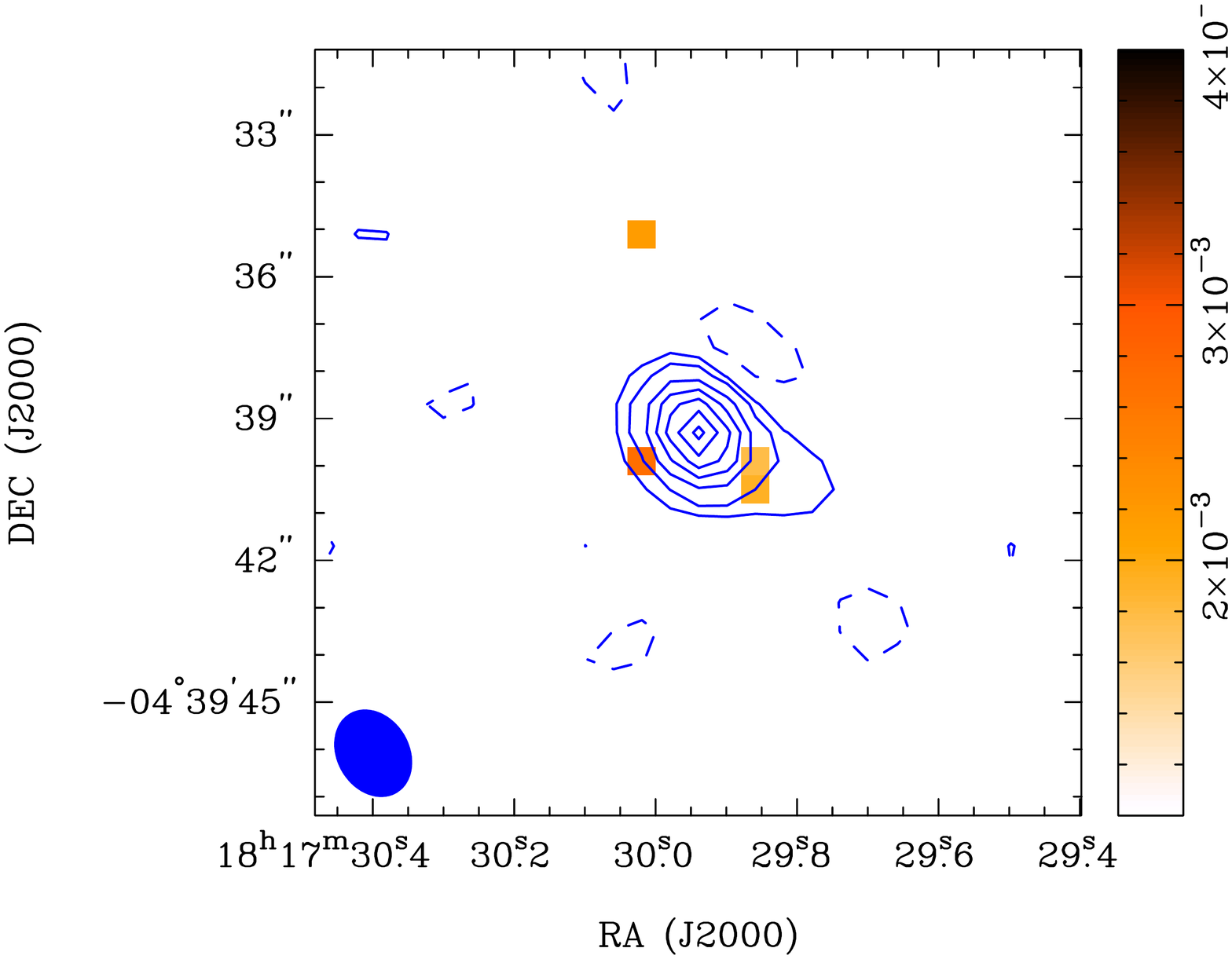} &
\hspace{-10pt}\vspace{-15pt}\includegraphics[width=5cm, angle=-90]{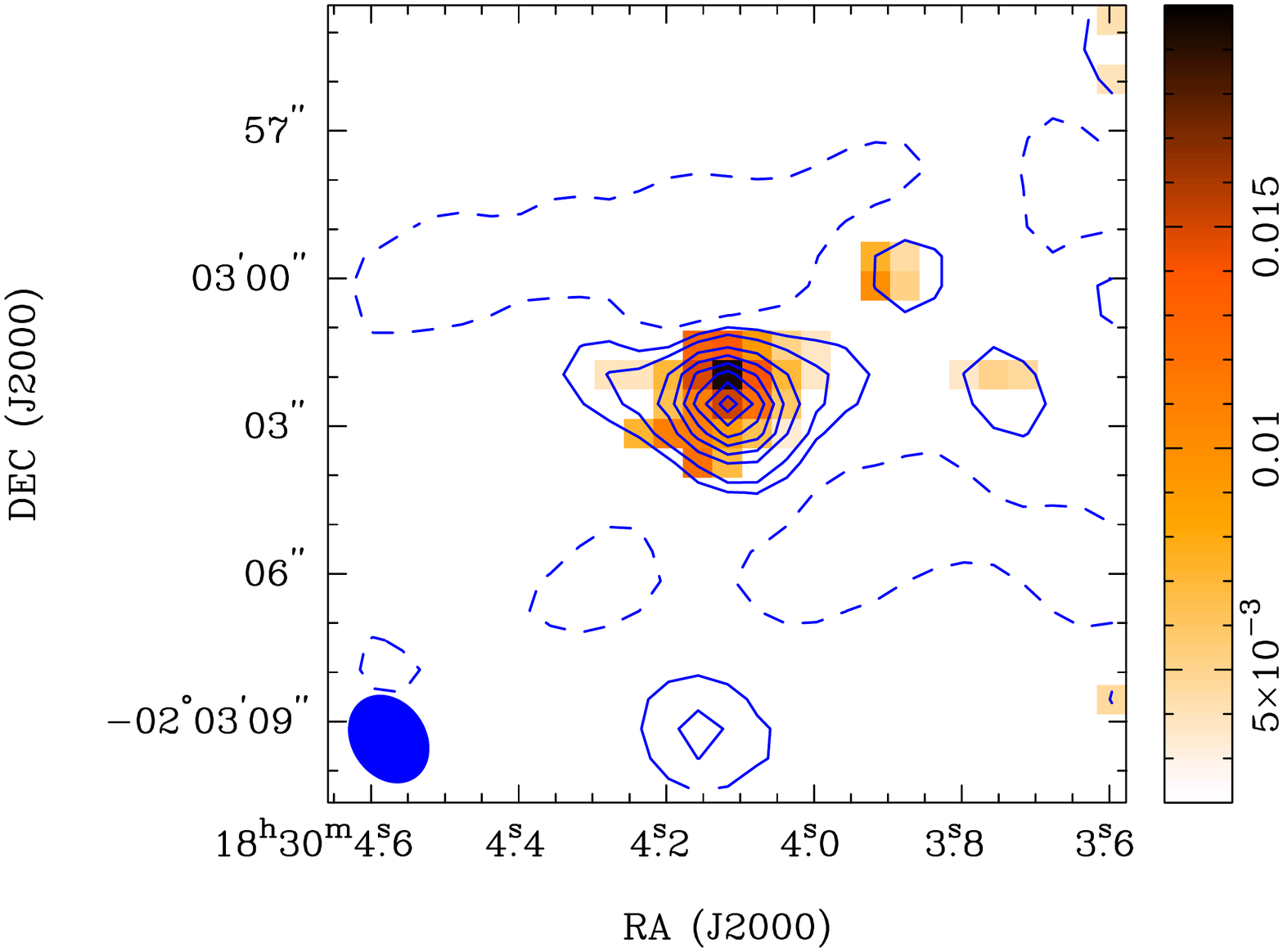} & \\
& \\
\hspace{-10pt}\vspace{-15pt}\includegraphics[width=5cm, angle=-90]{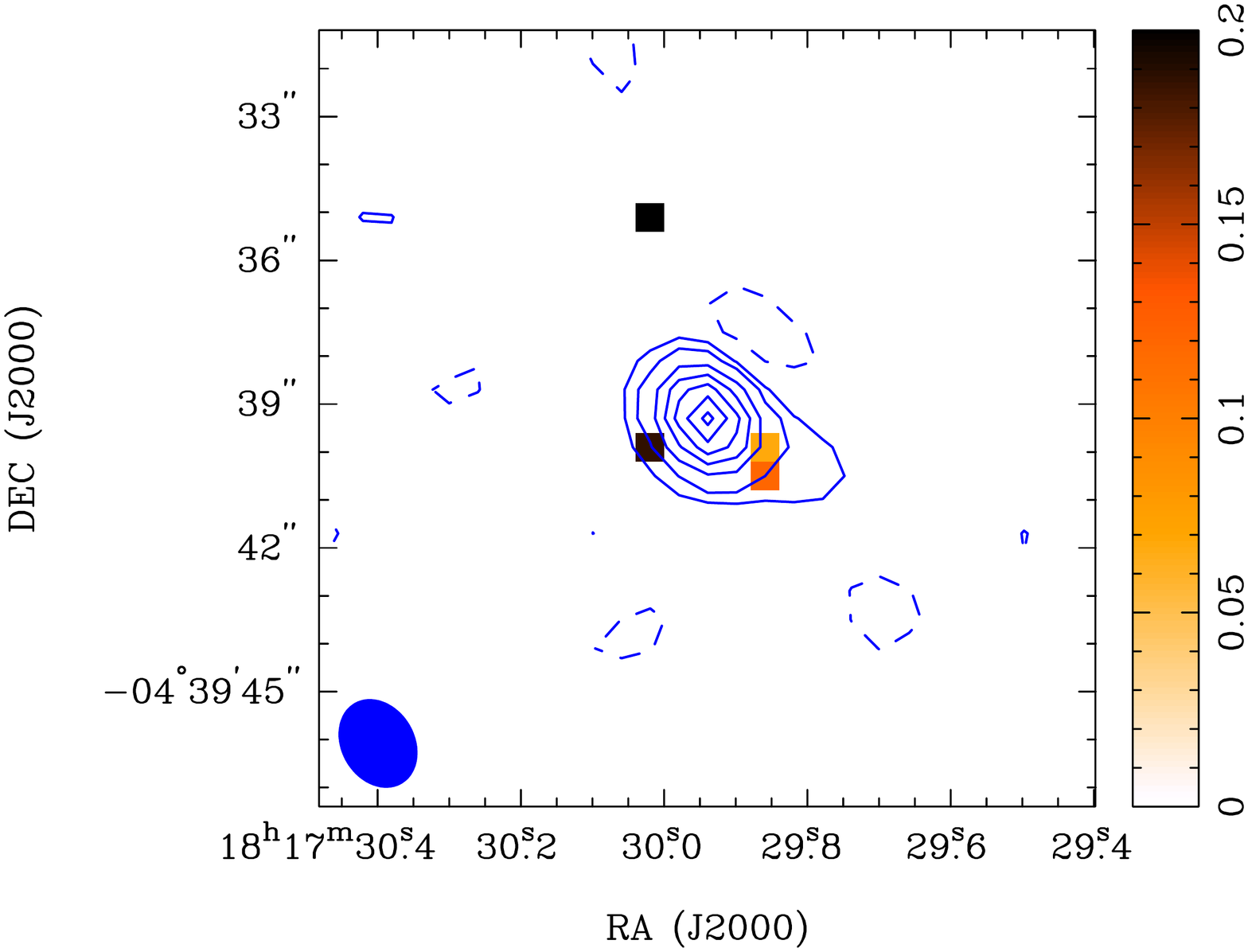} &
\hspace{-10pt}\vspace{-15pt}\includegraphics[width=5cm, angle=-90]{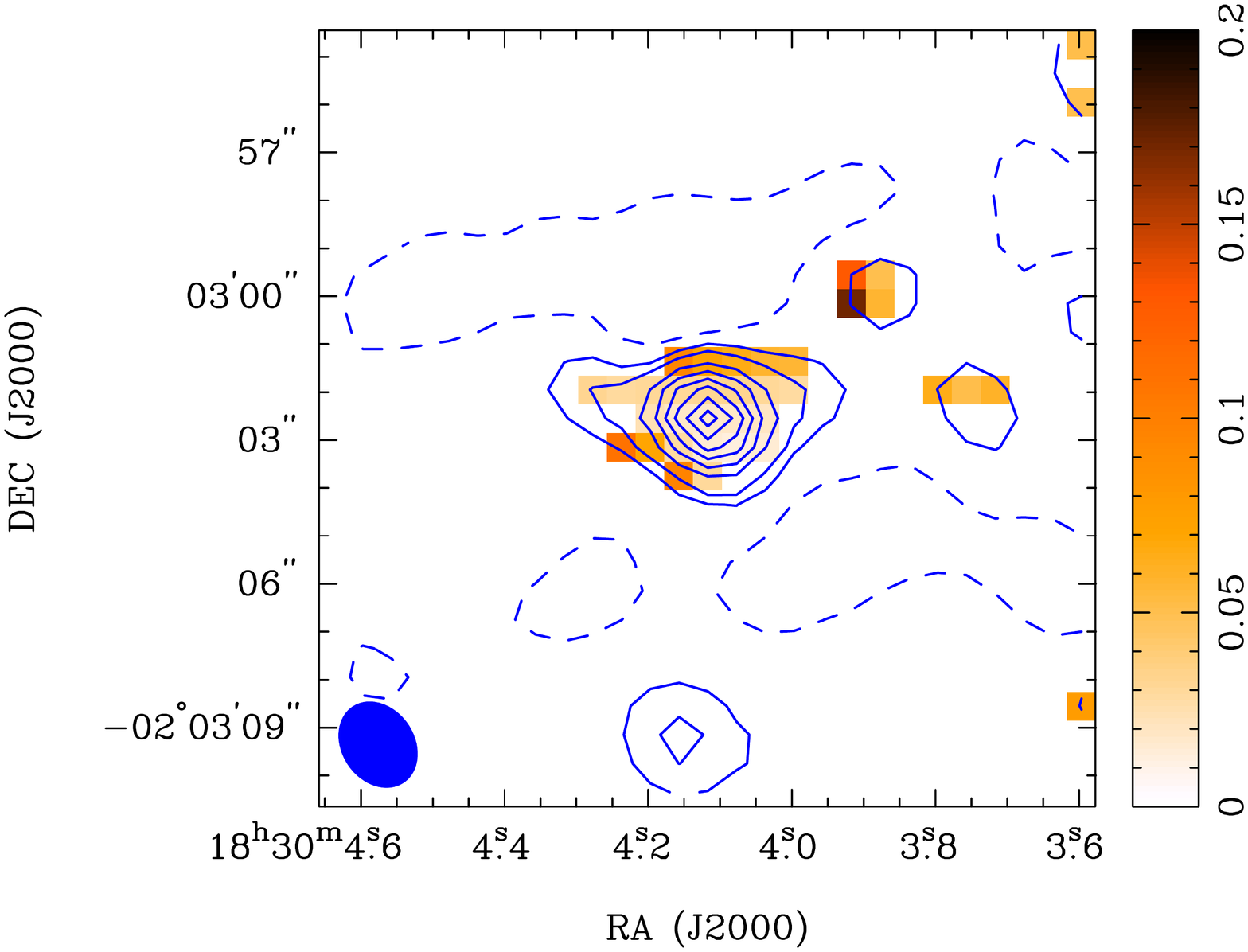} & \\
& \\
\end{tabular}
\vspace{10pt}
\caption{continued. }
\vspace{10pt}
\end{figure*}
%%%%%%%%%%%% Stokes Q, U, poli, pfrac maps %%%%%%%%%%%%%%

\newpage
\clearpage
\section{Velocity maps}

%%%%%%%%%%%%  Vel maps  %%%%%%%%%%%%%%
\begin{figure}[!h]
\centering
\vspace{20pt}
\begin{tabular}{cc}
\hspace{0.5cm}{\Large \bf B1-b} {\large \bf (N$_2$D$^+$)}& \hspace{0.5cm}{\Large \bf B1-c} {\large \bf (N$_2$H$^+$)} \\
\includegraphics[width=7.5cm]{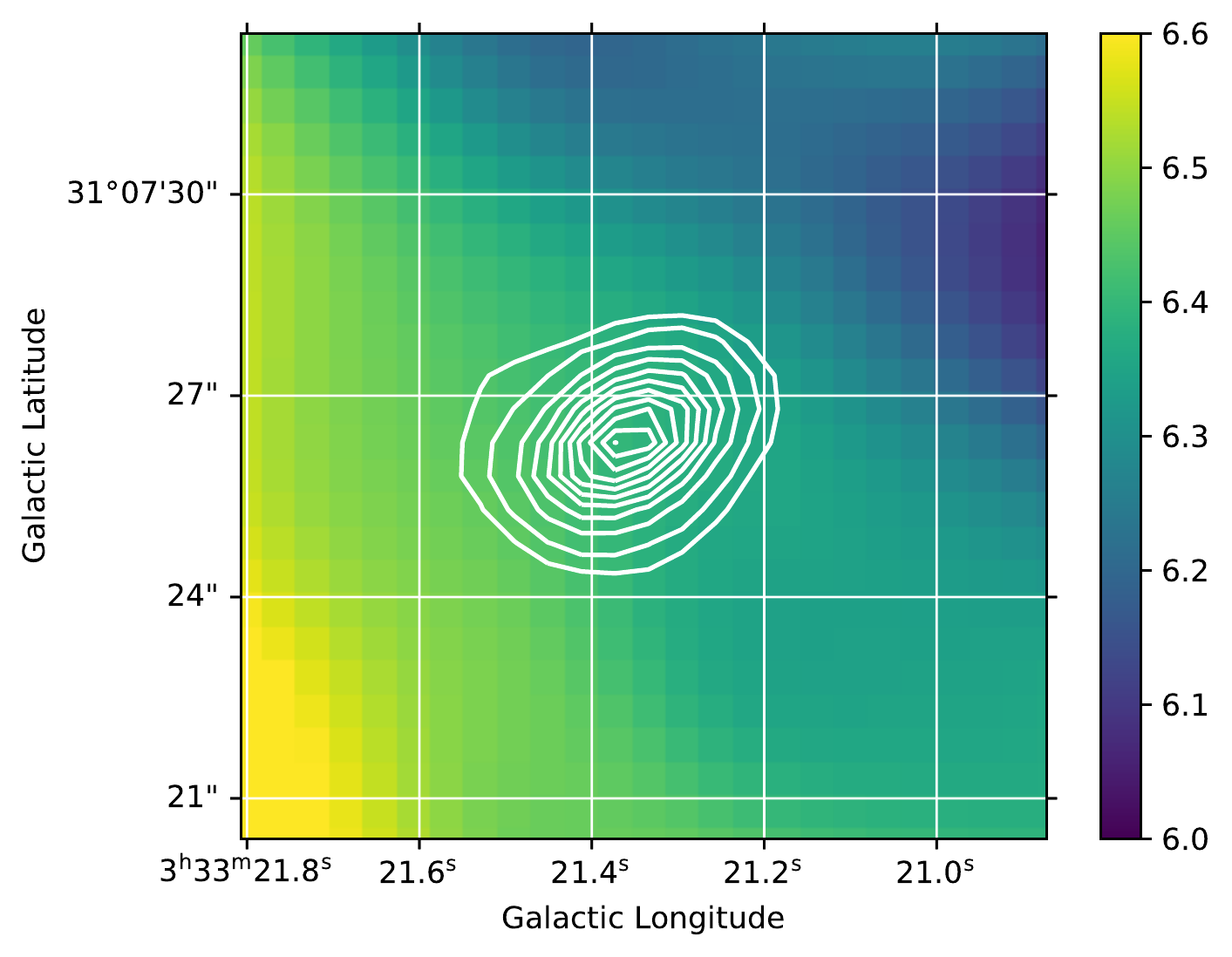} & 
\hspace{-10pt}\includegraphics[width=7.5cm]{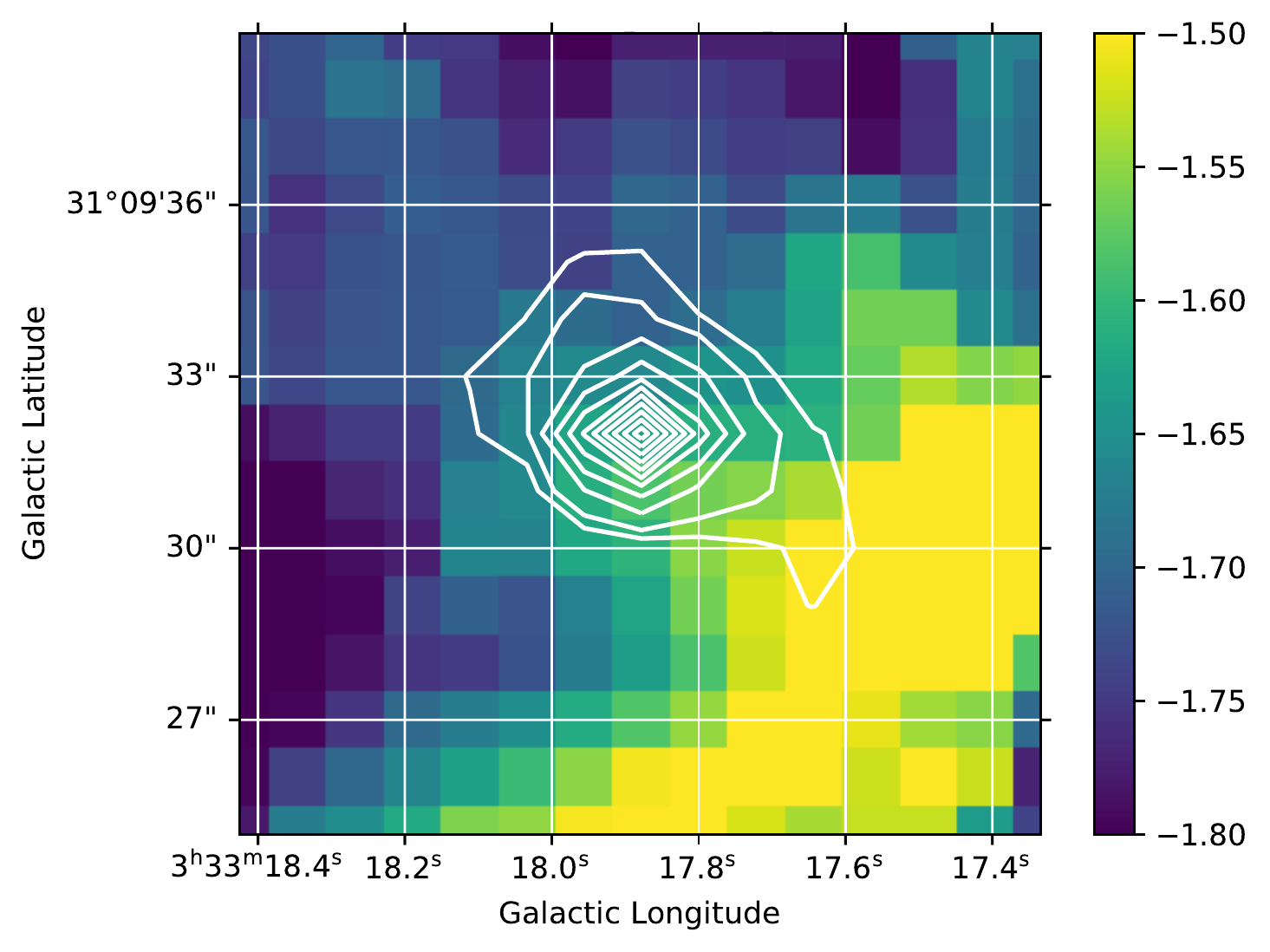} \\
&\\
\hspace{0.5cm}{\Large \bf BHR7-MMS} {\large \bf (N$_2$D$^+$)} & \hspace{0.5cm}{\Large \bf SSMM18 {\large \bf (N$_2$H$^+$)}} \\
\includegraphics[width=7.5cm]{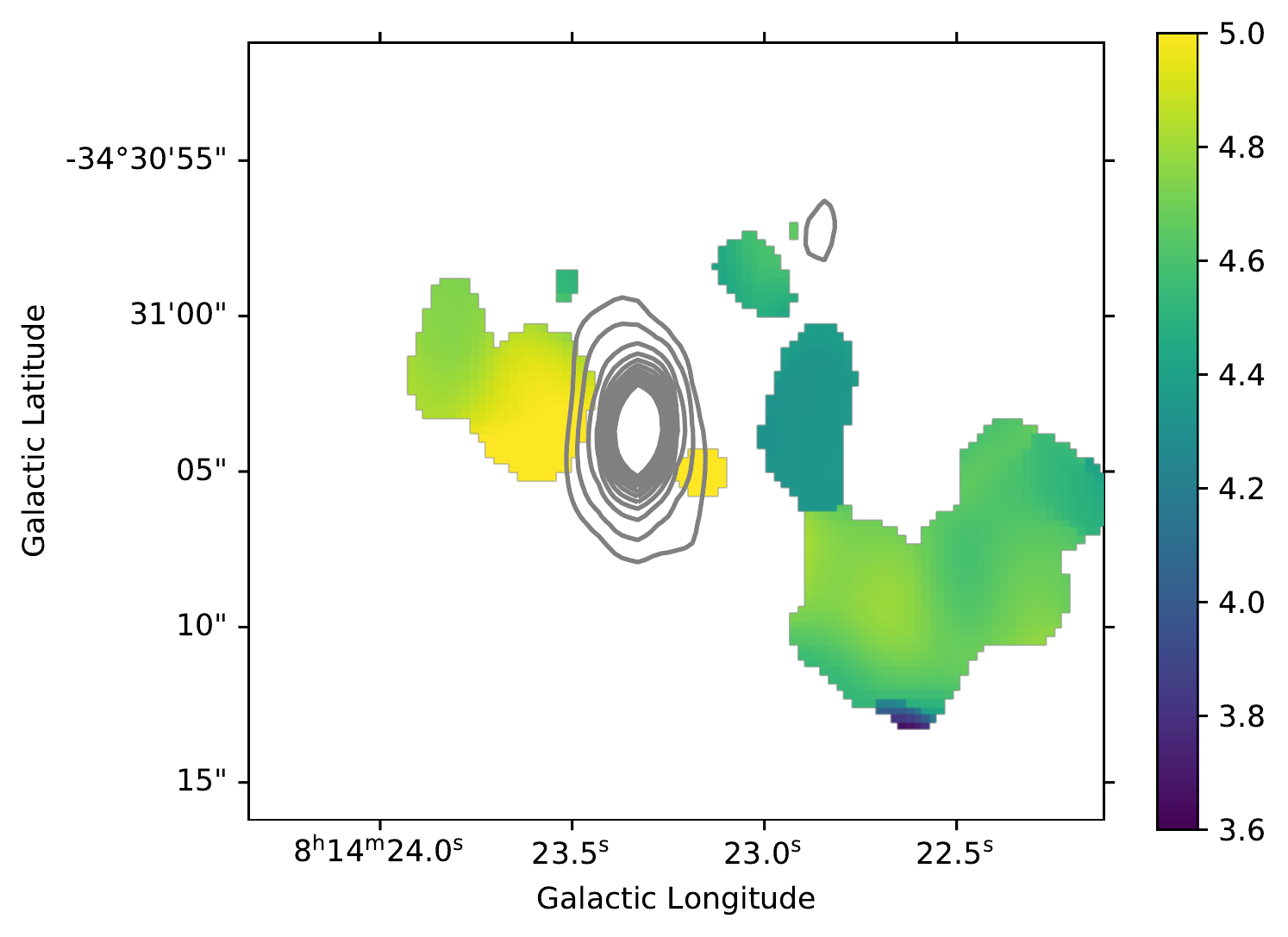} &
\hspace{-10pt}\includegraphics[width=7.5cm]{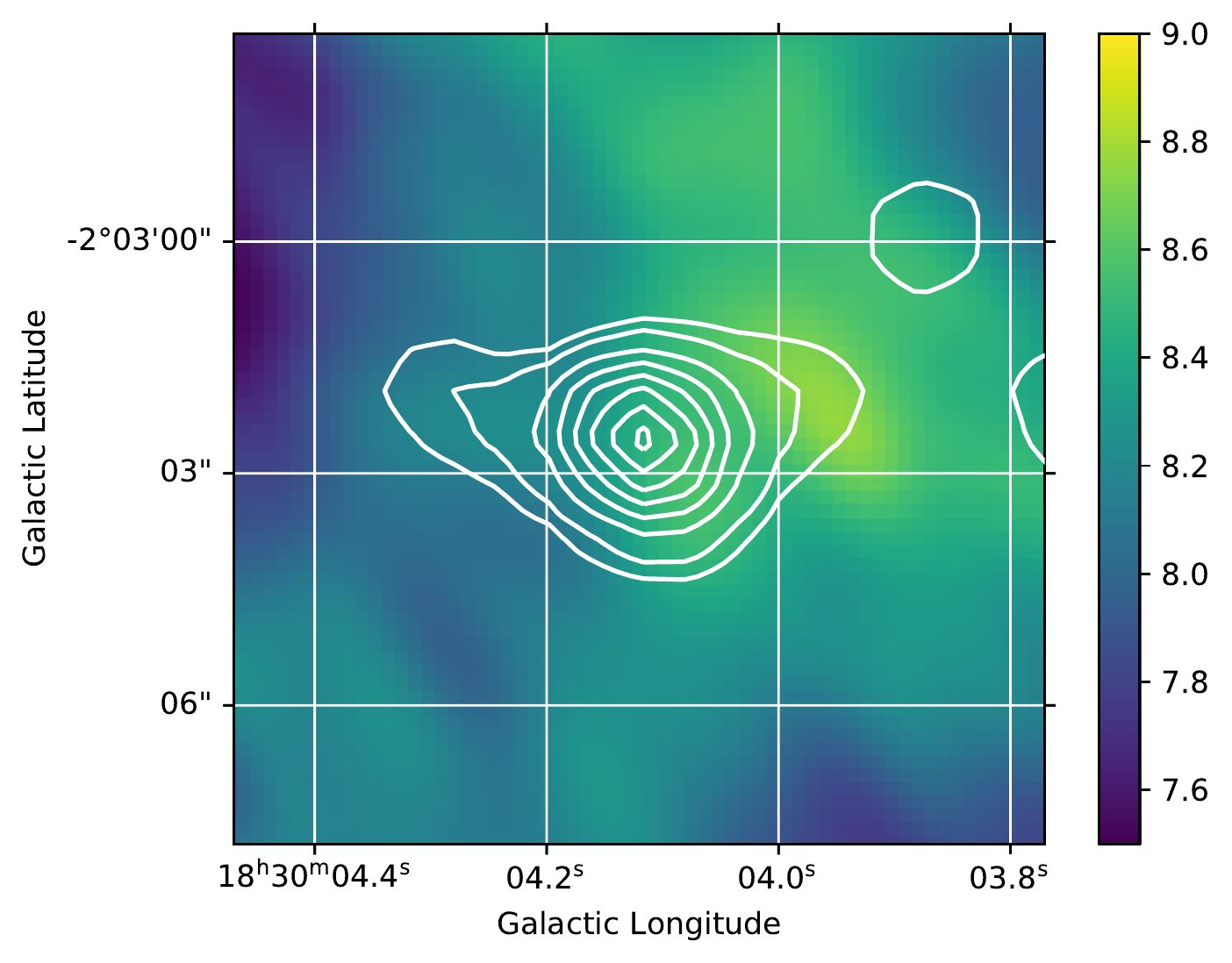}  \\
\end{tabular}
\caption{Velocity maps derived from N$_2$H$^+$ for B1-c and SSMM18,  N$_2$D$^+$ for B1-b and BHR7. 
Data are presented in \citet{Huang2013}, \citet{Matthews2008}, and \citet{Tobin2018} for B1-b, B1-c, and BHR7, respectively. 
The Serpens South MM18 data come from the PdBI CALYPSO survey.}
\label{VelocityRecalculated}
\end{figure}
%%%%%%%%%%%% Vel maps %%%%%%%%%%%%%%

\end{document}